\documentclass[3p,times]{elsarticle}

\usepackage[T1]{fontenc}
\usepackage[utf8]{inputenc}

\usepackage{amssymb}
\usepackage{amsthm}
\usepackage{graphicx}
\usepackage{amsmath,amsbsy}
\usepackage{calligra,calrsfs,mathrsfs}
\usepackage{multirow}
\usepackage{ulem,cancel}
\usepackage{cases}
\usepackage{pgfplots}
\pgfplotsset{compat=1.18}
\usepackage{subcaption}
\usepackage{tikz}
\usepackage{tkz-euclide}
\usetikzlibrary[patterns]
\usepackage{dsfont}
\usepackage{caption}
\usepackage{epstopdf}
\usepackage{epsfig}
\usepackage{array}
\newcolumntype{P}[1]{>{\centering\arraybackslash}p{#1}}
\usepackage{enumerate}
\usepackage{float}
\usepackage{geometry}
\usepgfplotslibrary{fillbetween}
\usetikzlibrary{patterns}
\usepackage{tikz-dimline}
\usepackage{color}
\usepackage{setspace}
\usepackage{textcomp}
\usepackage{soul}
\usepackage[inline]{enumitem}
\usepackage{colortbl} 
\biboptions{sort&compress}
\usepackage{dirtytalk}
\usepackage{csvsimple}

\pgfplotsset{
/pgfplots/bar cycle list/.style={/pgfplots/cycle list={%
    {blue,fill=blue!30!white,mark=none},%
    {red,fill=red!30!white,mark=none},%
    {brown!60!black,fill=brown!30!white,mark=none},%
    {black,fill=gray,mark=none},%
    {violet!80!black,fill=violet,mark=none},%
    {orange,fill=orange!50!white,mark=none}%
    }
},
}
\usepackage[noend]{algpseudocode}
\usepackage{algorithm}
\usepackage{pgfplotstable}
\usepackage{booktabs}
\usepackage{csvsimple}

\usepackage{hyperref}
\hypersetup{
    colorlinks,
    linkcolor={magenta},
    citecolor={blue},
    urlcolor={blue!80!black},
    breaklinks=true,
	plainpages=true
}

\usetikzlibrary{shapes.geometric, arrows, chains, positioning}

\newcommand{\ifcomment}{\iffalse}
\newcommand{\bs}[1]{\boldsymbol{#1}}

\newcommand{\ti}[1]{\tilde{#1}}
\newcommand{\G}{\Gamma}

\newcommand{\Om}{\Omega}

\newcommand{\cT}{{\mathcal T}}
\newcommand{\tO}{\tilde{\Omega}_h}

\newcommand{\tG}{\tilde{\Gamma}_h}

\newcommand{\tGD}{\tilde{\Gamma}_{D,h}}
\newcommand{\tGN}{\tilde{\Gamma}_{N,h}}

\newcommand{\tx}{{\tilde{\bs{x}}}}

\newcommand{\pd}[2]{\frac{\partial #1}{\partial #2}} 
\newcommand{\pdd}[2]{\frac{\partial^2 #1}{\partial #2^2}} 
\newcommand{\Mw}{\boldsymbol{w}^h}
\newcommand{\Mu}{\boldsymbol{u}^h}

\newdefinition{rem}{Remark}

\newtheorem*{remark}{Remark}

\newcommand{\cref}[2]{\hyperref[#2]{#1~\ref*{#2}}}
\newcommand{\colref}[2]{\hyperref[#2]{#1~\ref*{#2}}}
\newcommand{\eqnref}[1]{\colref{Eq.}{#1}}
\newcommand{\figref}[1]{\colref{Figure}{#1}}

\newcommand{\secref}[1]{\colref{Section}{#1}}
\newcommand{\tabref}[1]{\colref{Table}{#1}}

\newcommand{\Algref}[1]{\hyperref[#1]{Algorithm~\ref*{#1}}}

\newcolumntype{M}[1]{>{\centering\arraybackslash}m{#1}}
\newcolumntype{L}[1]{>{\raggedright\arraybackslash}m{#1}}
\newcolumntype{R}[1]{>{\raggedleft\arraybackslash}m{#1}}

\definecolor{ActiveElement}{RGB}{147,194,74}
\definecolor{InterceptedElement}{RGB}{255,208,48}
\definecolor{FalseInterceptedElement}{RGB}{92,91,255}

\newcommand{\dendrokt}{\textsc{Dendro-KT}}

\newcommand{\Intercepted}{\textsc{Intercepted}}

\newcommand{\petsc}{\href{https://petsc.org/release/}{\textsc{PETSc}}}
\newcommand{\Frontera}{\href{https://www.tacc.utexas.edu/systems/frontera}{Frontera}}

\definecolor{cpu1}{HTML}{4CAF50}
\definecolor{cpu2}{HTML}{FFC107}
\definecolor{cpu3}{HTML}{F44336}
\definecolor{cpu4}{HTML}{2196F3}
\definecolor{cpu5}{HTML}{9932CC}

\definecolor{gpu1}{HTML}{A5D6A7}
\definecolor{gpu2}{HTML}{FFE082}
\definecolor{gpu3}{HTML}{EF9A9A}
\definecolor{gpu4}{HTML}{90CAF9}

\definecolor{sq_b1}{RGB}{37,52,148}
\definecolor{sq_b2}{RGB}{44,127,184}
\definecolor{sq_b3}{RGB}{65,182,196}
\definecolor{sq_b4}{RGB}{127,205,187}
\definecolor{sq_b5}{RGB}{199,233,180}
\definecolor{sq_b6}{RGB}{255,255,204}

\definecolor{sq_r1}{RGB}{189,0,38}
\definecolor{sq_r2}{RGB}{240,59,32}
\definecolor{sq_r3}{RGB}{253,141,60}
\definecolor{sq_r4}{RGB}{254,178,76}
\definecolor{sq_r5}{RGB}{254,217,118}
\definecolor{sq_r6}{RGB}{255,255,178}

\definecolor{sq_g1}{RGB}{0,104,55}
\definecolor{sq_g2}{RGB}{49,163,84}
\definecolor{sq_g3}{RGB}{120,198,121}
\definecolor{sq_g4}{RGB}{173,221,142}
\definecolor{sq_g5}{RGB}{217,240,163}
\definecolor{sq_g6}{RGB}{255,255,204}

\definecolor{div_c1}{RGB}{230,171,2}
\definecolor{div_c2}{RGB}{102,166,30}
\definecolor{div_c3}{RGB}{231,41,138}
\definecolor{div_c4}{RGB}{117,112,179}
\definecolor{div_c5}{RGB}{217,95,2}
\definecolor{div_c6}{RGB}{27,158,119}
\definecolor{div_c7}{RGB}{215,48,39}

\definecolor{div_d1}{RGB}{215,25,28}
\definecolor{div_d2}{RGB}{253,174,97}
\definecolor{div_d3}{RGB}{255,255,191}
\definecolor{div_d4}{RGB}{171,217,233}
\definecolor{div_d5}{RGB}{44,123,182}

\definecolor{lineclr}{RGB}{0,0,0}
\definecolor{utorange}{RGB}{0,0,255}
\definecolor{utsecblue}{RGB}{255,255,0}
\definecolor{utsecgreen}{RGB}{255,0,0}
\definecolor{red!15}{RGB}{0,255,255}
\definecolor{fillclr5}{RGB}{0,255,0}
\definecolor{fillclr6}{RGB}{255,0,255}
\definecolor{fillclr7}{RGB}{255,255,255}
\definecolor{fillclr8}{RGB}{0,0,0}

\definecolor{armygreen}{rgb}{0.29, 0.33, 0.13}
\definecolor{aurometalsaurus}{rgb}{0.43, 0.5, 0.5}
\definecolor{applegreen}{rgb}{0.55, 0.71, 0.0}
\definecolor{darkgreen}{rgb}{0.0, 0.4, 0.25}

\def\drawcubeI(#1,#2,#3,#4,#5){ 
\coordinate (O) at (#1,#2,#3);
\coordinate (A) at (#1,#2+#4,#3);
\coordinate (B) at (#1,#2+#4,#3+#4);
\coordinate (C) at (#1,#2,#3+#4);
\coordinate (D) at (#1+#4,#2,#3);
\coordinate (E) at (#1+#4,#2+#4,#3);
\coordinate (F) at (#1+#4,#2+#4,#3+#4);
\coordinate (G) at (#1+#4,#2,#3+#4);
\draw[#5] (O) -- (C) -- (G) -- (D) -- cycle;
\draw[#5] (O) -- (A) -- (E) -- (D) -- cycle;
\draw[#5] (O) -- (A) -- (B) -- (C) -- cycle;
\draw[#5] (D) -- (E) -- (F) -- (G) -- cycle;
\draw[#5] (C) -- (B) -- (F) -- (G) -- cycle;
\draw[#5] (A) -- (B) -- (F) -- (E) -- cycle;
}

\def\drawcubeII(#1,#2,#3,#4,#5,#6,#7){ 
\coordinate (O) at (#1,#2,#3);
\coordinate (A) at (#1,#2+#4,#3);
\coordinate (B) at (#1,#2+#4,#3+#4);
\coordinate (C) at (#1,#2,#3+#4);
\coordinate (D) at (#1+#4,#2,#3);
\coordinate (E) at (#1+#4,#2+#4,#3);
\coordinate (F) at (#1+#4,#2+#4,#3+#4);
\coordinate (G) at (#1+#4,#2,#3+#4);
\draw[#5,fill=#6,opacity=#7] (O) -- (C) -- (G) -- (D) -- cycle;
\draw[#5,fill=#6,opacity=#7] (O) -- (A) -- (E) -- (D) -- cycle;
\draw[#5,fill=#6,opacity=#7] (O) -- (A) -- (B) -- (C) -- cycle;
\draw[#5,fill=#6,opacity=#7] (D) -- (E) -- (F) -- (G) -- cycle;
\draw[#5,fill=#6,opacity=#7] (C) -- (B) -- (F) -- (G) -- cycle;
\draw[#5,fill=#6,opacity=#7] (A) -- (B) -- (F) -- (E) -- cycle;
}

\def\drawNodes(#1,#2,#3,#4,#5,#6,#7){ 
\foreach \x in {#1,#7,...,#2}{
	\foreach \y in {#3,#7,...,#4}{
		\foreach \z in {#5,#7,...,#6}{
				\draw[fill=red!60] (\x,\y,\z) circle (0.15);
				}
			}
	}				
		
}

\pgfplotsset{
  log x ticks with fixed point/.style={
      xticklabel={
        \pgfkeys{/pgf/fpu=true}
        \pgfmathparse{exp(\tick)}%
        \pgfmathprintnumber[fixed relative, precision=3]{\pgfmathresult}
        \pgfkeys{/pgf/fpu=false}
      }
  },
  log y ticks with fixed point/.style={
      yticklabel={
        \pgfkeys{/pgf/fpu=true}
        \pgfmathparse{exp(\tick)}%
        \pgfmathprintnumber[fixed relative, precision=3]{\pgfmathresult}
        \pgfkeys{/pgf/fpu=false}
      }
  }
}

\makeatletter
\newcommand\resetstackedplots{
\makeatletter
\pgfplots@stacked@isfirstplottrue
\makeatother
\addplot [forget plot,draw=none] coordinates{(48,0) (96,0) (192,0) (384,0) (768,0) (1536,0) (3072,0) (6144,0)};
}
\makeatother

\makeatletter
\newcommand\resetstackedplotsOne{
\makeatletter
\pgfplots@stacked@isfirstplottrue
\makeatother
\addplot [forget plot,draw=none] coordinates{(384,0) (768,0) (1536,0) (3072,0) (6144,0)};
}
\makeatother

\makeatletter
\newcommand\resetstackedplotsTwo{
\makeatletter
\pgfplots@stacked@isfirstplottrue
\makeatother
\addplot [forget plot,draw=none] coordinates{(16,0) (32,0) (64,0) (128,0) (256,0) (512,0) (1024,0) (2048,0) (4096,0) (8192,0) (16384,0) (32768,0)};
}
\makeatother

\makeatletter
\newcommand\resetstackedplotsThree{
\makeatletter
\pgfplots@stacked@isfirstplottrue
\makeatother
\addplot [forget plot,draw=none] coordinates{(2,0) (4,0) (8,0) (16,0) (32,0) (64,0)};
}
\makeatother

\makeatletter
\newcommand\resetstackedplotsFour{
\makeatletter
\pgfplots@stacked@isfirstplottrue
\makeatother
\addplot [forget plot,draw=none] coordinates{(4,0) (8,0) (16,0) (32,0) (64,0)};
}
\makeatother

\makeatletter
\newcommand\resetstackedplotsFive{
\makeatletter
\pgfplots@stacked@isfirstplottrue
\makeatother
\addplot [forget plot,draw=none] coordinates{(1,0) (2,0) (4,0) (8,0) (16,0) (32,0) (64,0) (128,0)};
}
\makeatother

\makeatletter
\newcommand\resetstackedplotsSix{
\makeatletter
\pgfplots@stacked@isfirstplottrue
\makeatother
\addplot [forget plot,draw=none] coordinates{(2,0) (4,0) (8,0) (16,0) (32,0) (64,0) (128,0)};
}
\makeatother

\graphicspath{{Figures/}}
\DeclareGraphicsExtensions{.pdf,.png,.jpg}

\begin{document}
	
	
\begin{frontmatter}

\title{A Shifted Boundary Method for Thermal Flows}

\author[ISU]{Cheng-Hau Yang}
\ead{chenghau@iastate.edu}
\author[Duke]{Guglielmo Scovazzi}
\ead{guglielmo.scovazzi@duke.edu}
\author[ISU]{Adarsh Krishnamurthy}
\ead{adarsh@iastate.edu}
\author[ISU]{Baskar Ganapathysubramanian\texorpdfstring{\corref{cor}}{}}
\ead{baskarg@iastate.edu}
\cortext[cor]{Corresponding authors}
\address[ISU]{Iowa State University, Ames, IA}
\address[Duke]{Department of Civil and Environmental Engineering, Duke University, Durham, North Carolina 27708, USA}

\begin{abstract}
This paper presents an incomplete Octree mesh implementation of the Shifted Boundary Method (Octree-SBM) for multiphysics simulations of coupled flow and heat transfer. 
Specifically, a semi-implicit formulation of the thermal Navier-Stokes equations is used to accelerate the simulations while maintaining accuracy. The SBM enables precise enforcement of field and derivative boundary conditions on cut (intercepted) elements, allowing for accurate flux calculations near complex geometries, when using non-boundary fitted meshes. Both Dirichlet and Neumann boundary conditions are implemented within the SBM framework, with results demonstrating that the SBM ensures precise enforcement of Neumann boundary conditions on Octree-based meshes. We illustrate this approach by simulating flows across different regimes, spanning several orders of magnitude in both the Rayleigh number ($Ra \sim 10^3$--$10^9$) and the Reynolds number ($Re \sim 10^0$--$10^4$), and covering the laminar, transitional, and turbulent flow regimes. Coupled thermal-flow phenomena and their statistics across all these regimes are accurately captured without any additional numerical treatments, beyond a Residual-based Variational Multiscale formulation (RB-VMS). This approach offers a reliable and efficient solution for complex geometries, boundary conditions and flow regimes in computational multiphysics simulations.

\end{abstract}


\end{frontmatter}

\section{Introduction}
\label{Sec:Intro}

Natural and forced convection are fundamental mechanisms in heat transfer, influencing a wide range of engineering applications, such as the optimization of the design of thermal exchangers~\citep{bhutta2012cfd,abeykoon2020compact}. Beyond industrial contexts, convection plays a key role in addressing climate-related challenges. For example, the urban heat island effect has become increasingly prominent due to rising global temperatures and urbanization, making the understanding and management of convection in such environmental flows more critical than ever~\citep{priyadarsini2008microclimatic,allegrini2018simulations}. In sustainable building design, natural ventilation has emerged as a promising strategy for improving energy efficiency by harnessing wind and thermal energy~\citep{chenari2016towards,zhang2021critical}. Furthermore, accurately modeling the interaction between airflow and temperature in built environments is essential for ensuring indoor comfort and safety. This is particularly relevant in the context of public health, where effective airflow control is crucial in preventing the spread of infectious diseases~\citep{bhattacharyya2020novel,li2021probable,foster2021estimating,saurabh2021scalable,tan2023computational,rayegan2023review}.

Simulating these processes, however, often involves geometrically complex domains, such as urban layouts in heat island studies, human anatomy in aerosolized virus transmission modeling, or intricate configurations in thermal exchangers. Generating boundary-fitted meshes for such geometries is time-consuming and labor-intensive. Moreover, during simulations, engineers frequently discover that some regions require finer resolution, while others can afford coarser discretization. Adjusting these resolutions typically necessitates revisiting meshing procedures, producing impractical situations in workflows demanding rapid iterations. Another example of such situations is the increased interest in creating large computational datasets of flow past complex geometries for training AI-driven algorithms~\citep{tali2024flowbench}. Here again, the bottleneck is the creation of good body-fitted meshes and accurate imposition of boundary conditions. 

The immersed boundary method (IBM)~\citep{peskin1972flow, mittal2005immersed,colonius2008fast,wang2004extended,zhang2004immersed,borazjani2008curvilinear,zhao2022enriched} offers an alternative, as it avoids boundary-fitted meshes, significantly simplifying the meshing process. This approach decouples the computational grid from geometric shapes, allowing greater efficiency in testing and simulation. However, traditional IBM implementations, such as the Finite Cell Method (FCM)~\citep{Parvizian:07.1,Duester:08.1,schillinger2013review,stavrev2016geometrically,de2017condition, jomo2021hierarchical} and immersogeometric analysis (IMGA)~\citep{kamensky2015immersogeometric,xu2016tetrahedral, wang2017rapid, HOANG2019421, DEPRENTER2019604, ZhuQiming201911, XU2021103604, xu2019immersogeometric, kamensky2021open,jaiswal2024mesh}, face inherent challenges. Issues such as the small-cut cell problem and load balancing inefficiencies arise because elements intersected by the geometry often require a disproportionately large number of integration points, leading to uneven computational loads across processors. This remains an area of active development. 

To address these limitations, a recent development in IBMs, the Shifted Boundary Method (SBM)~\citep{main2018shifted,Main2018TheSB,KARATZAS2020113273,atallah2020second,atallah2021shifted,atallah2021analysis,colomes2021weighted,saurabh2021scalable,atallah2022high,ZENG2022115143,heisler2023generating,yang2024optimal} proposes to integrate the variational forms over a surrogate domain instead of directly working on the cut (or \texttt{Intercepted}) elements. The SBM has demonstrated its versatility in various applications, including fluid dynamics~\citep{MainS18a}, structural simulations~\citep{atallah2021shifted}, free surface flows~\citep{osti_1851595}, and one-way coupled fluid-structure interaction (FSI)~\citep{xu2024weighted}. By eliminating the need for boundary-fitted meshing, the SBM significantly reduces preprocessing time while maintaining high accuracy. However, efficiently and automatically generating non-boundary-fitted meshes for the SBM or IBM simulations remains challenging. To overcome this, there have been recent efforts to employ Octree meshes, which enable faster and parallelized generation of non-boundary-fitted meshes. 
Octree meshes stand out due to their favorable aspect ratios, intrinsic hierarchical structure, and compatibility with parallel computing frameworks~\citep{popinet2003gerris,losasso2004simulating,chen2009numerical,theillard2013second,papac2013level,guittet2015stable,sousa2019finite,egan2021direct,saurabh2021industrial,bayat2022sharp,van2022fourth,yu2022multi,kim2023super,blomquist2024stable}. These properties minimize inter-processor communication overhead by localizing the required neighborhood element information, making them highly efficient for large-scale, distributed simulations. The Octree-SBM framework, which combines Octree meshes with the SBM, has been applied to a variety of PDEs~\citep{yang2024optimal,yang2024simulating}. Together, they provide robust capabilities for handling geometrically complex domains while maintaining computational efficiency and accuracy, making them a promising approach for modern simulation challenges. 

Modeling thermal flows using the SBM is, therefore, a promising avenue for a variety of applications, but has been largely unexplored. Existing research has been limited to a single example involving a one-dimensional convection-diffusion equation~\citep{main2018shifted}. Motivated by this gap, this paper investigates the efficacy of the SBM for thermal incompressible flow simulations, with a focus on a coupled solver framework for the Navier-Stokes and energy (convection-diffusion) equations. The simulation framework developed in this study leverages Octree discretization, a linear semi-implicit Navier-Stokes model, a Variational Multiscale (VMS) formulation, a two-way coupling mechanism between the Navier-Stokes equations and heat transfer, backflow stabilization techniques for both Navier-Stokes and convection-diffusion equations, and the implementation of the SBM for efficiently handling complex geometries. This comprehensive approach enables accurate and efficient simulations of thermal incompressible flows, addressing key challenges in both computational efficiency and geometric flexibility. Our key contributions are:
\begin{itemize}
    \item \textit{The application of the SBM for thermal flow simulations}, with an Octree-based discretization for efficient handling of complex geometries.
    \item \textit{The development of a linear semi-implicit Navier-Stokes and fully implicit Heat Transfer (NS-HT) solver} for the fast and accurate solution of coupled thermal flow simulations.
    \item \textit{A comprehensive validation} across diverse geometries in two and three dimensions, spanning multiple flow regimes and boundary conditions.
\end{itemize}

This paper is structured as follows: In \secref{sec:Math}, we present the governing equations, that is the Navier-Stokes and the heat transfer subproblems. In \secref{sec:implement}, we discuss the coupling of Navier-Stokes and Heat Transfer, introducing the block-iterative strategy. In \secref{sec:results}, we perform various simulations, including both two-dimensional and three-dimensional cases; mixed, forced, and natural convection; and scenarios with Neumann and Dirichlet boundary conditions. Finally, in \secref{sec:Conclusions}, we summarize our findings and suggest directions for future work.

\section{Mathematical Formulation}\label{sec:Math}
\subsection{Strong form of the thermal Navier-Stokes equations}\label{sec:Math_NS}

Using Einstein's repeated index notation, the strong form of the (non-dimensional) incompressible Navier-Stokes equations can be written as follows:
\begin{align}
\label{eq:governing_NS}
\text{momentum equation:} 
 & \quad  \pd{u_i}{t} + u_j \, \pd{u_i}{x_j} - \nu \, \pdd{u_i}{x_j} + \pd{p}{x_i} - f_i(\theta) = 0 \; , \\ 
 \text{continuity equation:}
 & \quad \pd{u_i}{x_i} = 0 \; ,
 \end{align}
where $u_i$ is the $i$th component of the (non-dimensional) velocity $\bs{u}$, $p$ is the (non-dimensional) pressure, and $\theta$ is the (non-dimensional) temperature. 
The parameter $\nu$ in the non-dimensional momentum equation can be written in natural, forced, and mixed convection regimes:
\begin{align}
\label{eq:ns_coe_forced_natural}
\nu \, = \, 
\begin{cases}
\displaystyle \sqrt{\frac{Pr}{Ra}} = \sqrt{\frac{1}{Gr}} \; , & \mbox{for natural convection} \; , \\[.3cm]
\displaystyle \frac{1}{Re} \; , & \mbox{for forced or mixed convection} \; ,
\end{cases}
\end{align}
where $Ra = \frac{\hat{g} \, \beta \, \Delta T \, L_0^3}{\nu^* \, \alpha^*}$ is the Rayleigh number, $Gr = \frac{\hat{g} \, \beta \, \Delta T L_0^3}{\nu^{*2}}$ is the Grashof number, $Pr = \frac{\nu^*}{\alpha^*}$ is the Prandtl number, and $Re = \frac{\rho \, u_0 \, L_0}{\mu}$ is the Reynolds number. Here, $\hat{g}$ denotes the gravity acceleration, $\beta$ the coefficient of thermal expansion, and $\Delta T = T_h - T_c$, where $T_h$ and $T_c$ are, respectively, the highest and lowest temperature that are imposed as boundary conditions in the problem at hand. The parameter $L_0$ is the characteristic length scale of the system, $\mu$ (and $\nu^*$) is the dynamic (and kinematic) viscosity, $\alpha^*$ is the thermal diffusivity, $\rho$ is the density of the fluid.
The force term components $f_i(\theta)$ vary across three different scenarios:
\begin{equation}  
\label{eq:naturalorforced_f}
f_i(\theta) 
\, = \, 
\begin{cases}
\theta \, \delta_{im} \; , &  \mbox{for natural convection} \; , \\
0 \; ,   & \mbox{for forced convection} \; , \\
Ri \, \delta_{im} \; ,   & \mbox{for mixed convection} \; ,
\end{cases}
\end{equation}
where $m$ is the direction of the gravity (assumed to be aligned with one of the coordinate axes) and $Ri =  \frac{Gr}{Re^{2}}$ is the Richardson number.

The Navier-Stokes equations described in~\eqnref{eq:governing_NS} need to be complemented with boundary conditions. Typical boundary conditions for the velocity $\bs{u}$ are no-slip boundary conditions, that is $\bs{u}=\bs{0}$ on the portion $\G_D$ of the boundary. More generally, we can write $\bs{u}=\bs{u}_D$ with $\bs{u}_D$ given on $\G_D$, and this form also includes the case of inflow boundary conditions (i.e., a specified inflow velocity).

The portion $\G_N$ of the boundary is complementary to $\G_D$, that is $\partial \Om = \G = \G_D \cup \G_N$ and $\G_D \cap \G_N = \emptyset$.
A typical boundary conditions on $\G_N$ is the outflow boundary condition
\begin{equation}
\label{outflow-bc}
-p n_i + \nu \pd{u_i}{x_j} n_j = 0  \; ,
\end{equation}
where $n_i$ is the $i$th component of the outward point normal $\bs{n}$ to the boundary $\G=\partial \Om$.
This condition requires the sum of the pressure and viscous stress to vanish and allows the fluid to freely exit the computational domain. Outflow boundary conditions are applied weakly through the variational form of the Navier-Stokes and heat transfer equations.
The weak imposition of~\eqnref{outflow-bc} is sometimes complemented by the strong imposition of homogeneous pressure at the outflow boundaries.
This somewhat less orthodox way of imposing outflow conditions is used in some benchmark tests~\cite{bharti2007numerical,sun2019forced} found in the literature and, for the sake of close comparison, we use this strategy in those cases.


The Navier-Stokes equations are coupled to the energy transport equation, in the form of a (non-dimensional) convection-diffusion equation for the temperature field $\theta$: 
\begin{align}\label{eq:Convection-Diffusion-Strong}
& \pd{\theta}{t} + u_j \, \pd{\theta}{x_j} - \alpha \, \pdd{\theta}{x_{j}}  = 0 \; , 
\end{align}  
where $\alpha$ for natural and forced convection scenarios is given as:
\begin{align}
\label{eq:ht_coe_forced_natural}
\alpha \, = \, 
\begin{cases}
\displaystyle \sqrt{\frac{1}{Pr \, Ra}} \; ,&  \mbox{for natural convection} \; , \\[.3cm]
\displaystyle \frac{1}{Pe} \; ,   & \mbox{for forced or mixed convection} \; .
\end{cases}
\end{align}
Here, $Pe = Re \times Pr$ represents the P\'{e}clet number and set $Pr = 0.7$, corresponding to air.
\tabref{tab:ForcedOrNaturalOrMixed} shows how $\nu$, $f_i(\theta)$, and $\alpha$ are chosen based on different convection types (summarizing \eqnref{eq:ns_coe_forced_natural}, \eqnref{eq:naturalorforced_f}, and \eqnref{eq:ht_coe_forced_natural}). 
\begin{table}[!h]
  \centering
    \caption{Summary of non-dimensional parameters and forcing terms for natural, forced, and mixed convection scenarios.}
  \label{tab:ForcedOrNaturalOrMixed}
  \renewcommand{\arraystretch}{1.5} 
  \begin{tabular}{>{\centering\arraybackslash}p{2cm}|>{\centering\arraybackslash}p{3cm}|>{\centering\arraybackslash}p{3cm}|>{\centering\arraybackslash}p{3cm}|}
     & Natural & Forced & Mixed \\
    \hline
    $\nu$ & $\sqrt{\frac{Pr}{Ra}} = \sqrt{\frac{1}{Gr}}$ & $\frac{1}{Re}$ & $\frac{1}{Re}$ \\
    \hline
    $\alpha$ & $\sqrt{\frac{1}{Pr \, Ra}}$ & $\frac{1}{Pe}$ & $\frac{1}{Pe}$ \\
    \hline
    $f_i(\theta)$ & $\theta \, \delta_{im}$ & $0$ & $Ri \; \theta \, \delta_{im}$\\
    \hline
  \end{tabular}
\end{table}
The typical boundary conditions associated with~\eqnref{eq:Convection-Diffusion-Strong} are:
\begin{equation}  \label{eq:Convection-Diffusion-BC}
\begin{cases}
\theta = \theta_D  \; , & \mbox{on } \Gamma_D^\theta \; , \\[.1cm]
\nabla \theta \cdot \bs{n} = h_T  \; , & \mbox{on } \Gamma_N^\theta  \; ,
\end{cases}
\end{equation}
which, respectively, represent the Dirichlet condition on the temperature and a normalized heat-flow (Neumann) boundary condition, respectively.

\subsection{Time integration}\label{sec:time_integr}
We consider an implicit, second-order accurate in time discretization of $\pd{u_i}{t}$, using the Backward Difference Formula (BDF). Unlike our previous work, here we consider the case of a varying time step, that is the possibility that successive time-steps, $\Delta t^n = t^{n+1} - t^n$ and $\Delta t^{n-1} = t^n - t^{n-1}$ may not be necessarily equal. Due to these variable time steps, the time derivative term $\pd{u_i^{n+1}}{t}$ at $t_{n+1}$ can be expressed using the second-order Backward Difference Formula (BDF2):
\[
\pd{u_i^{n+1}}{t} \approx \gamma_0 u_i^{n+1} + \gamma_1 u_i^{n} + \gamma_2 u_i^{n-1} \; ,
\]
where coefficients $\gamma_0$, $\gamma_1$, and $\gamma_2$ are provided in \autoref{tab:BDF_coefficients}.
The BDF2 is an implicit time integrator, and a verification of its accuracy is presented in \ref{sec:NS_MMS} using the method of manufactured solutions (MMS).
In particular, these simulations highlight the importance of applying the correct BDF2 coefficients when using variable time steps.
In practical engineering applications, variable or adaptive time-stepping strategies can be useful when simulating the startup of the flow or when, in complex multi-physics problems, nonlinearities can require smaller time-steps at certain instants of the computation.
\begin{table}[!h]
  \centering
  \caption{Coefficients for the BDF2 (second order in time) and Backward Euler (BDF1, first order in time) implicit methods are provided, along with BDF2 coefficients for non-uniform timesteps. The validation results using the non-uniform timestep BDF2 coefficients are presented in \figref{fig:VariableTimeMMS}.}
  \label{tab:BDF_coefficients}
  \renewcommand{\arraystretch}{1.5} 
  \begin{tabular}{|>{\centering\arraybackslash}p{4cm}|>{\centering\arraybackslash}p{3cm}|>{\centering\arraybackslash}p{3cm}|>{\centering\arraybackslash}p{3cm}|}
    \hline
     & ${\gamma_0}$ & ${\gamma_1}$ & ${\gamma_2}$ \\
    \hline
    BDF2, non-uniform $\Delta t$ & $\frac{1}{\Delta t^n} \left(1+\frac{\Delta t^n}{\Delta t^n + \Delta t^{n-1}} \right)$ & $- \frac{\Delta t^n+\Delta t^{n-1}}{\Delta t^n \Delta t^{n-1}}$ & $\frac{1}{\Delta t^{n-1}} \left(\frac{\Delta t^n}{\Delta t^n + \Delta t^{n-1}}\right)$ \\
    \hline
    BDF2, uniform $\Delta t$ & $\frac{3}{2\Delta t}$ & $-\frac{2}{\Delta t}$ & $\frac{1}{2\Delta t}$ \\
    \hline
    BE (BDF1) & $\frac{1}{\Delta t^n}$ & $-\frac{1}{\Delta t^n}$ & 0 \\
    \hline
  \end{tabular}
\end{table}

\subsection{Body-fitted, stabilized, finite element formulation}\label{sec:var_forms_TNS}

We will first consider a body-fitted variational formulation of the thermal Navier-Stokes equations.
This formulation will serve as a benchmark in the numerical tests and as a starting point in the development of the proposed Octree-SBM.

\subsubsection{Weak for of the momentum and continuity equations}\label{weak-momentum-continuity}
We start with the momentum and continuity equations, in which we assume that the Dirichlet boundary condition on $\bs{u}$ is strongly enforced on $\G_D$, and the outflow condition~(\eqnref{outflow-bc}) is imposed weakly on $\G_N$.
Assume that $\cT_h(\Om)$ is a discretization of the computational domain $\Om$ into finite elements and that $T \in \cT_h(\Om)$ is an element of the discretization.
Hence, we introduce the scalar and vector discrete function spaces:
\begin{align}
S^h(\Om) = V^h(\Om) &:=\; \left\{ q^h \mid q^h \in C^0(\Om) \cap \mathcal{Q}^1(T) \,, \mbox{ with } T \in \cT_h(\Om) \right\}  \, ,
\\
\bs{S}^h(\Om) &:=\; \left\{ \Mu \mid \Mu \in (C^0(\Om))^d \cap (\mathcal{Q}^1(T))^d \,, \mbox{ with } T \in \cT_h(\Om) \,,  \Mu_{\mid \G_D} = \bs{u}_D \right\} \, ,
\\
\bs{V}^h(\Om) &:=\; \left\{ \Mu \mid \Mu \in (C^0(\Om))^d \cap (\mathcal{Q}^1(T))^d \,, \mbox{ with } T \in \cT_h(\Om) \,,  \Mu_{\mid \G_D} = \bs{0} \right\} \, ,
\end{align}
where $\mathcal{Q}^1(T)$ is the set of functions, defined over $T$, that are tensor product of linear polynomials along each of the coordinate directions.
The weak form of the governing equations, incorporating Variational Multiscale Stabilization (VMS) terms, can be written as follows:\\[.2cm]
Find $\bs{u} \in \bs{S}^h(\Om)$ and $p \in S^h(\Om)$, such that, for any $\bs{w} \in \bs{V}^h(\Om)$ and $q \in V^h(\Om)$,
\begin{align}
\mbox{Momentum:  } \quad & 
\mathcal{M}[\Om](\bs{u},p,\theta;\bs{w})
\, = \, 0 \; ,
\label{eqn:nav_stokes} \\
\mbox{Continuity:  } \quad & 
\mathcal{C}[\Om](\bs{u},p,\theta;q)
\, = \, 0 \; , 
\label{eqn:solenoidality}
\end{align}
where
\begin{align}
\mathcal{M}[\Om](\bs{u},p,\theta;\bs{w})
:= & \; 
\underbrace{\bigg(w_i,\gamma_0 u_i^{c;n+1} + \gamma_1 u_i^{c;n} + \gamma_2 u_i^{c;n-1}\bigg)_{\Om}}_{\mathrm{Time \; derivative \; term}} \nonumber + \underbrace{\bigg(w_i,u_j^*\pd{u_i^{c;n+1}}{x_j}\bigg)_{\Om}}_{\mathrm{Coarse \; convection \; term}} - \underbrace{\bigg(u_j^*\pd{w_i}{x_j},u_i^{f;n+1} \bigg)_{\Om}}_{\mathrm{Fine \; convection \; term}} \nonumber \\
& + \underbrace{\nu\bigg(\pd{w_i}{x_j},\pd{u_i^{c;n+1}}{x_j}\bigg)_{\Om}}_{\mathrm{Diffusion \; term}} - \underbrace{\bigg(\pd{w_i}{x_i},p^{c;n+1}\bigg)_{\Om}}_{\mathrm{Coarse \; pressure \; term}} - \underbrace{\bigg(\pd{w_i}{x_i},p_i^{f;n+1} \bigg)_{\Om}}_{\mathrm{Fine \; pressure \; term}} - \underbrace{\bigg(w_i,f_i^{n+1}\bigg)_{\Om}}_{\mathrm{Forcing \; term}} \; , 
\\[.2cm]
\mathcal{C}[\Om](u_i,p,\theta,q)
:= & \;
\underbrace{\bigg(q,\pd{u_i^{c;n+1}}{x_i}\bigg)_{\Om}}_{\mathrm{Coarse \; scale}} - \underbrace{\bigg(\pd{q}{x_i},u_i^{f;n+1} \bigg)_{\Om}}_{\mathrm{Fine \; scale}} \; , 
\end{align}
with 
\begin{align}
u_i^{f;n+1} 
& = \;   
-\tau_m(u_i^{c;n+1}) \, r_M(u_i^{c;n+1},p^{c;n+1}) 
\nonumber \\ 
& = \; 
-\tau_m(u_i^{c;n+1}) \, \bigg(\underbrace{\gamma_0 u_i^{c;n+1} + \gamma_1 u_i^{c;n} + \gamma_2 u_i^{c;n-1}}_{\mathrm{BDF2}}
+ u_j^*\pd{u_i^{c;n+1}}{x_j} - \nu\pdd{u_i^{c;n+1}}{x_j} + \pd{p^{c;n+1}}{x_i} - f_i^{n+1}\bigg)  \; ,
\label{eqn:u_f}
\\[.1cm]
p_i^{f;n+1}
& = \; 
-\tau_c(u_i^{c;n+1}) \, r_C(u_i^{c;n+1})= 
-\tau_c(u_i^{c;n+1}) \, \bigg(\pd{u_i^{c;n+1}}{x_i}\bigg)   \; ,
\label{eqn:p_f}
\\[.1cm]
u_i^n 
& = \; 
u_i^{c;n} + u_i^{f;n} 
\nonumber \\ 
& = \; 
u_i^{c, n} -\tau_m(u_i^{c;n}) \, \bigg(\underbrace{\gamma_0 u_i^{c;n} + \gamma_1 u_i^{c;n-1} + \gamma_2 u_i^{c;n-2}}_{\mathrm{BDF2}}+ u_j^*\pd{u_i^{c;n}}{x_j} - \nu\pdd{u_i^{c;n}}{x_j} + \pd{p^{c;n}}{x_i} - f_i^{n}\bigg)  \; ,
\label{u_n} \\[.1cm]
u_i^{n-1} 
& = \; 
u_i^{c;n-1} + u_i^{f;n-1} 
\nonumber \\ 
& = \; 
u_i^{c;n-1} -\tau_m(u_i^{c;n-1}) \, \bigg(\underbrace{{\gamma_0 u_i^{c;n-1} + \gamma_1 u_i^{c;n-2}}}_{\mathrm{BDF1}} + u_j^*\pd{u_i^{c;n-1}}{x_j} - \nu\pdd{u_i^{c;n-1}}{x_j} + \pd{p^{c;n-1}}{x_i} - f_i^{n-1}\bigg)  \; ,
\label{u_n_minus} 
\\[.1cm]
u_i^* 
& = \; 
\frac{(\Delta t_{n-1} + \Delta t_{n})u_i^{n} - \Delta t_{n} u_i^{n-1}}{\Delta t_{n-1}}  \; , 
\label{u_convection} 
\end{align}
and
\begin{align}
\tau_m(u)
& = \; 
\Bigg(\frac{4}{\Delta t^2} + u_j G_{ij} u_i + \frac{C_M}{Re^2}G_{ij} G_{ij}\Bigg)^{-\frac{1}{2}}  \; , 
\\[.1cm]
\tau_c(u)
& = \; 
(\tau_m(u) g_j g_j)^{-1}  \; ,
\\[.1cm]
G_{ij}
& = \; 
\frac{\partial \xi_k}{\partial x_i} \frac{\partial \xi_k}{\partial x_j}  \; , 
\\[.1cm]
g_i 
& = \; 
\sum_{j=1}^d\frac{\partial \xi_j}{\partial x_i}  \;  .
\end{align}
The quantities $G_{ij}$ and $g_i$ are related to mapping physical elements to their isoparametric elements, and $C_M$ and $C_E$ are chosen as 36.

In \eqnref{eqn:nav_stokes} and \eqnref{eqn:solenoidality}, the superscripts $c$ and $f$ are used to distinguish the coarse-scale and fine-scale variables, respectively. The superscripts $n+1$, $n$, and $n-1$ following $c$ and $f$ denote the time steps of the simulation: \( n+1 \) corresponds to the time step being solved, \( n \) represents the current time step, and \( n-1 \) refers to the previous time step. In \eqnref{eqn:nav_stokes}, an approximate velocity ($u_i^*$) is computed through an extrapolation based on the previous two time steps ($u_i^n$ and $u_i^{n-1}$), as detailed in equations \eqnref{u_n}, \eqnref{u_n_minus}, and \eqnref{u_convection}. 
{Hence, the convected velocity in the advection term is taken from the time step being solved (implicit) and the convecting velocity is extrapolated from the current and previous time steps (explicit), resulting in a linear semi-implicit Navier-Stokes solver. By replacing the nonlinear convective term with a linearized approximation, this approach simplifies the computational solution strategy.}
Note that in the limit of a vanishing time step, the linear {semi-implicit} form of the Navier-Stokes equations converges to the infinite-dimensional form of the same equations, so that consistency is preserved by the proposed approach. {
More generally, a formulation can be categorized as linear semi-implicit or non-linear fully implicit, depending on the treatment of the convection term in the Navier-Stokes equations:
\begin{align}
\label{eq:semi-implicit-fully-implicit}
\begin{cases}
\displaystyle u_j^*\pd{u_i^{c;n+1}}{x_j} \; ,&  \mbox{linear semi-implicit} \; , \\[.3cm]
\displaystyle u_j^{c;n+1}\pd{u_i^{c;n+1}}{x_j} \; ,   & \mbox{non-linear fully implicit} \; .
\end{cases}
\end{align}

A comparison of simulations based on these two approaches is provided in \secref{subsec:FPC}. Apart from this comparison, all simulations presented in this paper are based on the linear semi-implicit discretization of the Navier-Stokes equations.}

Introducing the function spaces
\begin{align}
S^h(\Om) &:=\; \left\{ \varphi^h \mid \varphi^h \in C^0(\Om) \cap \mathcal{Q}^1(T) \,, \mbox{ with } T \in \cT_h(\Om) \,,  \varphi^h_{\mid \G_D^{\theta}} = \theta_D \right\} \, ,
\\
V^h(\Om) &:=\; \left\{ \varphi^h \mid \varphi^h \in C^0(\Om) \cap \mathcal{Q}^1(T) \,, \mbox{ with } T \in \cT_h(\Om) \,,  \varphi^h_{\mid \G_D^{\theta}} = 0 \right\} \, ,
\end{align}
the weak form of the energy equation can be stated as: \\[.2cm]
Find $\theta \in S^h(\Om)$, such that $\forall \phi \in V^h(\Om)$,
\begin{align}
\mathcal{H}[\Om](\bs{u},\theta;\phi)
& = 0
\; ,
\label{eqn:convection_diffusion_eqn} 
\end{align}
where
\begin{align}
\mathcal{H}[\Om](\bs{u},\theta;\phi)
:= & \; 
\underbrace{\bigg(\phi,\gamma_0 \theta^{c;n+1} + \gamma_1 \theta^{c;n} + \gamma_2 \theta^{c;n-1}\bigg)_{\Om}}_{\mathrm{Time \; derivative \; term}} 
+ \underbrace{\bigg(\phi,u_j^{c;n}\pd{\theta^{c;n+1}}{x_j}\bigg)_{\Om}}_{\mathrm{Convection \; term}}  
+ \underbrace{\alpha\bigg(\pd{\phi}{x_j},\pd{\theta^{c;n+1}}{x_j}\bigg)_{\Om}}_{\mathrm{Diffusion \; term}} \nonumber \\
& + \underbrace{ \tau_{\text{SUPG}} \, \bigg(u_j^{c;n}\pd{\phi}{x_j}, \gamma_0 \theta^{c;n+1} + \gamma_1 \theta^{c;n} + \gamma_2 \theta^{c;n-1} + u_j^{c;n+1} \pd{\theta^{c;n+1}}{x_j} \bigg)_{\Om}}_{\mathrm{SUPG \; term}} 
\, . 
\label{eqn:convection_diffusion_eqn2} 
\end{align}
{For the convection term in \eqnref{eqn:convection_diffusion_eqn2}, we use only the coarse-scale velocity from the Navier-Stokes subproblem and neglect the fine-scale velocity. This approach is consistent with \citet{xu2018buoyancy}, and the omission of the fine-scale velocity has a minimal impact on the solution.} Similar to the Navier-Stokes equations, we use the BDF2 scheme $\gamma_0 \theta^{n+1} + \gamma_1 \theta^{n} + \gamma_2 \theta^{n-1} $ to approximate $\pd{\theta^{n+1}}{t}$, with the coefficients described in \tabref{tab:BDF_coefficients}. To address potential numerical instabilities in advection-dominated problems, we also have incorporated a Streamline-Upwind Petrov-Galerkin (SUPG) stabilization term~\citep{brooks1982streamline} in~\eqnref{eqn:convection_diffusion_eqn2} . This term is designed to enhance solution stability along the streamlines by skewing the test functions along streamlines. In implementing the SUPG term, we neglect the diffusion term given that we use a piecewise linear finite element basis. The SUPG stabilization parameter \( \tau_{\text{SUPG}} \) is defined as:
\begin{equation}
\tau_{\text{SUPG}} = \frac{h z}{2 \sqrt{u_i u_i}}
\; ,
\end{equation}
where \( u_i \) is the velocity vector and \( \sqrt{u_i u_i} \) its magnitude.
The element length \( h \) is:
\begin{equation}
h = \frac{2}{\sum_{A} \frac{\left| u_i \frac{\partial N_A}{\partial x_i} \right|}{\sqrt{u_i u_i}}}
\; ,
\end{equation}
with \( N_A \) as the shape function, \( \frac{\partial N_A}{\partial x_i} \) its gradient, and \( A \) running over all nodes of the element. The parameter \( z \) depends on the local Reynolds number:
\begin{equation}
Re_u = \frac{\sqrt{u_i u_i} h}{2 \nu}
\; ,
\end{equation}
The value of \( z \) is:
\begin{equation}
z = 
\begin{cases} 
1 & \text{if } Re_u > 3 \;, \\
\frac{Re_u}{3} & \text{if } Re_u \leq 3 \; .
\end{cases}
\end{equation}

\subsection{Backflow stabilization}\label{sec:backflow}
Instabilities arising from backflow at outflow or open boundaries can lead to solver divergence in thermal incompressible flow simulations. Backflow stabilization introduces a dissipative boundary term that activates in the presence of backflow, effectively maintaining stability. This approach is particularly suitable for simulations involving thermal incompressible flow problems characterized by strong flow recirculation or vortices impinging on the outlet boundary.

In both the backflow stabilization method~\citep{Moghadam2011} and the directional do-nothing (DDN) boundary condition~\citep{Braack2014}, a boundary term is introduced to the left-hand side of \eqnref{eqn:nav_stokes}:
\begin{align}
- \bigg< w_i \, , \beta_0 \, \min(0, u_j^{c;n} n_j) u_i^{c;n+1} \bigg>_{\Gamma_o}
\end{align}
where $\beta_0$ is the stabilization parameter for Navier-Stokes. Similarly, backflow stabilization is applied to the heat transfer equation, adding the following term to the left-hand side of \eqnref{eqn:convection_diffusion_eqn}:
\begin{align}
- \bigg< \phi \, , \beta_\theta \, \min(0, u_j^{c;n} n_j) \theta^{c;n+1} \bigg>_{\Gamma_o}
\end{align}
where $\beta_\theta$ is the stabilization parameter for heat transfer. In the simulations performed in this paper, we pick $\beta_0 = \beta_\theta = 0.5$.

\subsection{SBM preliminaries: The true domain, surrogate domain, and maps}
\label{sec:sbmDef}
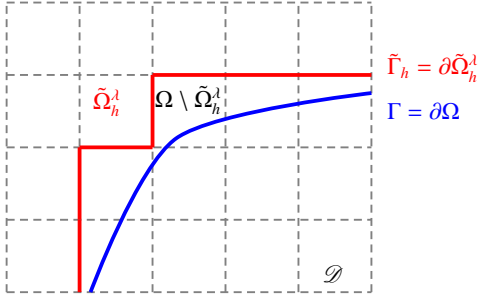
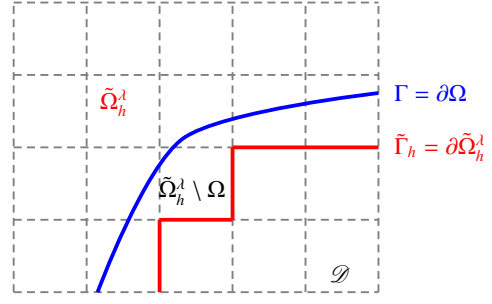
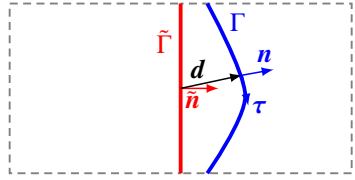
\begin{figure}[!t]
	\centering
     \begin{subfigure}[b]{.48\linewidth}\centering
    \begin{tikzpicture}[scale=0.48]
    
    \draw[line width = 0.25mm,densely dashed,gray] (-2,-4) -- (8,-4); 
    \draw[line width = 0.25mm,densely dashed,gray] (-2,-2) -- (8,-2); 
    \draw[line width = 0.25mm,densely dashed,gray] (-2,0) -- (8,0); 
    \draw[line width = 0.25mm,densely dashed,gray] (-2,2) -- (8,2); 
    \draw[line width = 0.25mm,densely dashed,gray] (-2,4) -- (8,4); 
    \draw[line width = 0.25mm,densely dashed,gray] (-2,-4) -- (-2,4); 
    \draw[line width = 0.25mm,densely dashed,gray] (0,-4) -- (0,4); 
    \draw[line width = 0.25mm,densely dashed,gray] (2,-4) -- (2,4); 
    \draw[line width = 0.25mm,densely dashed,gray] (4,-4) -- (4,4); 
    \draw[line width = 0.25mm,densely dashed,gray] (6,-4) -- (6,4); 
    \draw[line width = 0.25mm,densely dashed,gray] (8,-4) -- (8,4); 
    
    \draw[line width = 0.5mm,red] (8,2) -- (2,2);  
    \draw[line width = 0.5mm,red] (2,0) -- (2,2);  
    \draw[line width = 0.5mm,red] (0,0) -- (2,0);  
    \draw[line width = 0.5mm,red] (0,0) -- (0,-2);  
    \draw[line width = 0.5mm,red] (0,-2) -- (0,-4);  
    
    \draw [line width = 0.5mm,blue] plot[smooth] coordinates {(0.3,-4) (2.65,0.25) (8.0,1.5)};
    
    \node[text width=1.5cm] at (10.0,2.2) {\small${\color{red}\tG = \partial \tO^\lambda }$};
    \node[text width=3cm] at (3.5,1.25) {\small${\color{red}\tO^\lambda}$};
    \node[text width=1cm] at (7.7,-3.5) {\small${\color{black}{\mathcal D}}$};
    \node[text width=1.5cm] at (10.0,1.0) {\small${\color{blue}\G = \partial \Om}$};
    \node[text width=3cm] at (5.2,1.3) {\small$\Om \setminus \tO^\lambda $};
    \end{tikzpicture}
    \caption{The surrogate domain $\tO^\lambda \subset \Om$, the difference $\Om \setminus \tO^\lambda$ between the true and surrogate domains, true boundary $\G$, and the surrogate boundary $\tG$ ($\lambda=0$).}
    \label{fig:SBM}
\end{subfigure}
\hfill
\begin{subfigure}[b]{.48\linewidth}\centering
    \begin{tikzpicture}[scale=0.48]
    
    \draw[line width = 0.25mm,densely dashed,gray] (-2,-4) -- (8,-4); 
    \draw[line width = 0.25mm,densely dashed,gray] (-2,-2) -- (8,-2); 
    \draw[line width = 0.25mm,densely dashed,gray] (-2,0) -- (8,0); 
    \draw[line width = 0.25mm,densely dashed,gray] (-2,2) -- (8,2); 
    \draw[line width = 0.25mm,densely dashed,gray] (-2,4) -- (8,4); 
    \draw[line width = 0.25mm,densely dashed,gray] (-2,-4) -- (-2,4); 
    \draw[line width = 0.25mm,densely dashed,gray] (0,-4) -- (0,4); 
    \draw[line width = 0.25mm,densely dashed,gray] (2,-4) -- (2,4); 
    \draw[line width = 0.25mm,densely dashed,gray] (4,-4) -- (4,4); 
    \draw[line width = 0.25mm,densely dashed,gray] (6,-4) -- (6,4); 
    \draw[line width = 0.25mm,densely dashed,gray] (8,-4) -- (8,4); 
    
    \draw[line width = 0.5mm,red] (8,0) -- (4,0);  
    \draw[line width = 0.5mm,red] (4,0) -- (4,-2);  
    \draw[line width = 0.5mm,red] (4,-2) -- (2,-2);  
    \draw[line width = 0.5mm,red] (2,-2) -- (2,-4);  
    
    \draw [line width = 0.5mm,blue] plot[smooth] coordinates {(0.3,-4) (2.65,0.25) (8.0,1.5)};
    
    \node[text width=1.5cm] at (10.0,0.0) {\small${\color{red}\tG = \partial \tO^\lambda }$};
    \node[text width=3cm] at (3.5,1.25) {\small${\color{red}\tO^\lambda}$};
    \node[text width=1cm] at (7.7,-3.5) {\small${\color{black}{\mathcal D}}$};
    \node[text width=1.5cm] at (10.0,1.5) {\small${\color{blue}\G = \partial \Om}$};
    \node[text width=3cm] at (5.1,-1.2) {\small$\tO^\lambda \setminus \Om $};
    \end{tikzpicture}
    \caption{The surrogate domain $\Om \subset \tO^\lambda$, the difference $\tO^\lambda \setminus \Om$ between the true and surrogate domains, true boundary $\G$, and the surrogate boundary $\tG$ ($\lambda=1$).}
    \label{fig:SBM_lambda1p0}
\end{subfigure}\\
\vspace{0.4in}%
\begin{subfigure}[b]{.4\linewidth}\centering
	\begin{tikzpicture}[scale=0.5]
	\draw[line width = 0.25mm,densely dashed,gray] (0,0.5) rectangle (-4.5,5); 
	\draw[line width = 0.25mm,densely dashed,gray] (0,5) rectangle (4.5,0.5); 

	\draw [line width = 0.5mm,blue, name path=true] plot[smooth] coordinates {(0.7,0.5) (1.7,2.55) (0.7,5)};
	
	\draw[line width = 0.5mm,red] (0,0.5) -- (0,5);
	
	\node[text width=0.5cm] at (-0.125,4.0) {\normalsize${\color{red}\ti{\G}}$};
	\node[text width=0.5cm] at (1.8,4.5) {\normalsize${\color{blue}\G}$};
	\node[text width=0.5cm] at (0.75,3.25) {\normalsize$\bs{d}$};
	\node[text width=0.5cm,blue] at (2.5,3.6) {\normalsize$\bs{n}$};
	\node[text width=0.5cm,blue] at (2.4,2.25) {\normalsize$\bs{\tau}$};
	\node[text width=0.5cm,red] at (0.6,2.45) {\normalsize$\ti{\bs{n}}$}; 
	
	\draw[->,line width = 0.2mm,-latex,red] (0,2.75) -- (1,2.75); 
	\draw[->,line width = 0.2mm,-latex] (0,2.75) -- (1.62,3.1);
	\draw[->,line width = 0.2mm,-latex,blue] (1.62,3.1) -- (1.78,2.29);
	\draw[->,line width = 0.2mm,-latex,blue] (1.62,3.1) -- (2.45,3.25);
	
	\end{tikzpicture}
    \caption{The distance vector $\bs{d}$, the true normal $\bs{n}$, the true tangent $\bs{\tau}$, and the surrogate normal $\ti{\bs{n}}$ (horizontal).}
    \label{fig:ntd}
\end{subfigure}
    \caption{The surrogate domain, its boundary, and the distance vector $\bs{d}$.}
    \label{fig:surrogates}
\end{figure}

\figref{fig:SBM} illustrates a closed region ${\mathcal D}$, where $\text{clos}(\Om) \subseteq {\mathcal D}$ (with $\text{clos}(\Om)$ denoting the {\it closure} of $\Om$), along with the family $\cT_h({\mathcal D})$ of admissible, shape-regular discrete decompositions (meshes/grids) of ${\mathcal D}$.
In this study, we specifically focus on Octree grids that are aligned with the axes of the Cartesian coordinate system.
Each $\cT_h({\mathcal D})$ is then restricted by selecting only those elements $T \in \cT_h({\mathcal D})$ that satisfy the condition:
\begin{align} 
\label{eq:lambda_def} 
\mathrm{meas}(T \cap \Om) > (1-\lambda) , \mathrm{meas}(T) \; , \qquad \text{for some } \lambda \in [0,1] \; . 
\end{align}
In other words, these elements are those that intersect with the domain of interest $\Om$ and have an area or volume greater than $1-\lambda$ of their total area or volume, depending on whether the context is two- or three-dimensional.
For example, choosing $\lambda=0$ selects the elements that are strictly contained in the computational domain $\Om$ (see, e.g.,~\figref{fig:SBM}), choosing $\lambda =1$ selects the elements that have a non-empty intersection with $\Om$ (see, e.g.,~\figref{fig:SBM_lambda1p0}), and choosing $\lambda=0.5$ selects elements whose intersection with $\Om$ includes at least $50\%$ of their area/volume.

We define the family of grids that satisfies~\eqnref{eq:lambda_def} as  
$$
\ti{\cT}_h^{\lambda} := \{ T \in \cT_h({\mathcal D}) : \mathrm{meas}(T \cap \Om) > (1-\lambda) \, \mathrm{meas}(T) \}\,.
$$ 
This identifies the {\sl surrogate domain}
$$
\tO^{\lambda} := \text{int} \, \Bigl( \bigcup_{T \in \ti{\cT}_h^{\lambda}}  T \Bigr)  \,,
$$
or, more simply, $\tO$, with {\sl surrogate boundary} $\tG:=\partial \tO$ and outward-oriented unit normal vector $\ti{\bs{n}}$ to $\tG$. Obviously, $\ti{\cT}_h^{\lambda}$ is an admissible and shape-regular family of decompositions of $\tO$ (see again~\figref{fig:SBM}). Here, we choose $\lambda=1$, which is advantageous for thermal incompressible flow calculations, particularly for cases where the quantity of interest is the boundary thermal flux, or in dimensionless terms, the Nusselt number, which requires the first derivative at the cut (or \Intercepted{}) element. By selecting $\lambda=1$, no special implementation is needed to compute the first derivative at the cut (or \Intercepted{}) element. Instead, we can directly use the derivative of the shape function and nodal point values to interpolate and obtain the derivative on the true boundary ($\G$). 

The mapping sketched in \figref{fig:ntd} is defined as follows:  
\begin{subequations}\label{eq:defMmap}
\begin{align}
    \bs{M}_{h}:&\; \tG \to \G \;,  \\
    &\; \ti{\bs{x}} \mapsto \bs{x} \;, 
\end{align}
\end{subequations}
where $\bs{M}_{h}$ maps any point $\ti{\bs{x}} \in \tG$ on the surrogate boundary to a point $\bs{x} = \bs{M}_{h}(\ti{\bs{x}})$ on the physical boundary $\G$. 

In this study, $\bs{M}_{h}$ is defined as the closest-point projection of $\ti{\bs{x}}$ onto $\G$, as illustrated in \figref{fig:ntd}. Using this mapping, a distance vector function $\bs{d}_{\bs{M}_{h}}$ can be expressed as:
\begin{align}
    \label{eq:Mmap}
    \bs{d}_{\bs{M}_{h}} (\ti{\bs{x}})
    \, = \, 
    \bs{x} - \ti{\bs{x}}
    \, = \, 
    [ \bs{M} - \bs{I} ] (\ti{\bs{x}})
    \; ,
\end{align}
where $\bs{M}$ is the mapping operator, and $\bs{I}$ is the identity operator. For simplicity, we denote $\bs{d} = \bs{d}_{\bs{M}_{h}}$ and further decompose it as:
\[
    \bs{d} = \|\bs{d}\| \bs{\nu} \; ,
\]
where $\|\bs{d}\|$ is the magnitude of the distance vector, and $\bs{\nu}$ is a unit vector indicating the direction of the distance.

\begin{remark}
There are several strategies to define the mapping $\bs{M}_{h}$ and, consequently, the distance vector $\bs{d}$. The closest-point projection of $\ti{\bs{x}}$ onto $\G$ is a natural and widely used choice for determining $\bs{x}$ (and thus $\bs{M}_{h}$), provided it is uniquely defined. However, alternative approaches may be more suitable in specific scenarios. For example, a level-set representation of the true boundary can be employed, where $\bs{d}$ is defined using a distance function. For algorithms related to distance functions for complex geometries in three dimensions, see~\citep{yang2024optimal}. For further discussions, including considerations for domains with corners, refer to~\citep{atallah2021analysis,atallah2021shifted}.
\end{remark}

\subsection{Shifted boundary conditions}\label{sec:SBM_gen}

We begin by examining Dirichlet boundary conditions, which are particularly relevant as they encompass the no-slip boundary condition - a crucial constraint for fluid-solid interfaces.

Let us consider a surrogate Dirichlet boundary $\tilde{\Gamma}_D$ positioned near the actual Dirichlet boundary $\Gamma_D$. By utilizing the distance measure between these boundaries, we can express the velocity vector through its Taylor expansion:
\begin{align}
	\label{eq:SBM_Dirichlet1}
	\bs{u}(\tx)
	 + (\nabla \bs{u} \cdot \bs{d}) (\tx)  
	 + (\bs{R}_{D}(\bs{u}, \bs{d}))(\tx) 
	&=\;  
	\bs{u}_{D}(\bs{M}_h (\tx))
	\; , \quad  \mbox{ on } \tGD \; , 
\end{align}
Here, the remainder term $\bs{R}_{D}(\bs{u}, \bs{d})$ exhibits the property that 
$\| \bs{R}_{D}(\bs{u}, \bs{d}) \| = o(\| \bs{d}  \|^2)$ as $\| \bs{d} \, \| \to 0$. To formalize this relationship, we introduce two key operators on $\tGD$.
The \textit{extension} operator
    \begin{align}
    \label{eq:def-extS}
    \mathbb{E}\bs{u}_{D}(\tx) 	 
    &:= \; 
    \bs{u}_{D}(\bs{M}_h (\tx)) 
    \end{align}\\
and the \textit{shift} operator
    \begin{align}
    \label{eq:def-bndS}
    \bs{S}_{D,h} \, \bs{u} (\tx)	 
    \; := \; 
    \bs{u} (\tx)
    + \nabla \bs{u} (\tx) \, \bs{d} (\tx)
    \; , 
    \quad 
    \mbox{or, in index notation,}
    \quad
    S_{D,h} \, u_i (\tx)	 
    \; := \; 
    u_i(\tx) + \pd{u_i}{x_j}(\tx) \, d_j(\tx)
    \; .
    \end{align}
By disregarding the higher-order residual term in \eqnref{eq:SBM_Dirichlet1}, we arrive at the final form of the {\it shifted} boundary condition:
\begin{align}
\label{eq:SBM_Dirichlet}
\bs{S}_{D,h} \, \bs{u} 	
&=\;  
\mathbb{E}\bs{u}_{D}
\; , \qquad \mbox{ on } \tGD \; .
\end{align}

\subsection{SBM formulation for the {linear semi-implicit} Navier--Stokes equations}\label{sec:Math_NS_SBM}


The SBM formulation of the thermal Navier-Stokes equations relies on the general framework proposed by Nitsche~\citep{Nitsche:70.1} for the weak enforcement of boundary conditions. We refer the reader to~\citep{atallah2020second,main2018shifted_2} for a detailed description of how the SBM is developed starting from Nitsche's method.
Here we summarize the SBM equations in the specific context of the {linear semi-implicit} Navier-Stokes approach pursued here.
Introducing now the modified SBM scalar and vector discrete function spaces (which do not incorporate any boundary conditions constraints)
\begin{align}
\ti{V}^h(\tO^{\lambda}) &=\; \left\{ q^h \mid q^h \in C^0(\tO^{\lambda}) \cap \mathcal{Q}^1(T) \,, \mbox{ with } T \in \ti{\cT}_h^{\lambda} \right\}  \; ,
\\
\ti{\bs{V}}^h(\tO^{\lambda}) &=\; \left\{ \Mw \mid \Mw \in (C^0(\tO^{\lambda}))^d \cap (\mathcal{Q}^1(T))^d \,, \mbox{ with } T \in \ti{\cT}_h^{\lambda} \right\}\;,
\end{align}
we can define the SBM variational form of the momentum and continuity equations as \\[.2cm]

Find $\bs{u} \in \ti{\bs{V}}^h(\tO^{\lambda})$ and $p \in \ti{V}^h(\tO^{\lambda})$, such that, for any $\bs{w} \in \ti{\bs{V}}^h(\tO^{\lambda})$ and $q \in \ti{V}^h(\tO^{\lambda})$,
\begin{align}
0
= \; 
& 
\mathcal{M}[\tO^{\lambda}](\bs{u},p,\theta;\bs{w})
+\mathcal{C}[\tO^{\lambda}](\bs{u},p,\theta;q)
- \underbrace{ \bigg<w_i, \nu \left(\pd{u_i^{c;n+1}}{x_j}+\pd{u_j^{c;n+1}}{x_i} \right) \tilde{n}_j - p \tilde{n_i} \bigg>_{\tGD}}_{\mathrm{SBM \; consistency \; term}} 
\nonumber\\
&
- \underbrace{ \bigg<\nu \left(\pd{w_i}{x_j}+\pd{w_j}{x_i} \right) \tilde{n}_j  + q \tilde{n_i} , u_i^{c;n+1} + \pd{u_i^{c;n+1}}{x_j}d_j -u_{D;i} \bigg>_{\tGD}}_{\mathrm{SBM \; adjoint \; consistency \; term \; (Dirichlet)}}
\nonumber\\
&
+ \underbrace{ C^B_M \bigg< \frac{\nu }{h} \bigg( w_i +\pd{w_i}{x_j}d_j \bigg) , u_i^{c;n+1} + \pd{u_i^{c;n+1}}{x_j}d_j -u_{D;i}\bigg>_{\tGD}}_{\mathrm{SBM \; Penalty \; term \; (Dirichlet)}} 
,
\label{eq:ns_sbm}
\end{align}
where $C^B_M=200$ is a penalty parameter. If convergence issues occur during simulations, this penalty parameter is progressively increased by a factor of 10 until convergence is achieved.

\begin{remark}
In the limit of a body-fitted discretization, $\bs{d} \rightarrow \bs{0}$, $\tO^{\lambda} \rightarrow \Om$, $\tGD \rightarrow \G_D$, $\ti{\bs{n}} \rightarrow \bs{n}$ and~\eqnref{eq:ns_sbm} collapses to the standard Nitsche's formulation of the Navier-Stokes equations. See~\citep{main2018shifted_2} for a detailed derivation.
\end{remark}

\subsection{SBM formulation for the convection diffusion equation}\label{sec:Math_Energy_SBM}

%
The SBM variational statement of the heat transfer equation with weak enforcement of the Dirichlet and Neumann boundary conditions of the type described in~\eqnref{eq:Convection-Diffusion-BC} reads:\\[.2cm]
Find \( \theta \in \ti{V}^h(\tO^{\lambda}) \), such that \( \forall \phi \in \ti{V}^h(\tO^{\lambda}) \),
\begin{align}
0
= \; 
&\mathcal{H}[\tO^{\lambda}](\bs{u},\theta;\phi)
- \underbrace{ \bigg<\phi, \alpha \pd{\theta^{c;n+1}}{x_j} \tilde{n}_j \bigg>_{\ti{\G}_h^\theta}}_{\mathrm{SBM \; consistency \; term}} 
+  \underbrace{\bigg<(\tilde{n}_k n_k) \, \phi, \alpha \pd{\theta^{c;n+1}}{x_j} {n}_j - h_T \bigg>_{\tGN^\theta} }_{\mathrm{SBM \; Neumann \; term}} 
\nonumber\\
& 
- \underbrace{ \bigg< \alpha \pd{\phi}{x_j} \tilde{n}_j, \theta^{c;n+1} + \pd{\theta^{c;n+1}}{x_j}d_j - \theta_D \bigg>_{\tGD^\theta}}_{\mathrm{SBM \; adjoint \; consistency \; term \; (Dirichlet)}} 
+ \underbrace{ C^B_E  \bigg<\frac{ \alpha }{h} \bigg( \phi +\pd{\phi}{x_j}d_j \bigg) \, , \theta^{c;n+1} + \pd{\theta^{c;n+1}}{x_j}d_j - \theta_D \bigg>_{\tGD^\theta}}_{\mathrm{SBM \; Penalty \; term \; (Dirichlet)}} \, ,
\label{eq:HT-SBM}
\end{align}
where $C^B_E=400$ is the penalty parameter for the thermal (convection-diffusion) equation. If convergence issues arise during simulations, this penalty parameter is progressively increased by a factor 10 until convergence is achieved.
\begin{remark}
The derivation of the shifted Neumann conditions is done in analogy to~\citep{atallah2021shifted}, but we prefer to report the details here, for the sake of clarity.
In particular, a primal formulation is used here, instead of the mixed formulation discussed in~\citep{atallah2021shifted}.
More specifically, the gradient of the heat flux is not included in the shift operator, since piecewise linear interpolation is used.
{
This should, in principle, result in the loss of one order of accuracy in the convergence of the $L^2$-norm of the error, but we will show in the numerical examples section that the effect of our approach on the numerical results is not significant.
}%
Starting from the strong form of the diffusion term tested on the shape function $\phi$, we have
\begin{align}
\bigg(\phi, \frac{\partial}{\partial x_j} \bigg( - \alpha \pd{\theta^{c;n+1}}{x_j}\bigg) \bigg)_{\tO}
= \; 
&
\bigg(\pd{\phi}{x_j}, \alpha \, \pd{\theta^{c;n+1}}{x_j}\bigg)_{\tO}
- \bigg<\phi, \alpha \pd{\theta^{c;n+1}}{x_j} \tilde{n}_j \bigg>_{\ti{\G}_h^\theta}
\nonumber\\
= \; 
&
\bigg(\pd{\phi}{x_j}, \alpha \, \pd{\theta^{c;n+1}}{x_j}\bigg)_{\tO}
- \bigg<\phi, \alpha \pd{\theta^{c;n+1}}{x_j} \tilde{n}_j \bigg>_{\tGD^\theta}
- \bigg<\phi, \alpha \pd{\theta^{c;n+1}}{x_j} \left( (\tilde{n}_k n_k) n_j  + (\tilde{n}_k t_k) t_j \right) \bigg>_{\tGN^\theta}
\nonumber\\
= \; 
&
\bigg(\pd{\phi}{x_j}, \alpha \, \pd{\theta^{c;n+1}}{x_j}\bigg)_{\tO}
- \bigg<\phi, \alpha \pd{\theta^{c;n+1}}{x_j} \tilde{n}_j \bigg>_{\tGD^\theta}
- \big<\phi, (\tilde{n}_k n_k) \, h_T \big>_{\ti{\G}_h^\theta}
- \bigg<\phi, \alpha (\tilde{n}_k t_k) \pd{\theta^{c;n+1}}{x_j} \, t_j \bigg>_{\ti{\G}_h^\theta}
\nonumber\\
&
\pm \bigg<\phi, \alpha (\tilde{n}_k n_k) \, \pd{\theta^{c;n+1}}{x_j} \, n_j  \bigg>_{\tGN^\theta}
\nonumber\\
= \; 
&
\bigg(\pd{\phi}{x_j}, \alpha \, \pd{\theta^{c;n+1}}{x_j}\bigg)_{\tO}
- \bigg<\phi, \alpha \pd{\theta^{c;n+1}}{x_j} \tilde{n}_j \bigg>_{\ti{\G}_h^\theta}
+  \bigg<(\tilde{n}_k n_k) \, \phi , \alpha \pd{\theta^{c;n+1}}{x_j} \, n_j - h_T \bigg>_{\tGN^\theta}
\, .
\label{eq:Neumann_derivation_SBM}
\end{align}

Observe that the inner product $\ti{n}_k n_k = \ti{\bs{n}} \cdot \bs{n}$, found in the SBM Neumann term in~\eqnref{eq:HT-SBM}, represents an approximation to the projection of the area of the surrogate boundary onto the true boundary, and this term is responsible for a marked improvement in accuracy with respect to applying the shifted boundary conditions directly on $\tGN^\theta$, that is, replacing the last term in the last row of~\eqnref{eq:Neumann_derivation_SBM} with simply
$$
\bigg< \phi , \alpha \pd{\theta^{c;n+1}}{x_j} \, n_j - h_T \bigg>_{\tGN^\theta}
\; .
$$
\end{remark}
\begin{remark}
Also in this case, in the limit of a body-fitted discretization, $\bs{d} \rightarrow \bs{0}$, $\tO^{\lambda} \rightarrow \Om$, $\tGD \rightarrow \G_D$, $\ti{\bs{n}} \rightarrow \bs{n}$ and~\eqnref{eq:HT-SBM} collapses to the standard Nitsche's formulation of the equation of convective/diffusive transport of energy.
\end{remark}

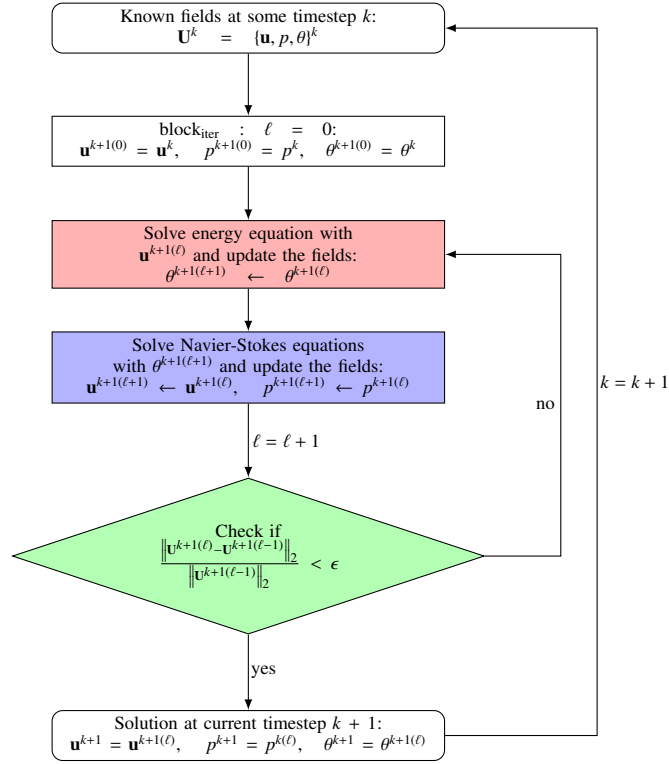
\begin{figure}[!h]
    \centering            
    \begin{tikzpicture}[%
    scale=1,transform shape,
    >=latex,              
    start chain=going right,    
    node distance= 15mm, 
    every join/.style={norm},   
    ]
    \scriptsize
    
    \tikzset{
        start/.style={rectangle, draw, text width=5cm, text centered, rounded corners, minimum height=3ex},
        base/.style={draw, on chain, on grid, align=center, minimum height=3ex, text width=3cm},
        proc/.style={base, rectangle, text width=5cm}, 
        inout/.style={base,trapezium,trapezium left angle=70,trapezium right angle=-70, text width=3.5cm},
        decision/.style={base, diamond, aspect=3.0, fill=green!30},
        norm/.style={->, draw},
    }
    
    \node [start] (b0) {Known fields at some timestep $k$:  \\
    $\mathbf{U}^{k} = \{\mathbf{u}, p, \theta\}^k$};
    \node [proc, below=of b0, fill=none] (b1) {$\text{block}_{\text{iter}}: \ell = 0$:\\
    $\mathbf{u}^{k+1(0)} = \mathbf{u}^{k}$, \quad $p^{k+1(0)} = p^{k}, \quad \theta^{k+1(0)} = \theta^{k}$};
    \node [proc, below=of b1, fill=red!30] (b2) {Solve energy equation with $\mathbf{u}^{k+1(\ell)}$ and update the fields: \\ $\theta^{k+1(\ell + 1)} \leftarrow \theta^{k+1(\ell)}$};
    \node [proc, below=of b2, fill=blue!30] (b3) {Solve Navier-Stokes equations with $\theta^{k+1(\ell + 1)}$ and update the fields: \\ $\mathbf{u}^{k+1(\ell + 1)} \leftarrow \mathbf{u}^{k+1(\ell)}$, \quad $p^{k+1(\ell + 1)} \leftarrow p^{k+1(\ell)}$};
    \node [decision, below=of b3, yshift=-1cm] (b4) {Check if $\frac{\left\|\mathbf{U}^{k+1(\ell)} - \mathbf{U}^{k+1(\ell -1)}\right\|_{2}}{\left\|\mathbf{U}^{k+1(\ell - 1)}\right\|_{2}} < \epsilon$};
    \node [start, below=of b4, yshift=0.5cm] (b5) {Solution at current timestep $k+1$: \\ $\mathbf{u}^{k+1} = \mathbf{u}^{k+1(\ell)}$, \quad $p^{k+1} = p^{k(\ell)}, \quad \theta^{k+1} = \theta^{k+1(\ell)}$};
    
    \draw [->] (b0) -- (b1);
    \draw [->] (b1) -- (b2);
    \draw [->] (b2) -- (b3);
    \draw [->] (b3) -- node[xshift = 0.5cm] {$\ell = \ell + 1$} (b4);
    \draw [->] (b4) -- node[xshift = 0.2cm] {yes} (b5);
    \draw [->] (b4.east) -- ++(1,0) |- node [near start,xshift = -0.2cm] {no} (b2.east);
    \draw [->] (b5.east) -- ++(2,0) |- node [near start,xshift = 0.5cm] {$k = k + 1$} (b0.east);
    \end{tikzpicture}
    \caption{Diagram illustrating the block iteration technique used to perform multiphysics simulations of thermal incompressible flow (NSHT).}
    \label{fig:diagram_block}        
\end{figure}
\section{Implementation details.}\label{sec:implement}

\subsection{Numerical implementations}

Our computational framework is built on two core components: \dendrokt~\citep{ishii2019solving,saurabh2021scalable} and \petsc, both of which play essential roles in enabling high-performance, large-scale scientific simulations across various domains. These tools work together to deliver an efficient and scalable solution for complex numerical problems. At the core of our numerical approach is \dendrokt, an in-house open source library that encapsulates Octree-based domain decomposition methods for parallel computing. In prior work unrelated to the SBM, this approach has been applied to various multiphysics applications, such as two-phase flow dynamics~\citep{khanwale2023projection}, electrokinetic transport phenomena~\citep{kim2024direct}, and computational risk assessments for disease transmission~\citep{tan2023computational}. We employ a block-iterative strategy to couple different PDEs, as mentioned in \secref{block-iter}. 
Key features of \dendrokt~include:
\begin{itemize}
\item[(a)] \textbf{Complex geometry handling}: in-out tests~\citep{haines1994point,borazjani2008curvilinear,saurabh2021industrial} are used to efficiently determine point locations within intricate structures, enabling the creation of incomplete Octrees for complex geometries. 
\item[(b)] \textbf{Adaptive Mesh Refinement (AMR)}: high resolution is ensured in critical regions (e.g., boundary layers and wake areas) while maintaining computational efficiency. This AMR framework can be applied not only to spatial discretization~\cite{khanwale2020simulating,khanwale2023projection,tan2023computational,kim2024direct} but also to temporal discretization~\cite{ishii2019solving}.
\item[(c)] \textbf{Load balancing}: space-filling curves (SFC) are used to optimally distribute computational tasks across processors in distributed memory environments. 
\item[(d)] \textbf{Efficient matrix assembly}: innovative traversal methods streamline the assembly process by eliminating the need for traditional element mappings. 
\item[(e)] \textbf{2:1 refinement balancing}: stability and accuracy are enhanced by ensuring that adjacent octants differ by only one refinement level~\citep{sundar2008bottom,fernando2017machine,ishii2019solving}.
\end{itemize}

Additionally, efficient distance function calculations play an essential role in the SBM. To address this, we use the k-d tree \texttt{nanoflann} library~\citep{blanco2014nanoflann}. For further details on the implementation of these components, readers are referred to~\citet{yang2024optimal}.

\subsection{Block-iterative strategy}\label{block-iter}

The block-iterative strategy we adopt has proven effective in addressing various multi-physics problems, such as fluid-structure interaction~\citep{tezduyar2006space}, thermal incompressible flow~\citep{xu2019residual}, Cahn-Hillard Navier-Stokes coupled two-phase flows~\citep{khanwale2020simulating, khanwale2023projection}, and thermal free-surface flows~\citep{xu2022finite}. In our framework, we also utilize a block-iterative strategy to couple energy and flow dynamics. The flowchart of this approach is illustrated in \figref{fig:diagram_block}. Within each block, we solve the convection-diffusion equation, passing the resulting temperature to the Navier-Stokes equation, which is then solved. We check for convergence within the block. If the solution is not converged, we pass the velocity obtained from solving the Navier-Stokes equation back to the convection-diffusion equation and follow the above loop again. This loop continues until the multi-field solution converges to below a user-defined tolerance.

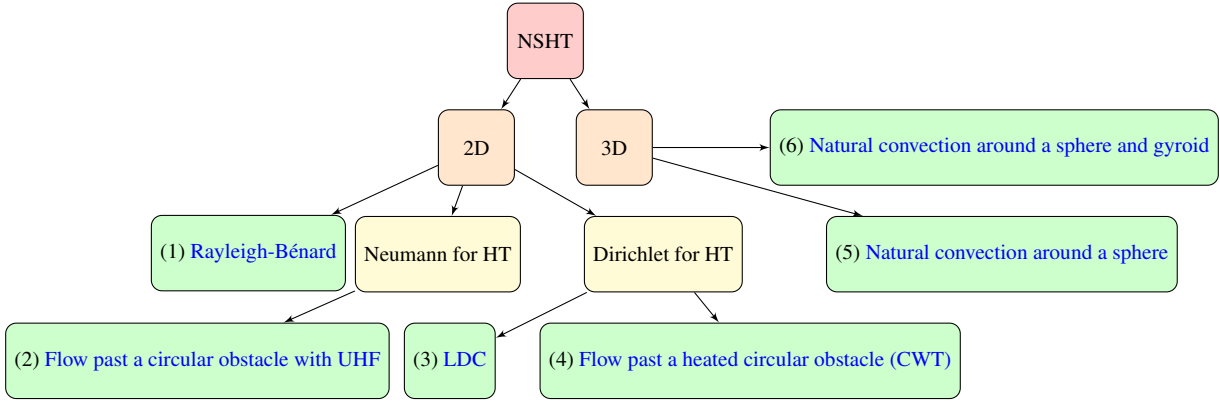
\begin{figure}[!t]
    \centering
\tikzstyle{box} = [rectangle, rounded corners, minimum width=1cm, minimum height=1cm, text centered, draw=black, fill=blue!20, font=\footnotesize]
\tikzstyle{lastbox} = [rectangle, rounded corners, minimum width=1cm, minimum height=1cm, text centered, draw=black, fill=green!20, font=\footnotesize]
\tikzstyle{line} = [draw, -latex']
\begin{tikzpicture}[node distance=2cm, auto]
  \node [box, anchor=east,fill=red!20] (nsht) {NSHT};
  \node [box, below right of=nsht, anchor=east,fill=orange!20] (3D) {3D};
  \node [box, below right of=3D, anchor=east, xshift=6cm,fill=green!20] (Robin_cube) {(5) \hyperref[Natural-3D]{Natural convection around a sphere}};
  \node [box, right of=3D, xshift=3cm, fill=green!20] (gyroid) {(6) \hyperref[Natural-3D-Gyroid]{Natural convection around a sphere and gyroid}};
  \node [box, below left of=nsht, anchor=west,fill=orange!20] (2D) {2D};
  \node [box, below right of=2D, xshift=0cm, anchor=west,fill=yellow!20] (dirichlet) {Dirichlet for HT};
  \node [box, below left of=2D, anchor=east,xshift=2cm,fill=yellow!20] (neumann) {Neumann for HT};
  \node [lastbox, below left of=neumann, xshift=-1.75cm] (uhf) {(2) \hyperref[subsub:UHF]{Flow past a circular obstacle with UHF}};
  \node [lastbox, below left of=dirichlet, xshift=-2cm, anchor=west] (ldc) {(3) \hyperref[subsub:Mixed_LDC]{LDC}};
  \node [lastbox, below right of=dirichlet, xshift=2.5cm, anchor=east] (cwt) {(4) \hyperref[subsub:CWT]{Flow past a heated circular obstacle (CWT)}};
  \node [lastbox, below of=2D, xshift=-3cm,yshift=0.6cm] (rb) {(1) \hyperref[sec:RB]{Rayleigh-B\'{e}nard}};
  \path [line] (nsht) -- (2D);
  \path [line] (nsht) -- (3D);
  \path [line] (3D) -- (Robin_cube);
  \path [line] (3D) -- (gyroid);
  \path [line] (2D) -- (dirichlet);
  \path [line] (2D) -- (neumann);
  \path [line] (dirichlet) -- (ldc);
  \path [line] (dirichlet) -- (cwt);
  \path [line] (neumann) -- (uhf);
  \path [line] (2D) -- (rb);
\end{tikzpicture}
    \caption{Schematic graph of various NSHT simulations performed in the paper.}
    \label{fig:2DNSHT}
\end{figure}

\section{Results}
\label{sec:results}

\figref{fig:2DNSHT} provides a bird's-eye view of the set of tests performed with the Linear semi-implicit Octree-SBM approach. Most of the simulation results are displayed in this section, except for the Rayleigh-B\'{e}nard results, which are discussed separately in \ref{sec:RB}. Additionally, we performed a numerical study using the method of manufactured solutions to validate the Linear semi-implicit Navier-Stokes solver in \ref{sec:NS_MMS}.

\subsection{Run-time comparison between {linear semi-implicit} and Standard (Non-Linear) {fully implicit} Navier-Stokes simulations}
\label{subsec:FPC}

\begin{figure}[!t]
    \centering
    \includegraphics[width=0.7\linewidth,trim=0 0 0 0,clip]{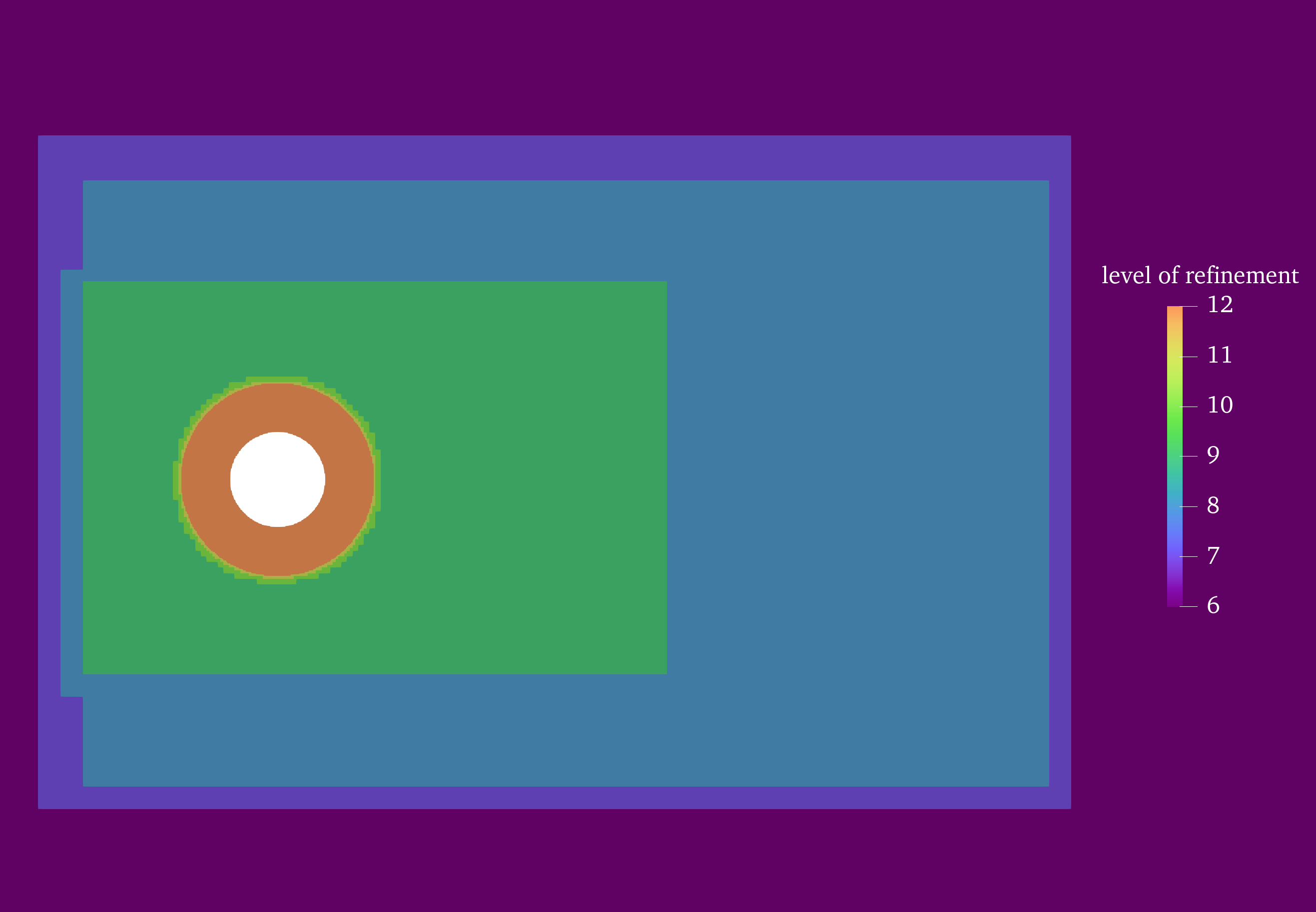}
    \caption{Two-dimensional flow past a cylinder (\secref{subsec:FPC}): plot of the various levels of mesh refinement. An increase of mesh resolution by 1 represents a reduction in element size by a factor of 2. Here, the finest element and the coarsest element vary by factor of 64 in length (and $64^2 = 4096$ in area).}
    \label{fig:LNS_NLNS_ElementSize}
\end{figure}

To demonstrate the benefit of using the linear {semi-implicit} Navier-Stokes equations in simulations, we selected a canonical problem in incompressible flow, that is the two-dimensional flow past a fixed cylinder at a Reynolds number of 100. {In this subsection, we use the terms linear semi-implicit and non-linear fully implicit approaches. Readers may refer to \secref{weak-momentum-continuity} for a detailed explanation of this terminology.} The Navier-Stokes equations are solved with $\petsc$ using the GMRES (Generalized Minimal Residual) method with a restart value of 1000, coupled with the Additive Schwarz Method (ASM) preconditioner configured with an overlap of 10. The simulations were conducted on the TACC~\Frontera~system, utilizing 10 nodes and a total of 560 processors. The simulation domain is $[0, 30] \times [0, 20]$, featuring a circular obstacle with a radius of 0.5, positioned at the coordinates $(10, 10)$. A non-dimensional freestream velocity of (1, 0) was applied to all boundary walls, except for the outlet wall ($x+$), {where the outflow condition~(\eqnref{outflow-bc}) is weakly imposed and complemented with strong enforcement of a homogeneous pressure}. Local mesh refinement was applied to ensure sufficient resolution in regions with complex flow behavior. A circular region, centered at (10, 10) with a radius of 1, was refined to level 12 (mesh size = $30 \cdot 2^{-12}$) to capture critical boundary-layer flow features. Additionally, two rectangular refinement regions were used. The first rectangle spans $[8, 14] \times [8, 12]$ and was refined to level 9 (mesh size = $30 \cdot 2^{-9}$). The second rectangle spans $[8, 18] \times [7, 13]$ and was refined to level 8 (mesh size = $30 \cdot 2^{-8}$). These refinement strategies are summarized in \figref{fig:LNS_NLNS_ElementSize}. We used both linear {semi-implicit} Navier-Stokes and standard (non-linear) {fully implicit} Navier-Stokes solvers to solve this problem with two different timesteps, 0.01 and 0.002.
Our findings, reported in~\tabref{tab:Flow2DCircle}, indicate that both solvers yielded comparable results in terms of the drag coefficient, regardless of the time step. For simulations up to the final non-dimensional time of 10, the linear {semi-implicit} Navier-Stokes solver demonstrates superior performance, reducing the total running time by nearly 60$\%$ compared to the non-linear {fully implicit} Navier-Stokes solver when using a timestep of $\Delta t = 0.002$. The solving time for the non-linear {fully implicit} Navier-Stokes equations is used to normalize the solving time for the linear {semi-implicit} Navier-Stokes solver, clearly highlighting the computational efficiency gained by using the linear {semi-implicit} Navier-Stokes approach in \tabref{tab:LNS-LNS-Speedup}.

\begin{table}[t!]
    \centering
    \caption{Two-dimensional flow past a cylinder at $Re = 100$: results using both the Linear Navier-Stokes (LNS) as well as the Non-Linear Navier-Stokes~\citep{yang2024simulating} (NLNS) frameworks.}
    \begin{tabular}{L{8cm}L{2cm}R{1cm}}
        \toprule
        \multicolumn{2}{l}{\textbf{Study}}  &  \textbf{Cd}   \\      
        \midrule
         \multicolumn{2}{l}{\citet{liu1998preconditioned}} &  1.350   \\      
         \multicolumn{2}{l}{\citet{posdziech2007systematic}} &  1.310 \\      
         \multicolumn{2}{l}{\citet{wu2009implicit}} & 1.364  \\
         \multicolumn{2}{l}{\citet{yang2009smoothing}}  & 1.393 \\
         \multicolumn{2}{l}{\citet{rajani2009numerical}}    &  1.340 \\   
         \multicolumn{2}{l}{\citet{Kamensky:2015ch}} &  1.386  \\      
         \multicolumn{2}{l}{\citet{main2018shifted_2} (triangular grid)} &  1.360   \\      
         \multirow{2}{*}{{linear semi-implicit} Navier-Stokes solver (LNS)} & $\Delta t = 0.01$ &  1.351  \\
                                        & $\Delta t = 0.002$ & 1.350 \\
        \multirow{2}{*}{{Non-linear fully implicit} Navier-Stokes solver (NLNS)} & $\Delta t = 0.01$ &  1.351 \\
                                        & $\Delta t = 0.002$ & 1.350 \\
        \bottomrule
    \end{tabular}
    \label{tab:Flow2DCircle}
\end{table}

\begin{table}[t!]
    \centering
    \caption{Two-dimensional flow past a cylinder at $Re = 100$: solution times for the {linear semi-implicit} Navier-Stokes (LNS) solver and the non-linear {fully implicit} Navier-Stokes (NLNS) solver for different time steps.}
    \label{tab:LNS-LNS-Speedup}
    \begin{tabular}{M{3.5cm}R{3cm}R{3cm}R{3cm}}
        \toprule
        \textbf{Time Step (\( \Delta t \))} & \textbf{NLNS Time (s)} & \textbf{LNS Time (s)} & \textbf{Speed-Up (\( \frac{\text{NLNS}}{\text{LNS}} \))} \\ 
        \midrule
        0.002 & 1180.0 & 483.5 & 2.44 \\ 
        0.010  & 271.1  & 136.3 & 2.00 \\ 
        \bottomrule
    \end{tabular}
\end{table}

\begin{table}[h!]
    \caption{Thermal lid-driven cavity test with one circular obstacle (\secref{subsub:Mixed_LDC}). Boundary conditions for mixed convection (\secref{subsub:Mixed_LDC}).}
    \centering
    \small
    \renewcommand{\arraystretch}{1.5} 
    \setlength{\tabcolsep}{10pt}      
    \begin{tabular}{|m{4cm}|>{\columncolor[HTML]{D9EAFD}}m{5cm}|>{\columncolor[HTML]{FEEBC8}}m{5cm}|}
        \hline
        \textbf{Boundary} & \textbf{Navier-Stokes (Velocity components)} & \textbf{Heat Transfer (Temperature, $\theta$)} \\
        \hline
        \textbf{Top Wall}    & $u_x = 1, \; u_y = 0$ & $\theta = 0$ \\
        \hline
        \textbf{Bottom Wall} & $u_x = 0, \; u_y = 0$ & $\theta = 1$ \\
        \hline
        \textbf{Left Wall}   & $u_x = 0, \; u_y = 0$ & Zero flux: $\frac{\partial \theta}{\partial x} = 0$ \\
        \hline
        \textbf{Right Wall}  & $u_x = 0, \; u_y = 0$ & Zero flux: $\frac{\partial \theta}{\partial x} = 0$ \\
        \hline
    \end{tabular}
    \label{tab:boundary_conditions_LDC}
\end{table}

\subsection{Dirichlet boundary condition for heat transfer in two dimensions}
\subsubsection{Mixed convection in lid-driven cavity thermal flows}\label{subsub:Mixed_LDC}

In this test, obstacles in the shape of circular disks are placed inside a lid-driven cavity and simulations of mixed convection (the combination of natural and forced convection) are performed. The Richardson number (\(Ri\)) characterizes the flow regime: when \(Ri \ll 1\), the flow is dominated by forced convection; when \(Ri \gg 1\), it is dominated by natural convection; and when $Ri$ close to 1, both natural and forced convection contribute significantly. In this study, we conducted two types of simulations: the first involved a single circular obstacle inside the cavity, and the second involved two circular obstacles. 

\begin{figure}[!t]
    \centering
    \begin{subfigure}{0.3\textwidth}
    \centering
        \includegraphics[width=\linewidth,trim=700 150 700 200,clip]{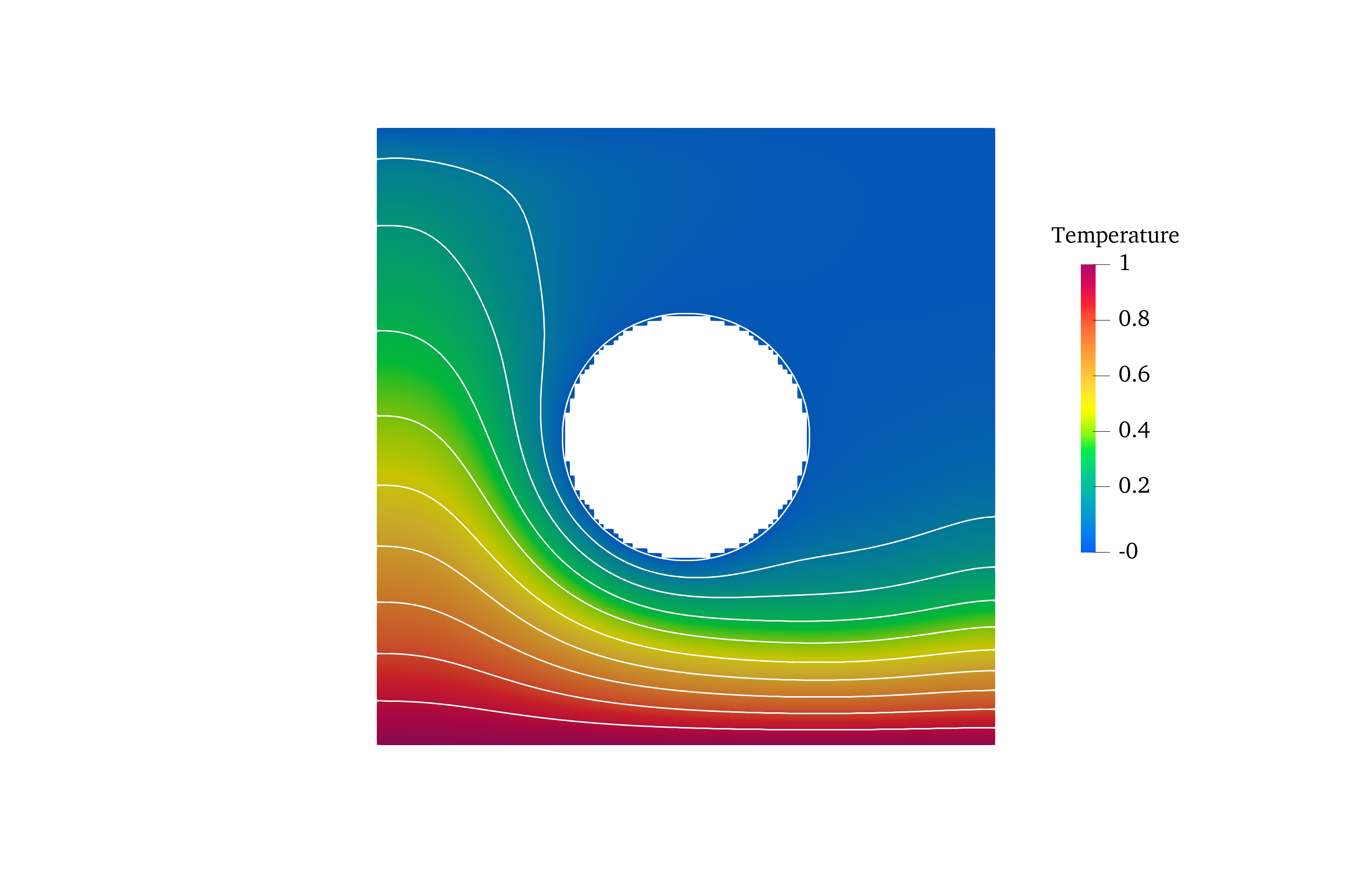}
        \caption{Ri = 0.01}
        \label{fig:LDC_oneCircle_Ri0p01}
    \end{subfigure}%
    \begin{subfigure}{0.3\textwidth}
    \centering
        \includegraphics[width=\linewidth,trim=700 150 700 200,clip]{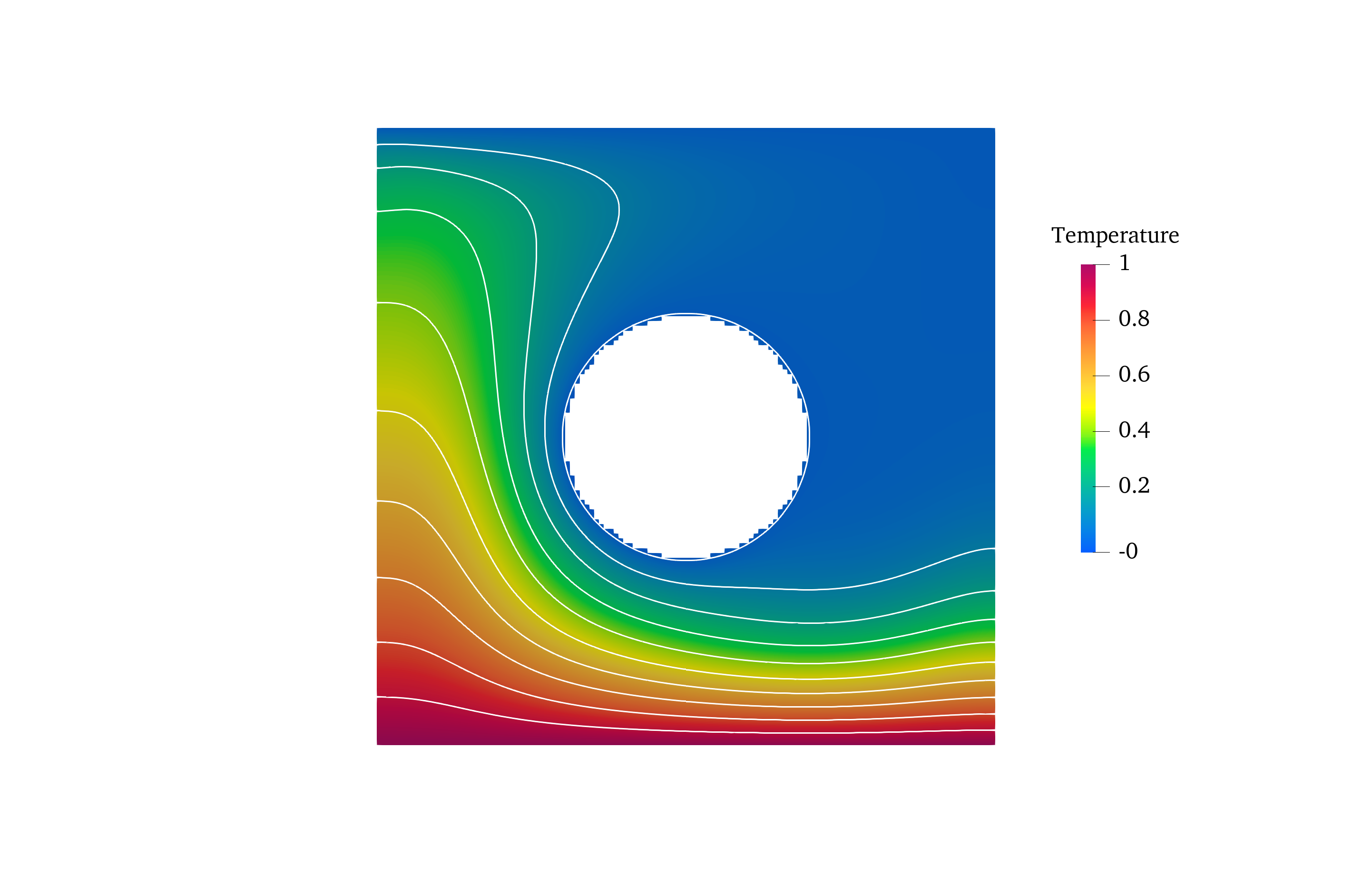}
        \caption{Ri = 1}
        \label{fig:LDC_oneCircle_Ri1}
    \end{subfigure}%
        \begin{subfigure}{0.3\textwidth}
        \centering
        \includegraphics[width=\linewidth,trim=700 150 700 200,clip]{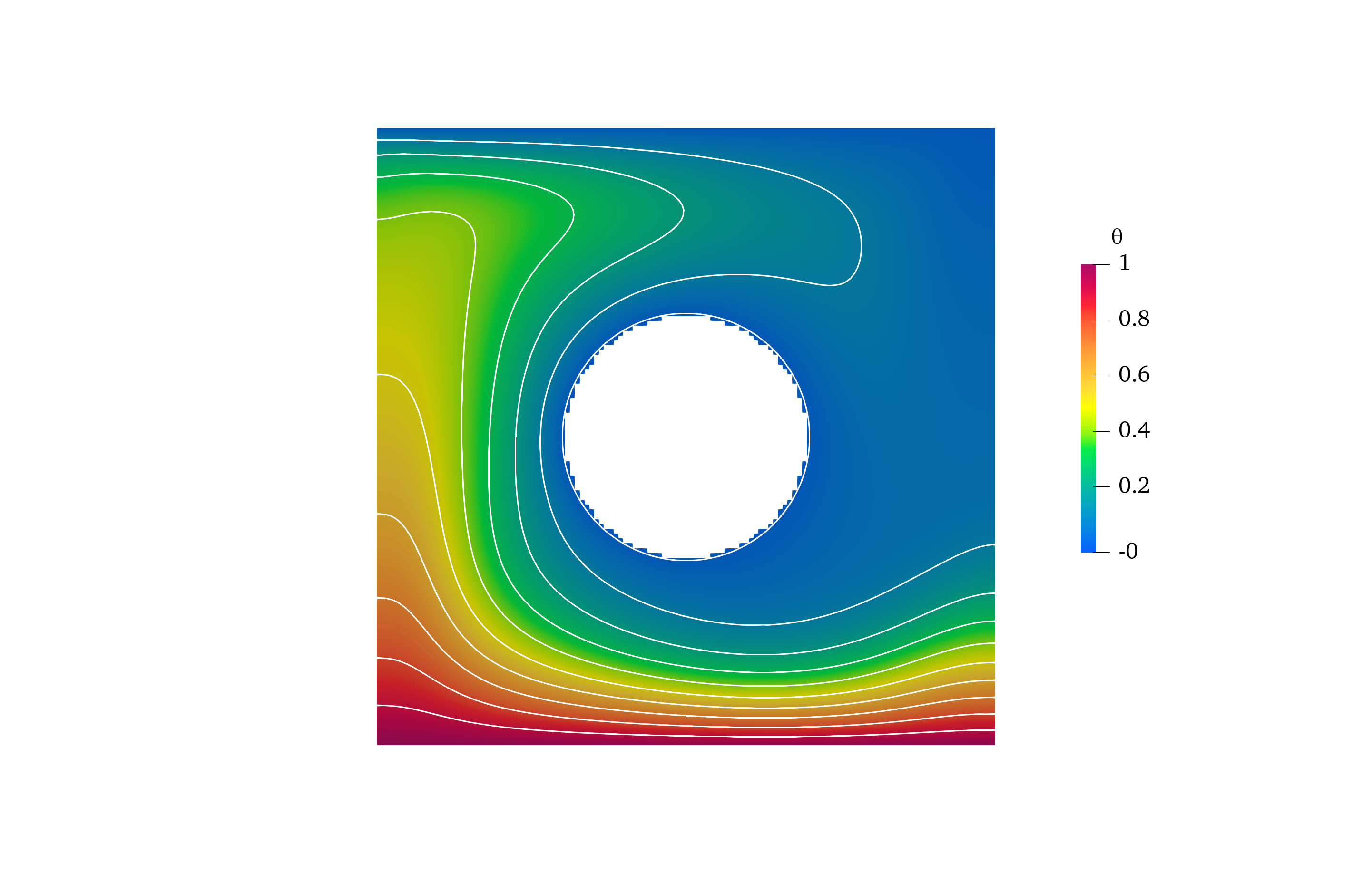}
        \caption{Ri = 5}
        \label{fig:LDC_oneCircle_Ri5}
    \end{subfigure}%
    \begin{subfigure}{0.09\linewidth}
        \centering
        \includegraphics[width=\linewidth,trim=2200 300 350 400,clip]{Ar2p5_Ri5_Paraview.png}
    \end{subfigure}%
    \caption{Thermal lid-driven cavity test with one circular obstacle (\secref{subsub:Mixed_LDC}): contour lines of the non-dimensional temperature at steady state are plotted from 0 to 1 at regular intervals of 0.1.}
    \label{fig:LDC_oneCircle}
\end{figure}

\begin{figure}[b!]
        \centering
        \includegraphics[width=0.5\linewidth,trim=0 0 0 0,clip]{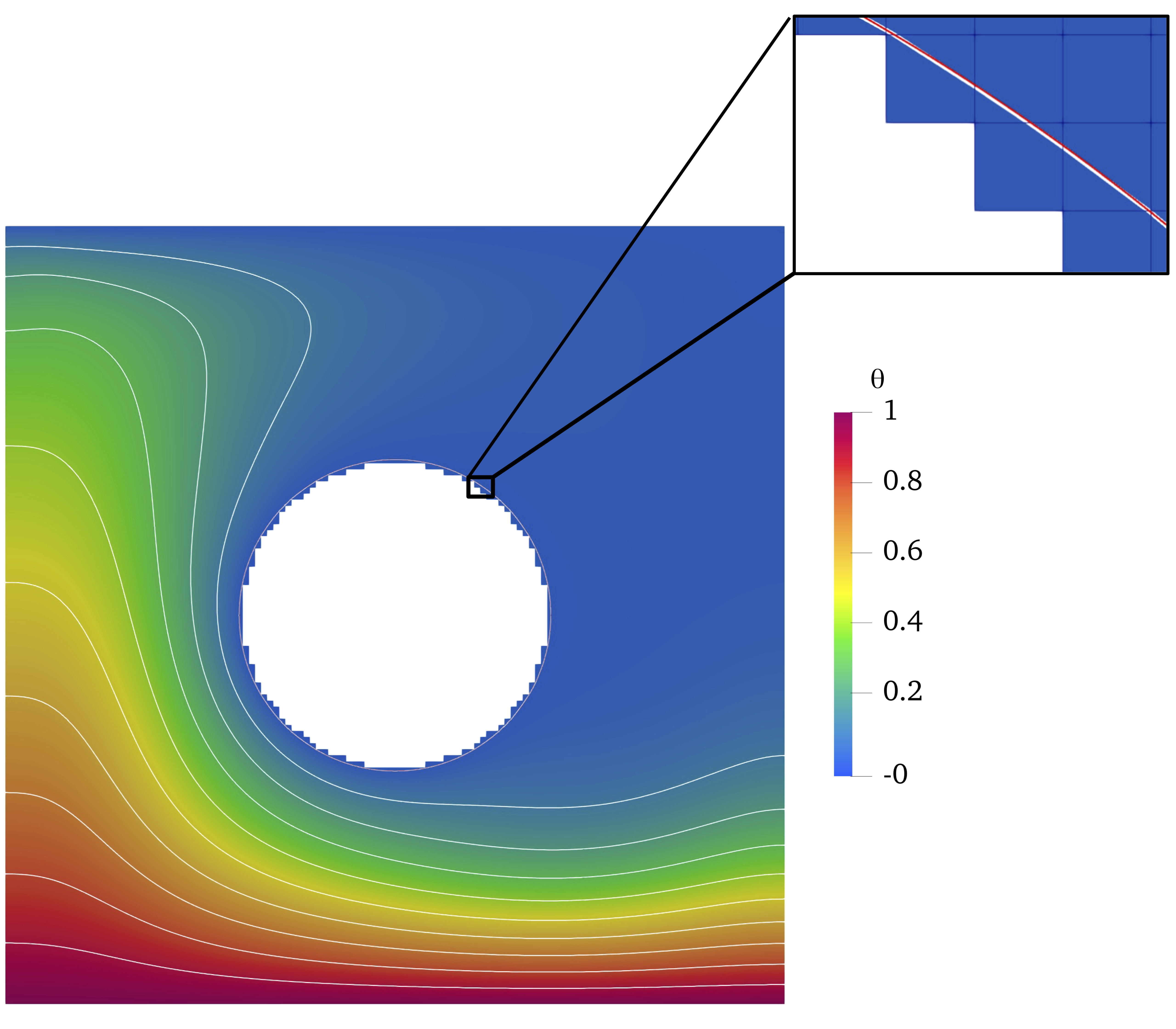}
        \caption{Thermal lid-driven cavity test with one circular obstacle (\secref{subsub:Mixed_LDC}): close-up of the temperature contour at steady state: the red line indicates the geometric boundary, while the white line represents the zero-temperature contour. The two closely match.}
    \label{fig:LDC_oneCircle_ZoomIn}
\end{figure}


\begin{figure}[b!]
\centering
\includegraphics[width=0.7\linewidth]{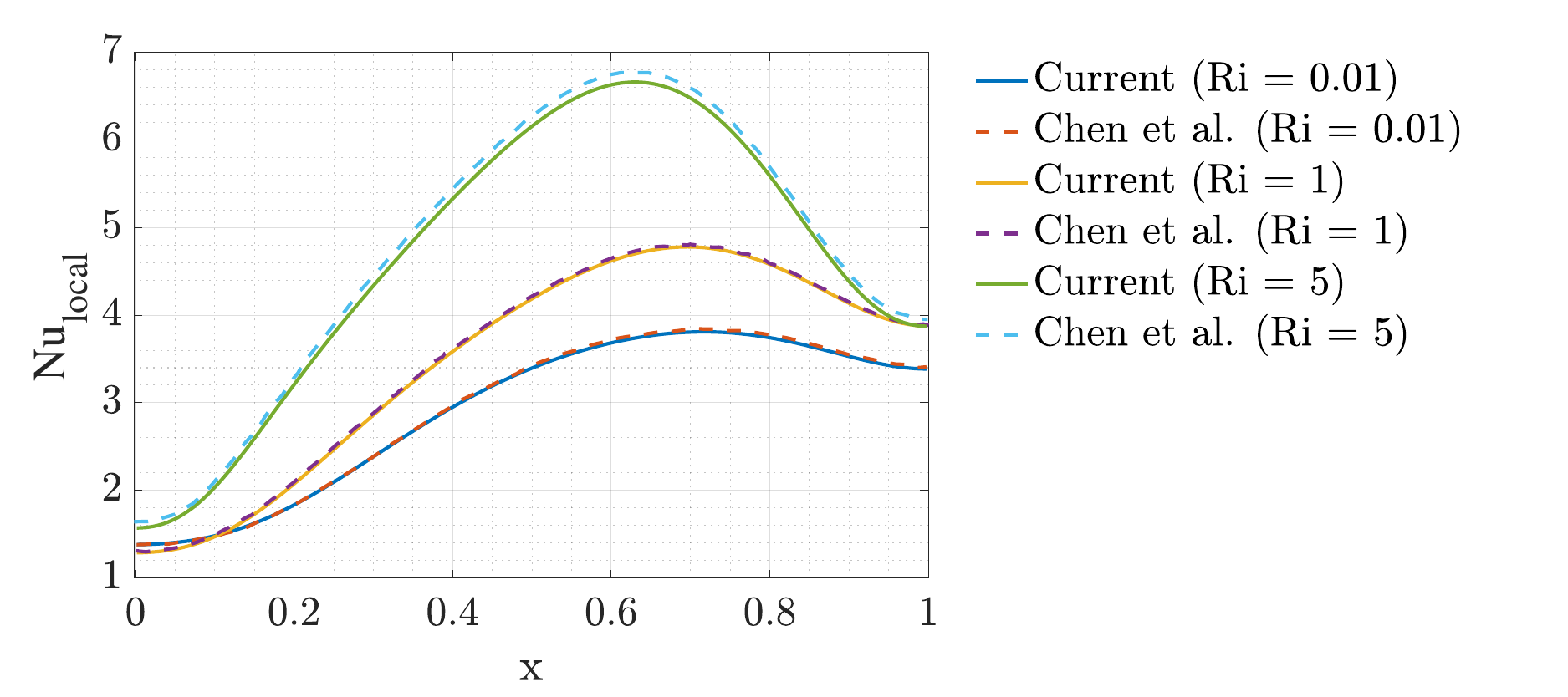}
\caption{Thermal lid-driven cavity test with one circular obstacle (\secref{subsub:Mixed_LDC}): comparison of the local Nusselt number along the bottom wall ($\nabla \theta \cdot \bs{n}$, i.e., the non-dimensional thermal flux), at steady state.
Here, $\bs{n}$ is the inward-facing normal vector directed toward the cylinder.
The local Nusselt number obtained with Octree-SBM simulations is compared against the results of~\citet{Chen2020}. }
\label{fig:LDC_NSHT_BottomNu}
\end{figure}

In the single-obstacle simulation, a disk with radius $0.2L$ is centered in a lid-driven cavity of unit length ($L = 1$). Using SBM, no-slip velocity and homogeneous temperature conditions are enforced on the disk, while the square domain's boundary conditions are detailed in \tabref{tab:boundary_conditions_LDC}.
The mesh resolution is \(128 \times 128\) (mesh size = $2^{-7}$). Since the problem we solved is a steady-state problem, we set the non-dimensional time step to 1 and run the simulation until the flow reaches steady state.

The non-dimensional parameters used are \(Re = 100\), \(Pr = 0.7\), and \(Ri\) values ranging from 0.01 to 5.0, covering the flow regime from forced convection to mixed convection and natural convection.
The temperature contours are shown in \figref{fig:LDC_oneCircle}. As \(Ri\) increases, buoyancy effects become stronger, causing the high-temperature region to extend further upward in the cavity. 

\begin{figure}[t!]
\centering
\includegraphics[width=1\linewidth]{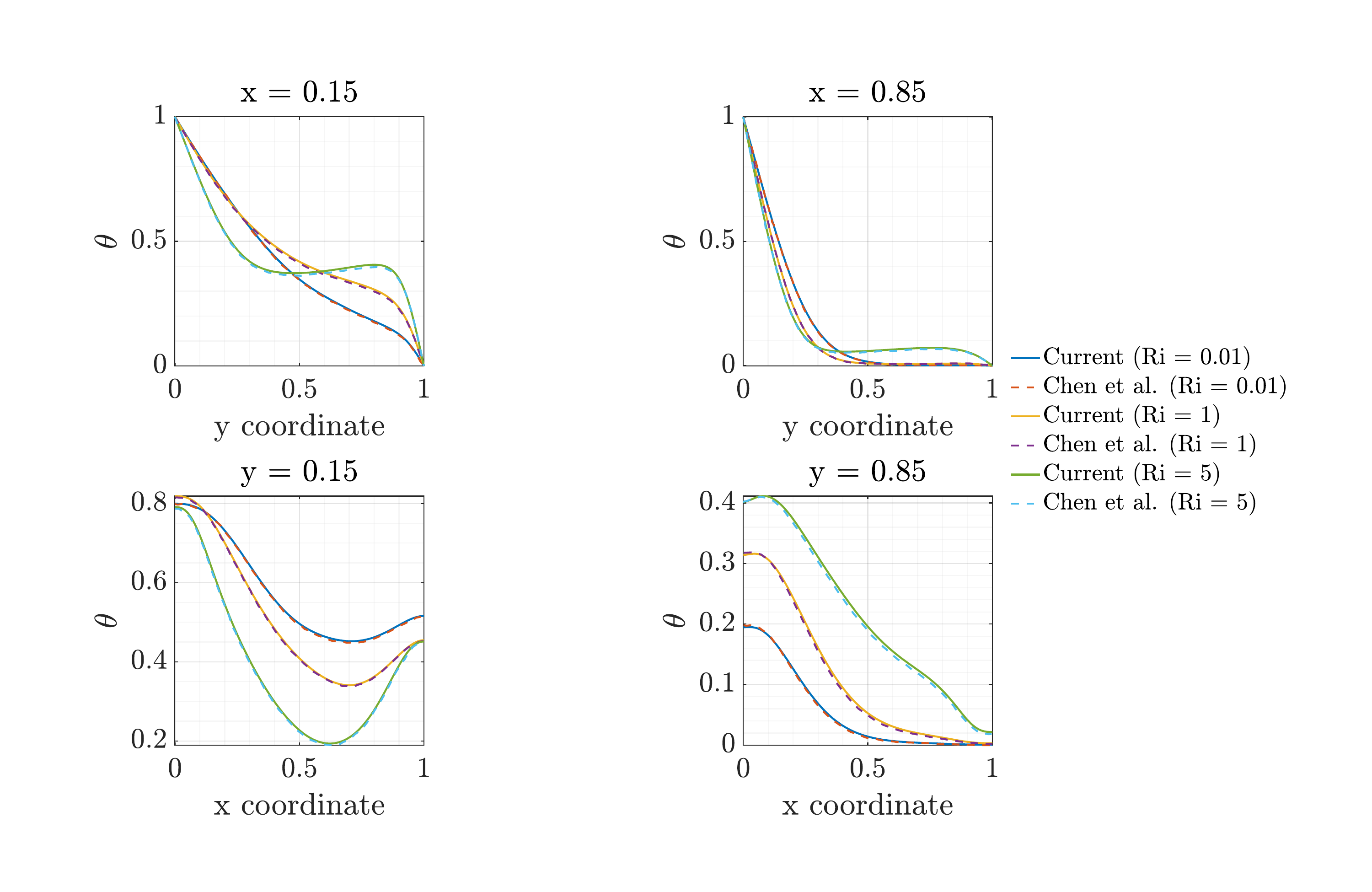}
\caption{Thermal lid-driven cavity test with one circular obstacle (\secref{subsub:Mixed_LDC}): comparison of the steady-state temperature distributions against the results of~\citet{Chen2020} at various Richardson numbers. The temperature profiles are shown along (a) \( x = 0.15 \), (b) \( x = 0.85 \), (c) \( y = 0.15 \), and (d) \( y = 0.85 \), illustrating the impact of Richardson numbers on convective patterns within the cavity.}
\label{fig:LDC_NSHT_Temperature}
\end{figure}

Notably, the contour line near the disk region in \figref{fig:LDC_oneCircle} represents the zero-temperature contour, closely following the shape of the disk. This observation demonstrates that the SBM effectively enforces the true boundary condition, even when applied at the surrogate Octree-based boundary. To further illustrate this, \figref{fig:LDC_oneCircle_ZoomIn} provides a zoomed view of the temperature contour at \(Ri = 1\), where the zero-temperature contour (white line) closely matches the circular boundary (red line). Nusselt number comparisons along the bottom wall are presented in \figref{fig:LDC_NSHT_BottomNu}. Additionally, we compared the temperature profiles along specific lines with data from the literature~\citep{Chen2020}, as shown in \figref{fig:LDC_NSHT_Temperature}.

Next, we analyze a lid-driven cavity with two circular obstacles, a test also described in~\citep{Khanafer2015}. 
The non-dimensional temperatures on the left and right circular obstacles are set to $\theta = 1$ and $\theta = 0$, respectively, and are enforced using the SBM. The boundary conditions for the walls in the Navier-Stokes equations remain identical to the single circular obstacle case. All walls of the square cavity have homogeneous temperature conditions.

\begin{figure}[!t]
    \centering
    \begin{subfigure}{0.28\linewidth}
    \centering
        \includegraphics[width=\linewidth,trim=650 250 650 250,clip]{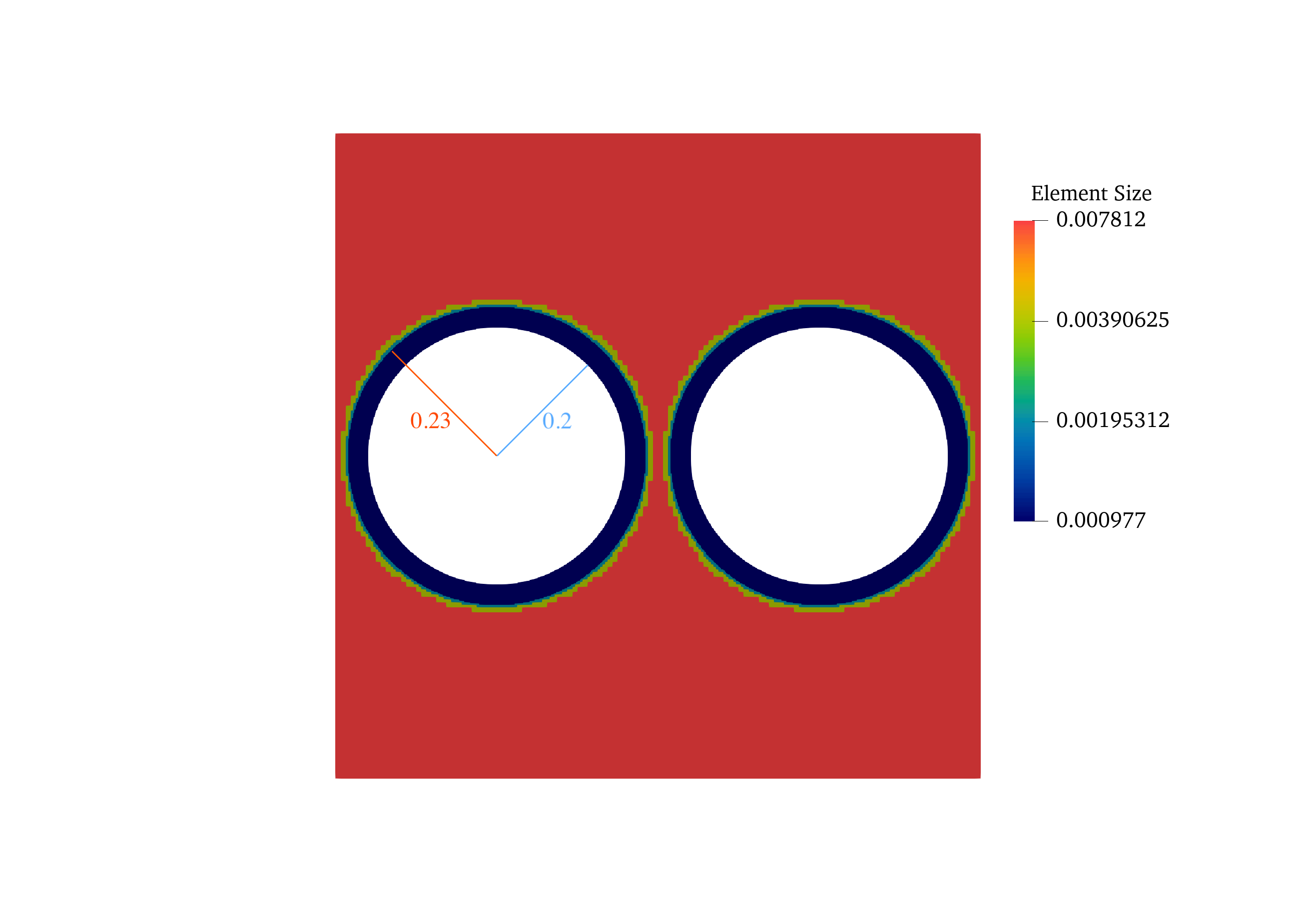}
        \caption{Element size}
        \label{fig:LDC_TwoCircle_ElementSize}
    \end{subfigure}%
    \begin{subfigure}{0.08\linewidth}
        \centering
        \includegraphics[width=\linewidth,trim=2000 250 290 250,clip]{ElementSize_TwoCircle.png}
    \end{subfigure}%
    \begin{subfigure}{0.28\linewidth}
    \centering
        \includegraphics[width=\linewidth,trim=650 250 650 250,clip]{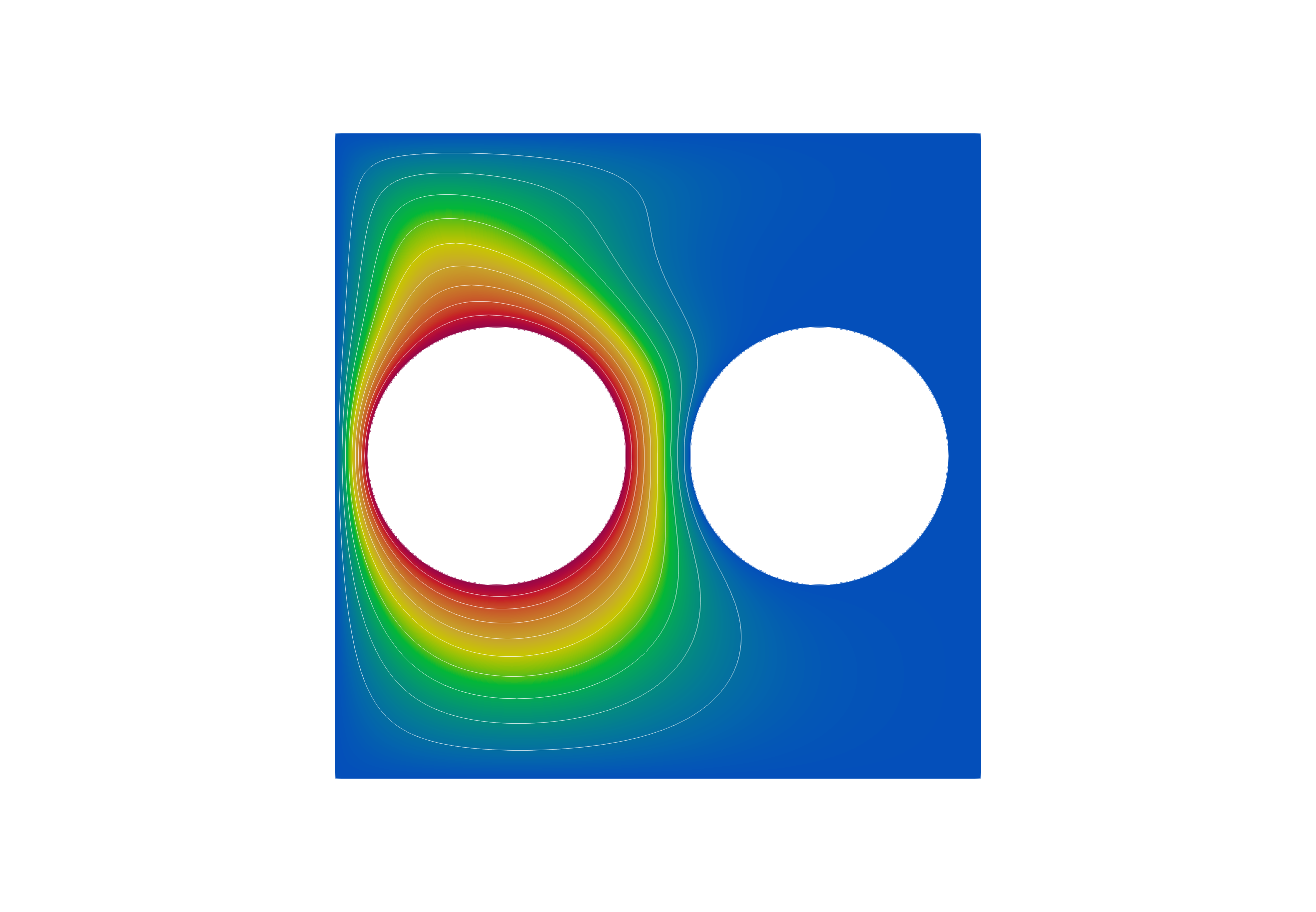}
        \caption{Ri = 0.01}
        \label{fig:LDC_TwoCircle_Ri0p01}
    \end{subfigure}%
    \begin{subfigure}{0.28\linewidth}
    \centering
        \includegraphics[width=\linewidth,trim=650 250 650 250,clip]{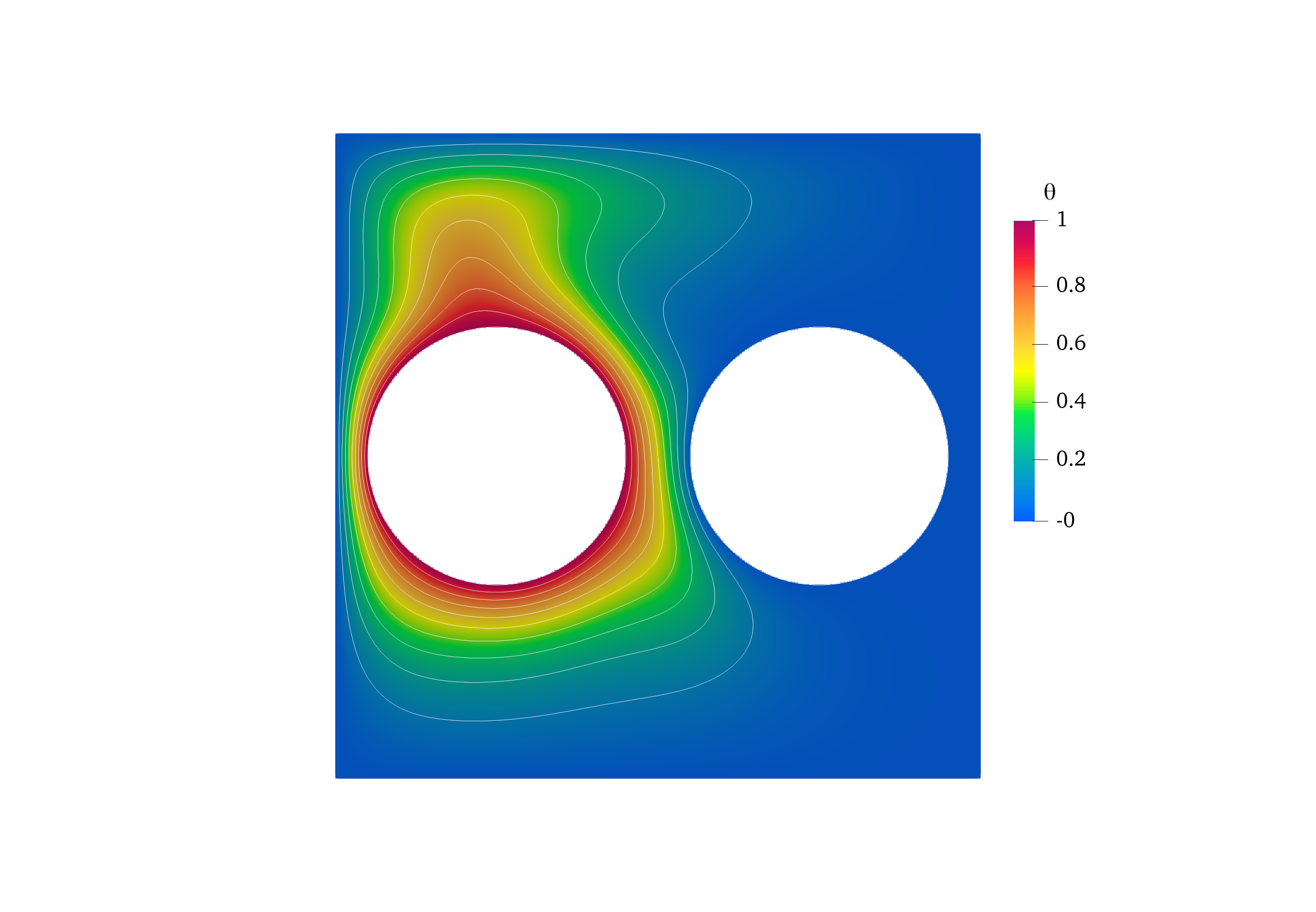}
        \caption{Ri = 10}
        \label{fig:LDC_TwoCircle_Ri1}
    \end{subfigure}%
    \begin{subfigure}{0.08\linewidth}
        \centering
        \includegraphics[width=0.9\linewidth,trim=2000 250 290 250,clip]{Ri10_match.png}
    \end{subfigure}%
    \caption{Thermal lid-driven cavity test with two circular obstacles (\secref{subsub:Mixed_LDC}): steady-state temperature contours and local mesh refinement. The contour lines represent non-dimensional temperatures ranging from 0 to 1, plotted at regular intervals of 0.1.}
    \label{fig:LDC_twoCircle}
\end{figure}

\begin{figure}[!t]
\centering
\begin{subfigure}{0.7\linewidth}
\centering
\includegraphics[width=0.99\linewidth]{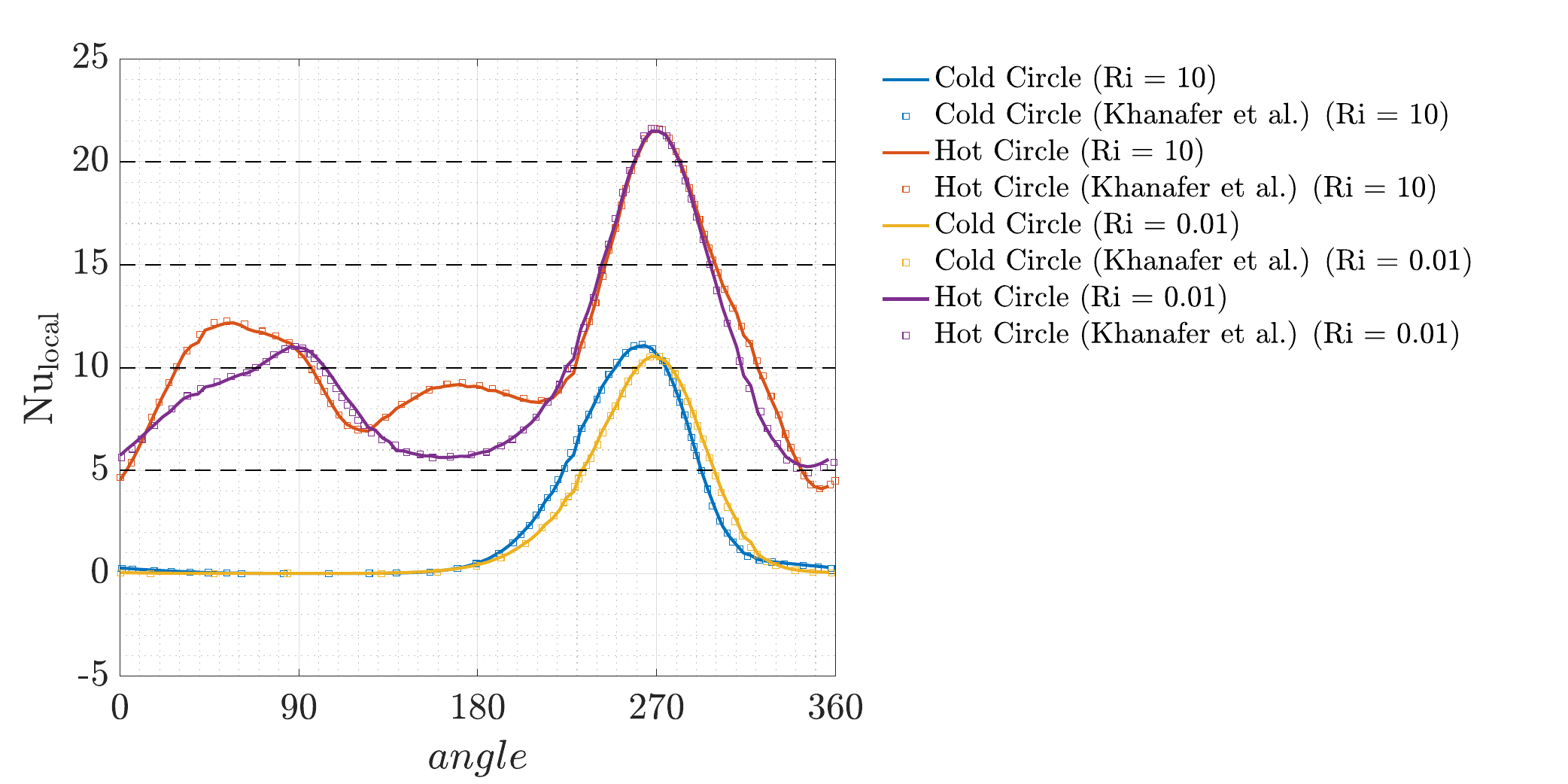}
\caption{Local Nusselt number $Nu$ over the true boundary}
\label{fig:sub1Nu}
\end{subfigure}%
\begin{subfigure}{0.29\linewidth}
\centering
\begin{tikzpicture}[scale=1]
\draw (-1.2,0) -- (1.2,0);
\draw (0,-1.2) -- (0,1.2);
\draw[dashed] (0,-1.1) -- (0,1.1);
\draw[-latex] (1,0) arc (0:-360:1);
\draw[dashed] (-1,0) -- (1,0);
\filldraw (0,1) circle (0.02) node[anchor=south west] {$0$};
\filldraw (1,0) circle (0.02) node[anchor=south west] {$\frac{\pi}{2}$};
\filldraw (0,-1) circle (0.02) node[anchor=north west] {$\pi$};
\filldraw (-1,0) circle (0.02) node[anchor=north east] {$\frac{3\pi}{2}$};
\node at (1,1) {$angle$};
\end{tikzpicture}
\caption{Angle notation}
\label{fig:angle_rot}
\end{subfigure}
\caption{Thermal lid-driven cavity test with two circular obstacles (\secref{subsub:Mixed_LDC}): distribution at steady state of the local Nusselt number ($Nu_{local} = \nabla \theta \cdot \bs{n}$, i.e., a non-dimensional thermal flux) and comparison against the simulations of~\citet{Khanafer2015}.}
\label{fig:TwoCircleTrueBoundary}
\end{figure}
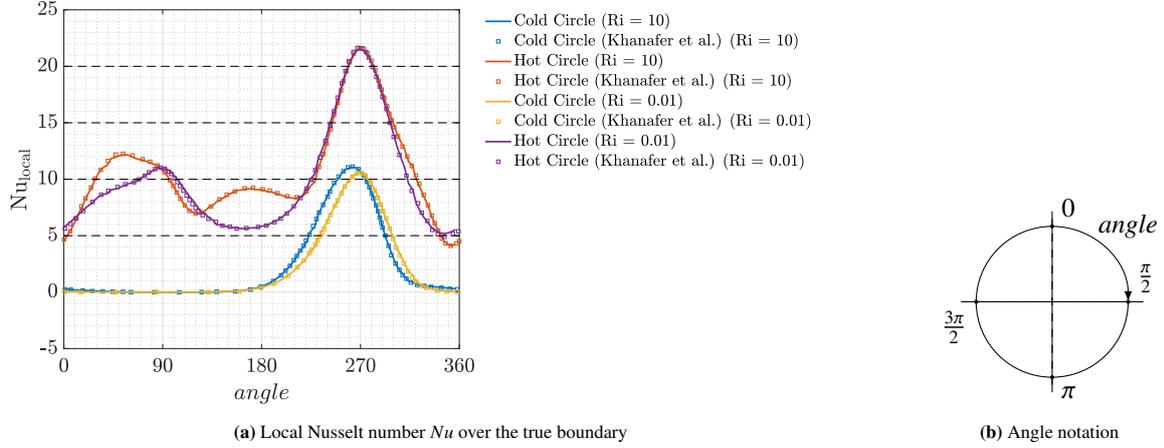

\begin{figure}[!t]
\centering
\begin{subfigure}{0.58\textwidth}
\centering
\includegraphics[width=\linewidth]{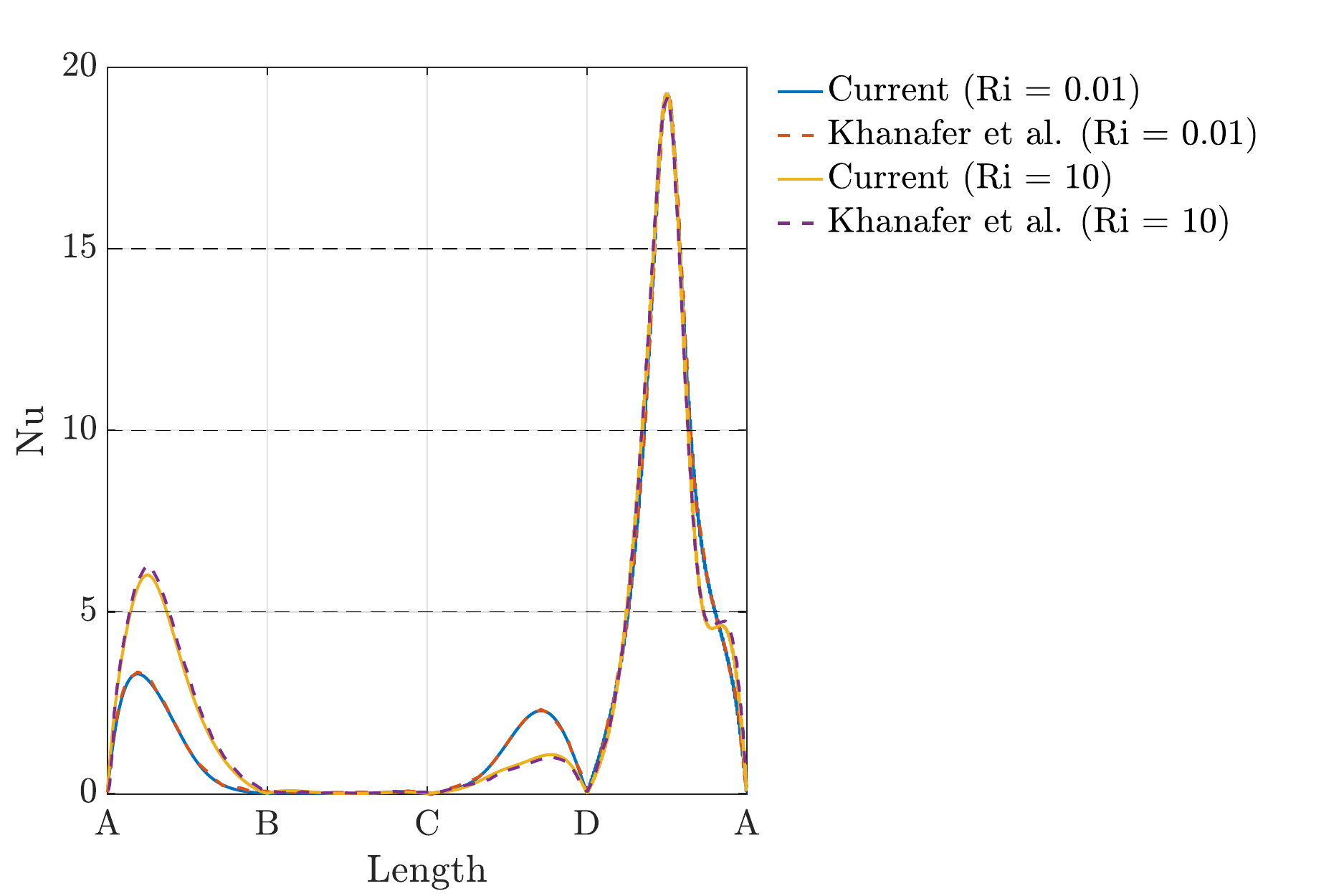}
\caption{Local Nu over walls}
\label{fig:sub1Walls}
\end{subfigure}%
\begin{subfigure}{0.3\textwidth}
\centering
\begin{tikzpicture}[thick]
\definecolor{myblue}{RGB}{68,114,196}
\definecolor{mygreen}{RGB}{85,168,104}

\draw[myblue] (0,0) node[below left] {$D$} --
              (0,2) node[above left] {$A$} --
              (2,2) node[above right] {$B$} --
              (2,0) node[below right] {$C$} --
              cycle;
      
\draw[-{Latex[length=2mm]}, mygreen] (0,0) -- (0,2);
\draw[-{Latex[length=2mm]}, mygreen] (0,2) -- (2,2);
\draw[-{Latex[length=2mm]}, mygreen] (2,2) -- (2,0);
\draw[-{Latex[length=2mm]}, mygreen] (2,0) -- (0,0);

\foreach \x/\y in {0/0,0/2,2/2,2/0}{
    \filldraw (\x,\y) circle (2pt);
}
\end{tikzpicture}
\caption{Labels of ABCD}
\label{fig:labels_ABCD}
\end{subfigure}
\caption{Thermal lid-driven cavity test with two circular obstacles (\secref{subsub:Mixed_LDC}): steady-state local Nusselt number distribution (\(Nu_{local} = \nabla \theta \cdot \bs{n}\), non-dimensional thermal flux) along the cavity walls, compared with~\citet{Khanafer2015}.}
\label{fig:Nu4side}
\end{figure}

The base refinement level is set to 7, corresponding to a mesh size of $2^{-7}$.
To accurately capture the Nusselt number on the boundaries of circular obstacles, we apply local mesh refinement with a refinement level of 10 (mesh size = $2^{-10}$), in the circular crown between the radii 0.2 and 0.23, as shown in \figref{fig:LDC_TwoCircle_ElementSize}. 
The nondimensional parameters are $Re = 100$, $Pr = 0.7$, and $Ri = 0.01$ and $10.0$. The temperature contours for \( Ri = 0.01 \) and \( Ri = 10 \) are presented in \figref{fig:LDC_TwoCircle_Ri0p01} and \figref{fig:LDC_TwoCircle_Ri1}, respectively. The local Nusselt numbers on circular obstacles and wall boundaries are illustrated in \figref{fig:TwoCircleTrueBoundary} and \figref{fig:Nu4side}. In \figref{fig:TwoCircleTrueBoundary}, we compare the Nusselt number on the true boundary, which involves calculating the first derivative of our numerical results inside the \texttt{Intercepted} elements: the SBM simulations can accurately capture this quantity.

\subsubsection{Flow past a heated cylinder with constant wall temperature (CWT)}\label{subsub:CWT}

\begin{figure}[!t]
    \centering
    \includegraphics[width=\linewidth,trim=0 0 0 0,clip]{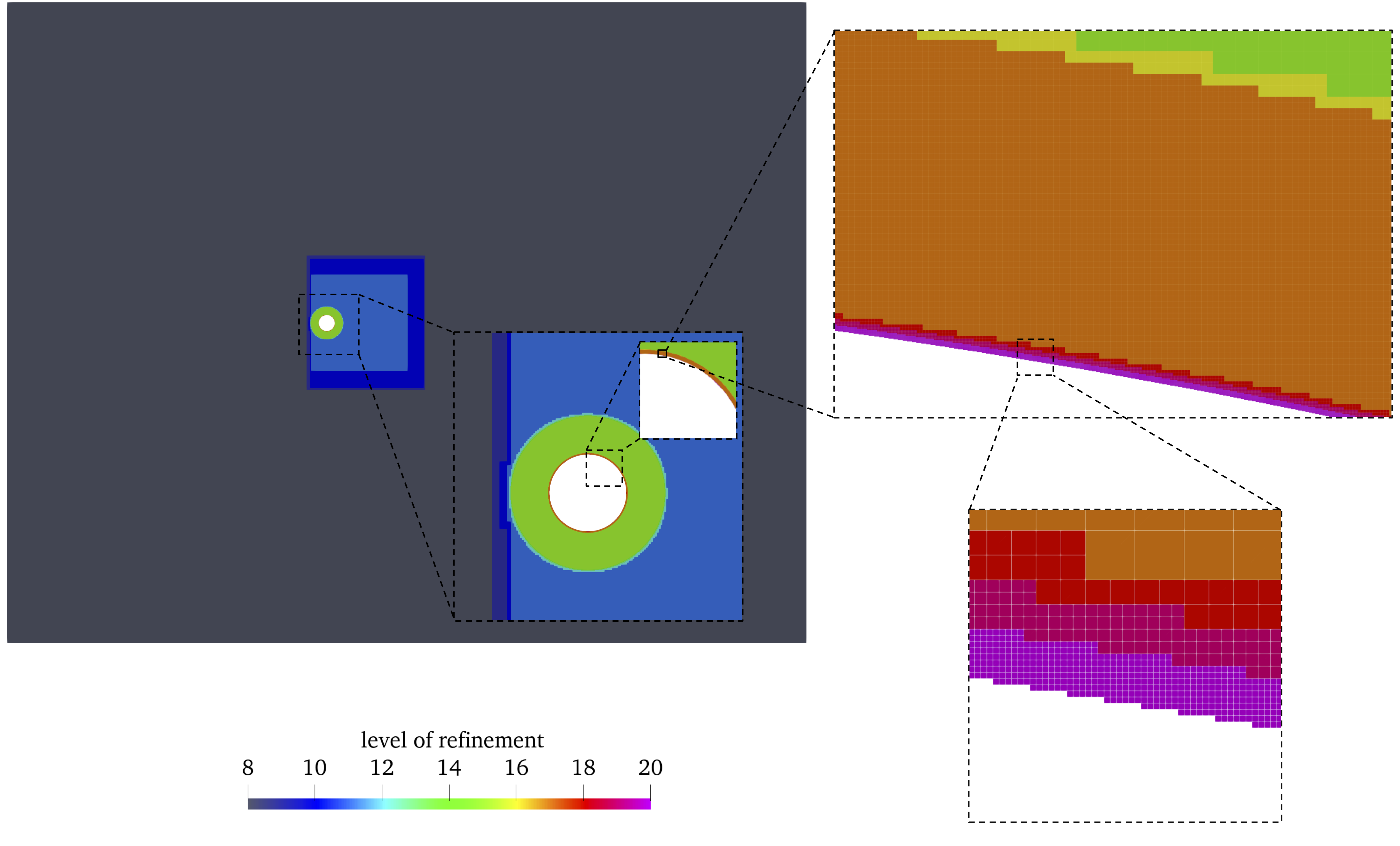}
    \caption{Flow past a heated cylinder with constant wall temperature (CWT, \secref{subsub:CWT}): distribution of the levels of mesh refinement. Note that the mesh is highly refined near the cylinder boundary, where the thermal boundary layer forms.}
    \label{fig:CWT_ElementSize}
\end{figure}

\begin{figure}[!t]
    \centering
    \includegraphics[width=0.5\linewidth,trim=0 0 0 0,clip]{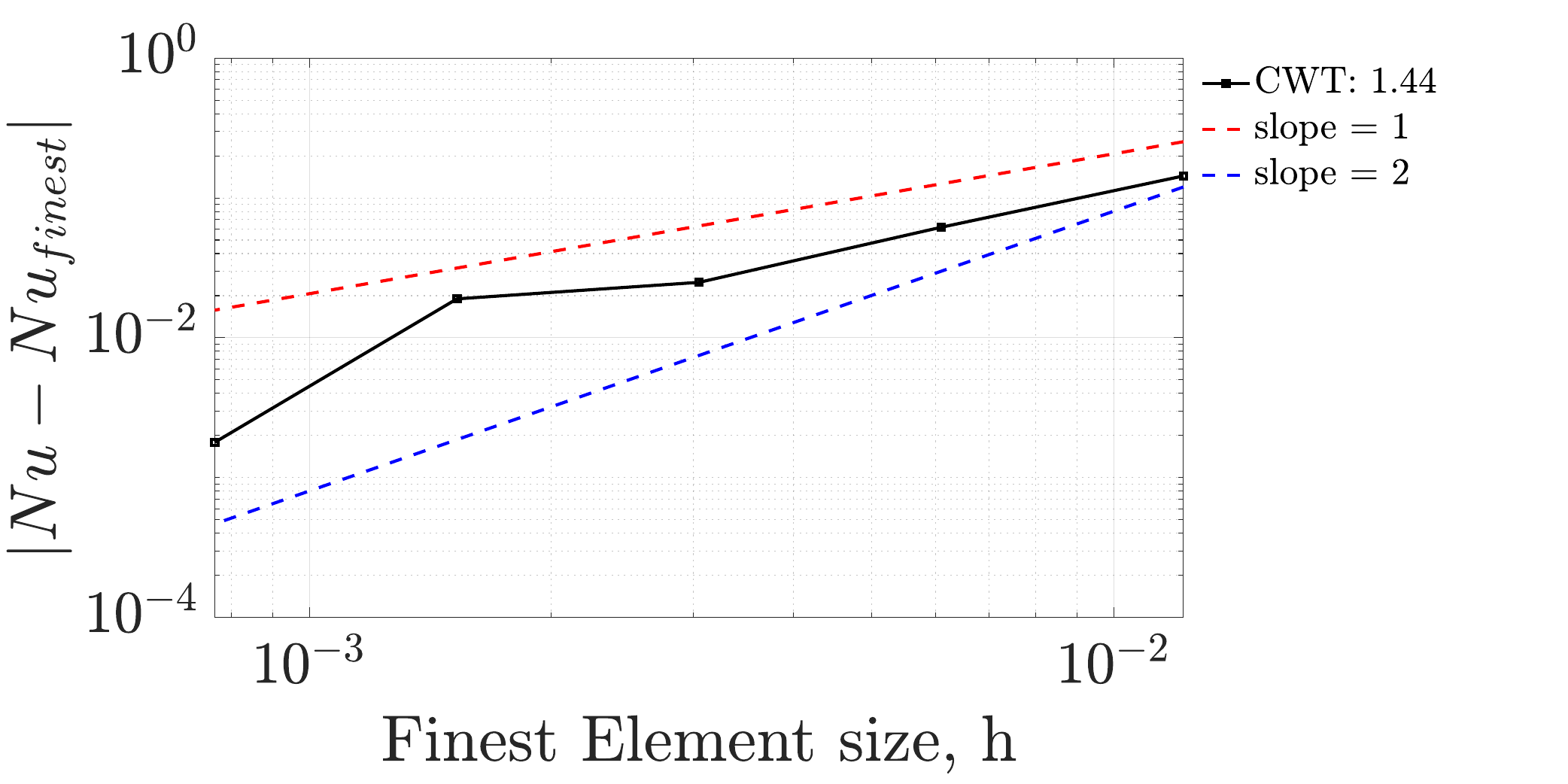}
    \caption{Mesh convergence study for the time-averaged global Nusselt number ($Nu = P^{-1} \, \int_\Gamma \nabla \theta \cdot \bs{n} \, d\Gamma$, where $P$ represents the perimeter of the circular cylinder) in the flow past a heated cylinder with constant wall temperature at $Re = 100$ (CWT, see \secref{subsub:CWT}).}
    \label{fig:CWT_Nu_MeshRefinementStudy}
\end{figure}

We next compare our results with several studies on flow past a circular obstacle with a constant wall temperature (CWT)~\citep{scholten1998unsteady, szczepanik2004numerical, nakamura2004variation, zhukauskasheat, pachpute2015numerical,hsu2021heat}. For these simulations, no-slip and $\theta = 1$ boundary conditions are applied on the boundary of the circular obstacle using the SBM. In addition to the circular obstacle, the boundary conditions on the domain walls are described in \tabref{tab:boundary_conditions_CWT}. Simulations encompass a wide range of Reynolds numbers (from 100 to 50,350) and backflow stabilized boundary conditions are employed as a precaution. The simulations are conducted under forced convection, involving one-way coupling where the heat transfer is influenced by the flow, but the flow remains unaffected by the heat. The simulation domain spans \([0, 50] \times [0, 40]\), with the circular obstacle positioned at the center \((20, 20)\). Details of the mesh configurations are provided in \tabref{table:MeshCWT}. Compared to the mesh size used in other studies addressing the forced convection problem at $Re = 100$, such as~\citet{golani2014fluid} (mesh size = $0.0015$), our finest mesh size near the circular disk is approximately $1.9073 \times 10^{-4}$, about 3.9 times smaller. Given that our approach employs a non-boundary-fitted mesh and linear basis functions, this ratio is reasonable and enables meaningful comparisons of both local and global Nusselt numbers with studies using boundary-fitted meshes.
{We conducted a mesh refinement study on the time-averaged global Nusselt number at $Re = 100$. 
The base mesh refinement level is 8 and the mesh refinement strategy is identical to that shown in \figref{fig:CWT_ElementSize}, except that the finest element refinement level is set to 17. Two circular refinement regions are defined, sharing the same center as the circular disk geometry: (a) a radius of $0.52$ with a refinement level of $17$, and (b) a radius of $1$ with a refinement level of $15$. Additionally, two rectangular refinement regions are defined: (a) $[19, 25] \times [17, 23]$ with a refinement level of $11$, and (b) $[19, 26] \times [16, 24]$ with a refinement level of $10$. The simulation was performed using a non-dimensional time of 0.05. The simulation with these mesh refinement levels is considered the ground truth. To perform the mesh convergence study shown in \figref{fig:CWT_Nu_MeshRefinementStudy}, we systematically reduced the refinement levels across the mesh. The resulting order of accuracy is approximately $1.44$, which exceeds the theoretical maximum of $1$ achievable with linear basis functions for the first derivative. This slightly higher order of accuracy may be attributed to two factors: (1) the "ground truth" solution is not the exact solution but rather the result of the highest refinement level simulation; and (2) the octree framework employed in this study represents every element as a rectangle in two dimensions, with potentially some super-convergence effects.}
The results of simulations demonstrate the effectiveness of the SBM combined with local mesh refinement, for a wide range of Reynolds numbers. \figref{fig:CWT_ElementSize} depicts the refinement level distribution and in particular the fine boundary mesh near the circular obstacle, required to accurately capture the thermal boundary layer. The simulation results are presented in \figref{fig:ReAndNuForCWT} and summarized in \tabref{table:NuandReCWT}.

Our results match closely those reported in the literature, including experimental and numerical studies. This agreement covers the range of Reynolds numbers from 100 to 50,350. Additionally, we computed the local Nusselt number (nondimensional heat flux) on the actual boundary of the circle, achieving an excellent match with the literature, as shown in \figref{fig:PolarNu_Re100and500}, which demonstrates that the SBM can accurately capture the first derivative of the temperature over the \Intercepted{} elements.

\begin{table}[t!]
    \caption{Flow past a heated cylinder with constant wall temperature (CWT, \secref{subsub:CWT}): boundary conditions.}
    \centering
    \renewcommand{\arraystretch}{1.5} 
    \setlength{\tabcolsep}{8pt}      
    \begin{tabular}{|m{3cm}|>{\columncolor[HTML]{D9EAFD}}m{4.5cm}|>{\columncolor[HTML]{FEEBC8}}m{7cm}|}
        \hline
        \textbf{Boundary} & \textbf{Navier-Stokes (Velocity)} & \textbf{Heat Transfer (Temperature, $\theta$)} \\
        \hline
        \textbf{Inlet (Left Side)}  & $u_x = 1, \; u_y = 0$ & $\theta = 0$ \\
        \hline
        \textbf{Outlet (Right Side)} & Backflow stabilization & Backflow stabilization for temperature \\
        \hline
        \textbf{Top Wall}           & $u_x = 1, \; u_y = 0$ & $\theta = 0$ \\
        \hline
        \textbf{Bottom Wall}        & $u_x = 1, \; u_y = 0$ & $\theta = 0$ \\
        \hline
    \end{tabular}
    \label{tab:boundary_conditions_CWT}
\end{table}

\begin{table}[t!]
    \centering
    \caption{Flow past a heated cylinder with constant wall temperature (CWT, \secref{subsub:CWT}): mesh specifications for varying Reynolds numbers.}
    \renewcommand{\arraystretch}{1.5} 
    \begin{tabular}{L{4.0cm}R{2.0cm}R{2.0cm}R{2.0cm}R{2.1cm}R{2.1cm}R{2.1cm}}
    \hline
    \textbf{Reynolds Number} & \textbf{100} & \textbf{500} & \textbf{7190} & \textbf{21580} & \textbf{35950} & \textbf{50350} \\
    \hline
    $\frac{\Tilde{D}}{h}$ & 2622 & 2622 & 10486 & 20971 & 20971 & 20971 \\
    \hline
    Total mesh nodes & 1573621 & 1573621 & 1849272 & 2252484 & 2252484 & 2252484 \\
    \hline
    \end{tabular}
    \label{table:MeshCWT}
\end{table}

\begin{table}[t!]
\centering
\setlength{\extrarowheight}{2pt}
\caption{Flow past a heated cylinder with constant wall temperature (CWT, \secref{subsub:CWT}): comparison of the time-averaged global Nusselt number against various literature sources (\secref{subsub:CWT}). The global Nusselt number is calculated based as $Nu = P^{-1} \, \int_\Gamma \nabla \theta \cdot \bs{n} \, d\Gamma$, where $P$ represents the perimeter of the circular cylinder. } 
\begin{tabular}{L{4.5cm}R{1.5cm}R{1.5cm}R{1.5cm}R{1.5cm}R{1.5cm}R{1.5cm}}
\hline
\textbf{Reynolds Number} & \textbf{100} & \textbf{500} & \textbf{7190} & \textbf{21580} & \textbf{35950} & \textbf{50350} \\
\hline
\citet{scholten1998unsteady} & \phantom{0} & \phantom{0} & 51.00 & 103.40 & 127.50 & 155.10 \\
\citet{szczepanik2004numerical} & \phantom{0} & \phantom{0} & 67.30 & 148.00 & \phantom{0} & 191.10 \\
\citet{nakamura2004variation} & 6.21 & 13.19 & 51.68 & 102.17 & \phantom{0} & \phantom{0} \\
\citet{zhukauskasheat} & 5.10 & 10.78 & 47.3 & 91.30 & 124.00 & 151.70 \\
\citet{pachpute2015numerical} & 5.18 & 12.17 & 55.52 & 111.45 & 142.86 & 171.28 \\
\citet{hsu2021heat} & 5.25 & 12.28 & \phantom{0} & \phantom{0} & \phantom{0} & \phantom{0} \\
\hline
Octree-SBM & 5.13 & 11.93 & 56.40 & 101.74 & 132.09 & 159.29 \\
\hline
\end{tabular}
\label{table:NuandReCWT}
\end{table}

\newcommand{\ft}{1.2} 
\newcommand{\legendft}{1.0} 
\begin{figure}[!t]
    \centering
    \begin{subfigure}[b]{0.65\textwidth} 
    \centering
    \begin{tikzpicture}
    \begin{loglogaxis}[
        width=\linewidth,
        scaled y ticks=true,
        xlabel={\scriptsize\scalebox{\ft}{$Re$}},
        ylabel={\scriptsize\scalebox{\ft}{$Nu$}},
        legend entries={
            \scriptsize\scalebox{\ft}{\citet{zhukauskasheat}},
            \scriptsize\scalebox{\ft}{\citet{pachpute2015numerical}},
            \scriptsize\scalebox{\ft}{Octree-SBM},
            \scriptsize\scalebox{\ft}{$slope = 2$}
        },
        legend style={
            at={(0.05,0.95)}, 
            anchor=north west, 
            nodes={scale=\legendft, transform shape} 
        },
        legend columns=1,
        xtick={100, 500, 7190, 21580, 35950, 50350},
        xticklabels={
            \scriptsize\scalebox{\ft}{100}, 
            \scriptsize\scalebox{\ft}{500}, 
            \scriptsize\scalebox{\ft}{7190}, 
            \scriptsize\scalebox{\ft}{21580}, 
            \scriptsize\scalebox{\ft}{35950}, 
            \scriptsize\scalebox{\ft}{50350}
        },
        xmin=80,
        xmax=60000,
        xticklabel style={
            rotate=45, 
            anchor=north east, 
            font=\scriptsize\scalebox{\ft}{}
        },
        grid,
        tick label style={font=\scriptsize\scalebox{\ft}{}},
        label style={font=\scriptsize\scalebox{\ft}{}}
    ]
    \addplot table [x={Re}, y={Zukauskas}, col sep=comma] {CWT.txt};
    \addplot +[dashed] table [x={Re}, y={Pachpute}, col sep=comma] {CWT.txt};
    \addplot table [x={Re}, y={Current}, col sep=comma] {CWT.txt};
    \end{loglogaxis}
    \end{tikzpicture}
    \caption{Time-averaged global Nusselt number ($Nu$) versus Reynolds number ($Re$).}
    \label{fig:ReAndNuForCWT}
    \end{subfigure}
    \hspace{0.1cm}
    \begin{subfigure}[b]{0.75\textwidth}
        \centering
        \includegraphics[width=\linewidth]{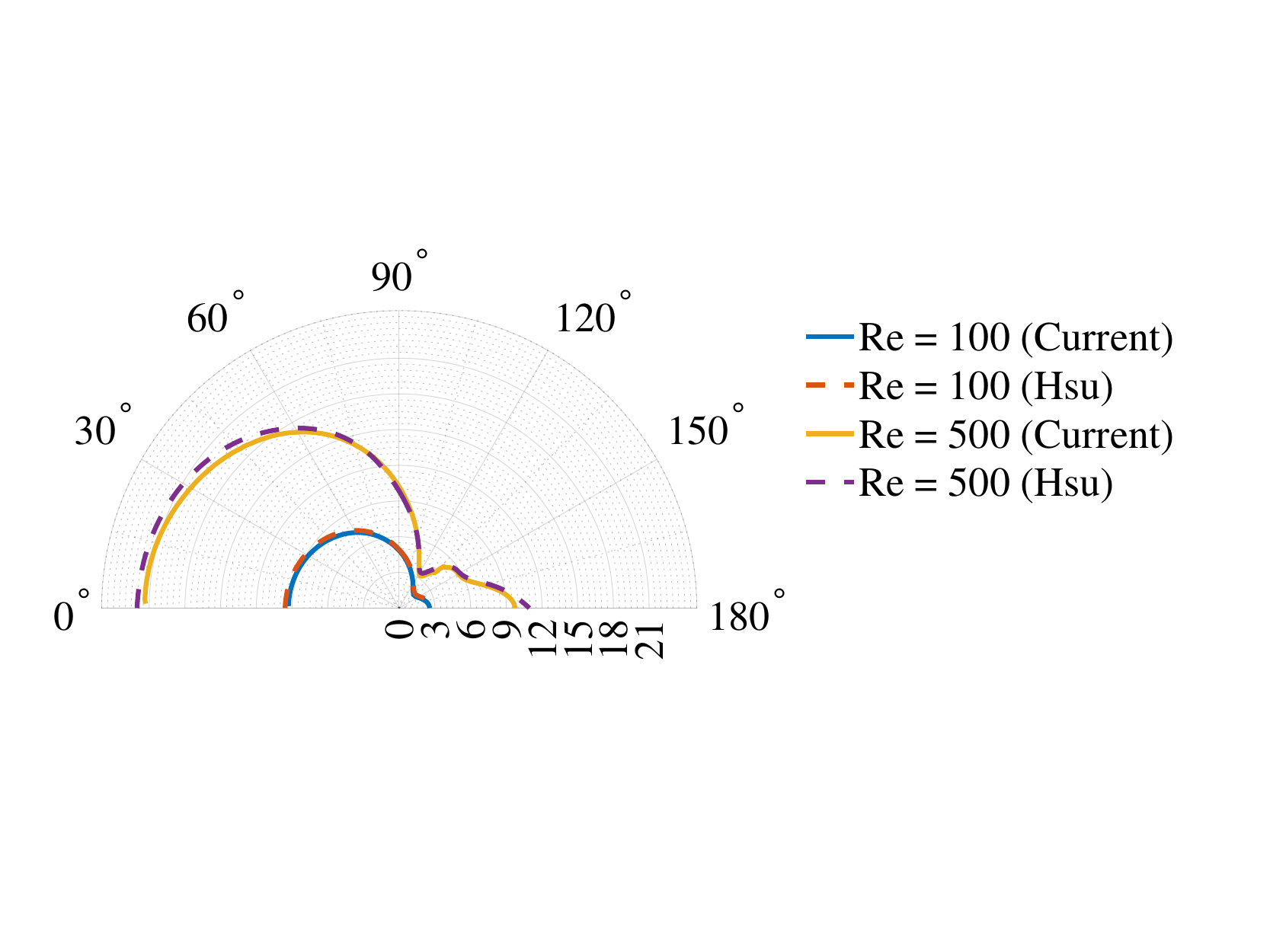}
        \caption{Angular distribution of time-averaged local Nusselt number along the cylinder boundary for $Re = 100$ and $Re = 500$.}
        \label{fig:PolarNu_Re100and500}
    \end{subfigure}
    \caption{Flow past a heated cylinder with constant wall temperature (CWT, \secref{subsub:CWT}): Comparison with~\citet{hsu2021heat} of the (time-averaged) global Nusselt number ($Nu = P^{-1} \, \int_\Gamma \nabla \theta \cdot \bs{n} \, d\Gamma$, where $P$ represents the perimeter of the circular cylinder) and the local Nusselt number ($\nabla \theta \cdot \bs{n}$) for various Reynolds numbers and angular positions.}
    \label{fig:CombinedCWT}
\end{figure}


\subsection{Neumann boundary conditions and SBM: Flow past a heated cylinder with constant wall heat (UHF)}\label{subsub:UHF}

\begin{figure}[!t]
    \centering
    \includegraphics[width=.86\linewidth,trim=0 0 0 0,clip]{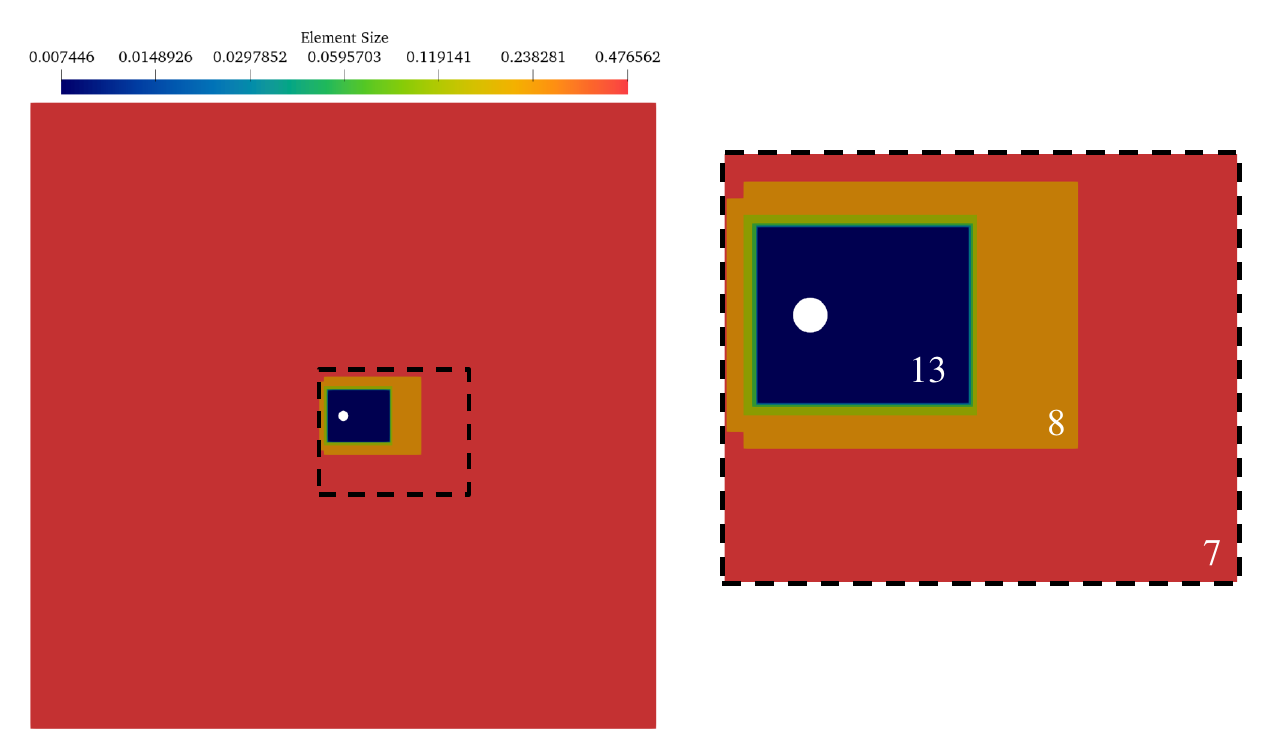}
    \caption{Flow past a heated cylinder with constant wall heat (UHF, \secref{subsub:UHF}): mesh refinement levels, with finer grids near the cylinder boundary.}
    \label{fig:UHF_ElementSize}
\end{figure}

In addition to the Dirichlet boundary conditions, Neumann boundary conditions (or flux boundary conditions) are common in thermal incompressible flow problems. We first consider a a forced convection problem similar to \secref{subsub:CWT}. 
A cylinder with an imposed uniform wall heat flux (UHF)~\citep{dennis1968steady,ahmad1992laminar,bharti2007numerical} is placed at the center of a fluid domain of size is $[0, 61] \times [0, 61]$. The velocity is set to $(1, 0)$ on the top, left, and bottom walls of the computational domain.
On the right wall, the outflow condition~(\eqnref{outflow-bc}) is weakly imposed and complemented with strong enforcement of a homogeneous pressure.
For the heat transfer problem, zero-flux boundary conditions are imposed at all external walls, except the left wall, where are homogeneous temperature condition has been set ($\theta = 0$). The non-dimensional uniform heat flux applied to the cylinder wall is $q^* = - \pd{\theta}{n}=-1$.
Note that $\bs{n}$ is directed inwards toward the cylinder and the fluid increases its energy due to $q^*$.
Three different levels of mesh refinement were applied.
The region closest to the circular obstacle is refined to level 13 (mesh size = $61 \cdot 2^{-13}$). Compared to the finest element size ($0.01$) used in the boundary-fitted mesh by~\citet{bharti2007numerical}, our finest mesh size is approximately $1.35$ times smaller. 
Surrounding this is a larger rectangular region with a refinement level of 8 (mesh size = $61 \cdot 2^{-8}$).
The base refinement level for the entire simulation is 7 (mesh size = $61 \cdot 2^{-7}$).
Due to the 2:1 balancing constraint in our Octree mesh framework, intermediate refinement levels are automatically introduced between levels 13 and 8.
The mesh is shown in \figref{fig:UHF_ElementSize}.
Since the problem in this section involves a steady-state solution and a low Reynolds number flow, we set the non-dimensional time step to 0.1 for this simulation.

\begin{figure}[!t]
    \centering

    \begin{subfigure}{0.49\linewidth}
        \includegraphics[width=\linewidth,trim=0 0 0 0,clip]{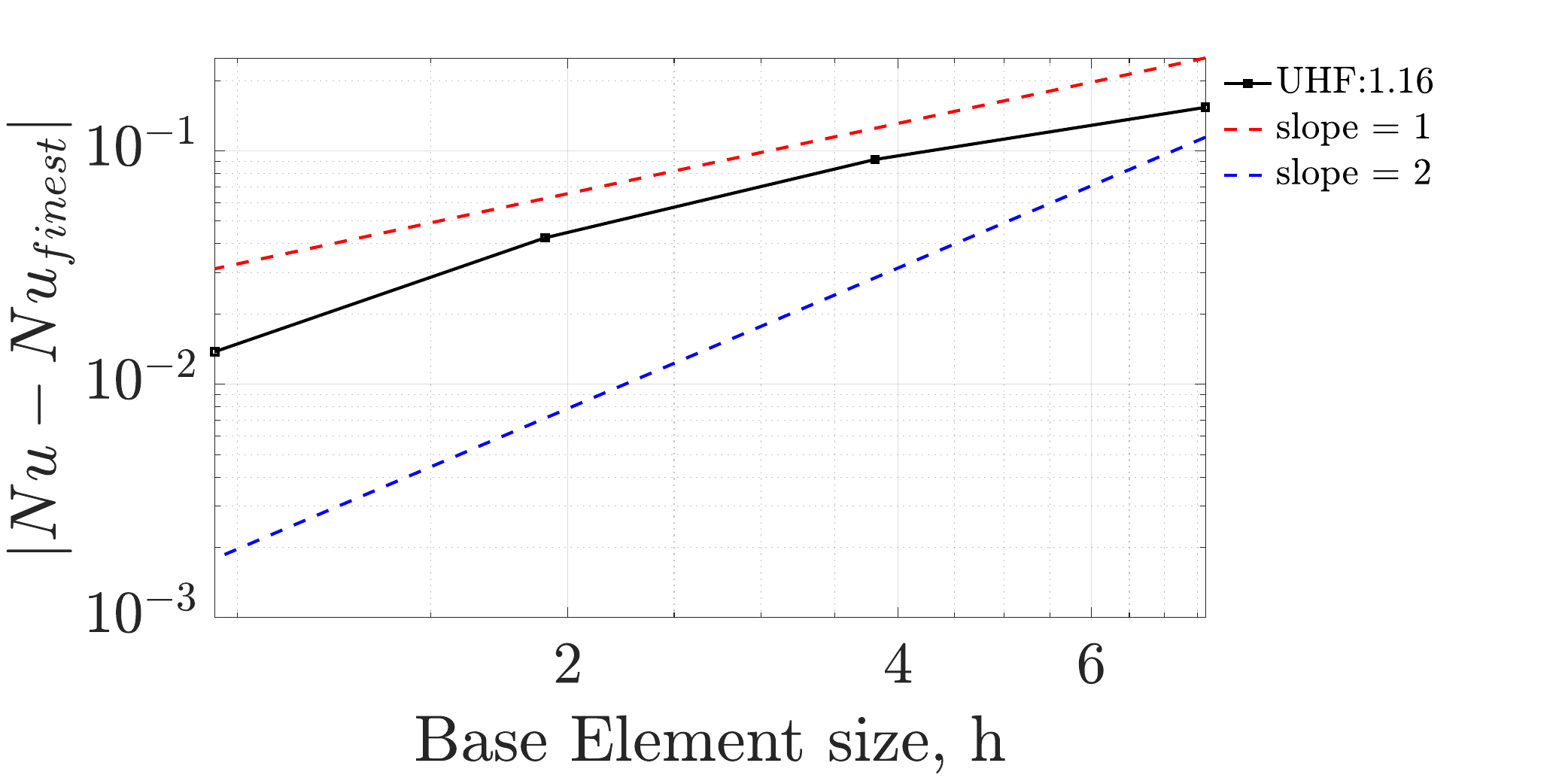}
        \caption{Mesh convergence study for the global Nusselt number ($Nu = P^{-1} \, \int \theta^{-1} \, d\Gamma$), where $P$ represents the perimeter of the circular cylinder.}
        \label{fig:UHF_Nu_Global}
    \end{subfigure}
    \hfill
    \begin{subfigure}{0.49\linewidth}
        \includegraphics[width=\linewidth,trim=0 0 0 0,clip]{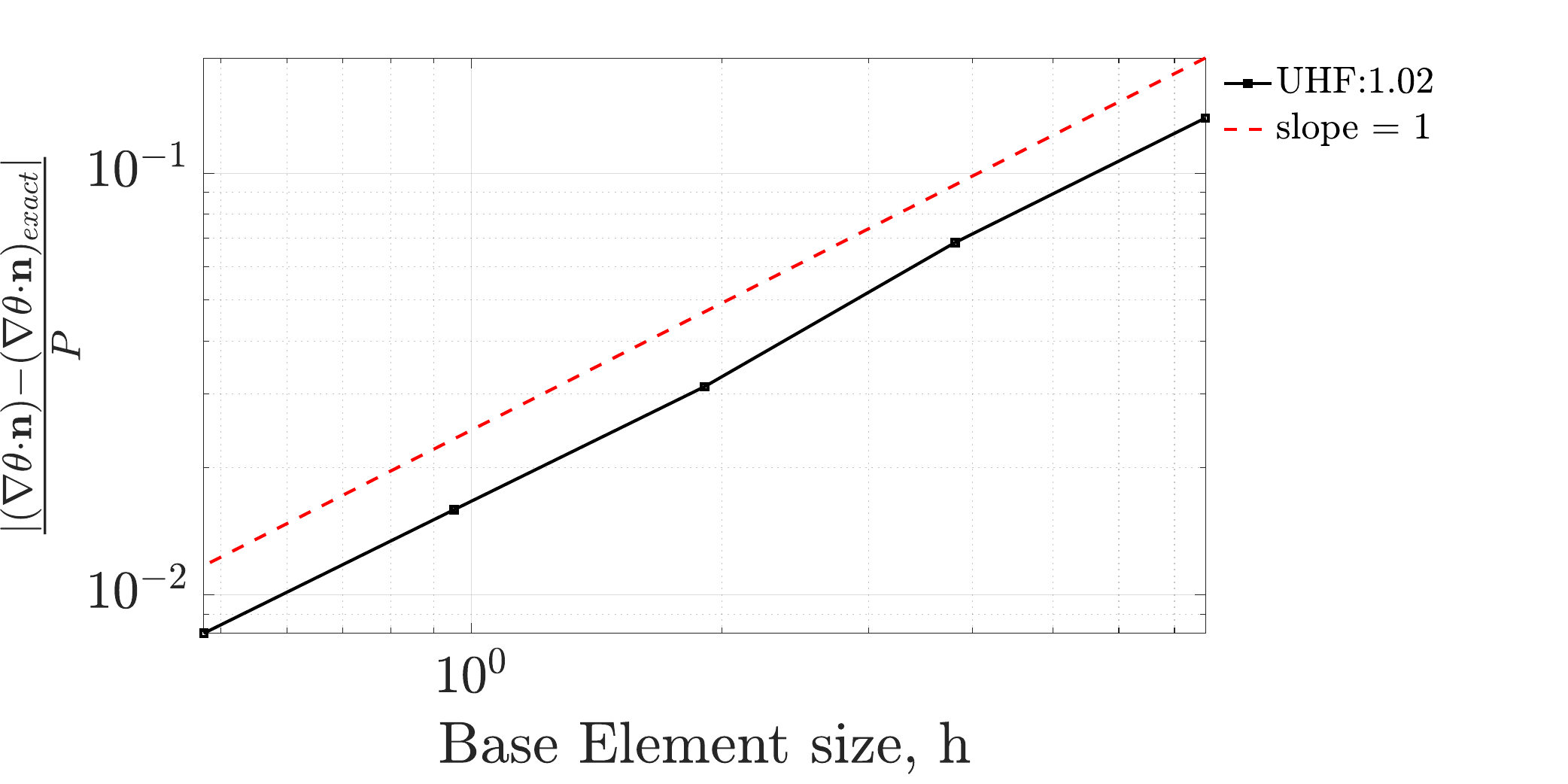}
        \caption{Mesh convergence study for the global non-dimensional flux ($Nu = P^{-1} \, \int \nabla \theta \cdot \mathbf{n} \, d\Gamma$), where $P$ represents the perimeter of the circular cylinder. The exact solution for this value corresponds to the non-dimensional Neumann boundary condition set for this problem ($q^* =-1$).}
        \label{fig:UHF_Flux_Global}
    \end{subfigure}
    
    \caption{Mesh convergence study for the global Nusselt number ($Nu = P^{-1} \, \int \theta^{-1} \, d\Gamma$) and the global non-dimensional flux ($Nu = P^{-1} \, \int \nabla \theta \cdot \mathbf{n} \, d\Gamma$) in the flow past a heated cylinder with uniform heat flux at $Re = 10$ (UHF, see \secref{subsub:UHF}).}
    \label{fig:UHF_Nu_MeshRefinementStudy}
\end{figure}

\begin{figure}[!t]
\centering
\begin{subfigure}{0.33\textwidth}
\centering
\includegraphics[width=\linewidth]{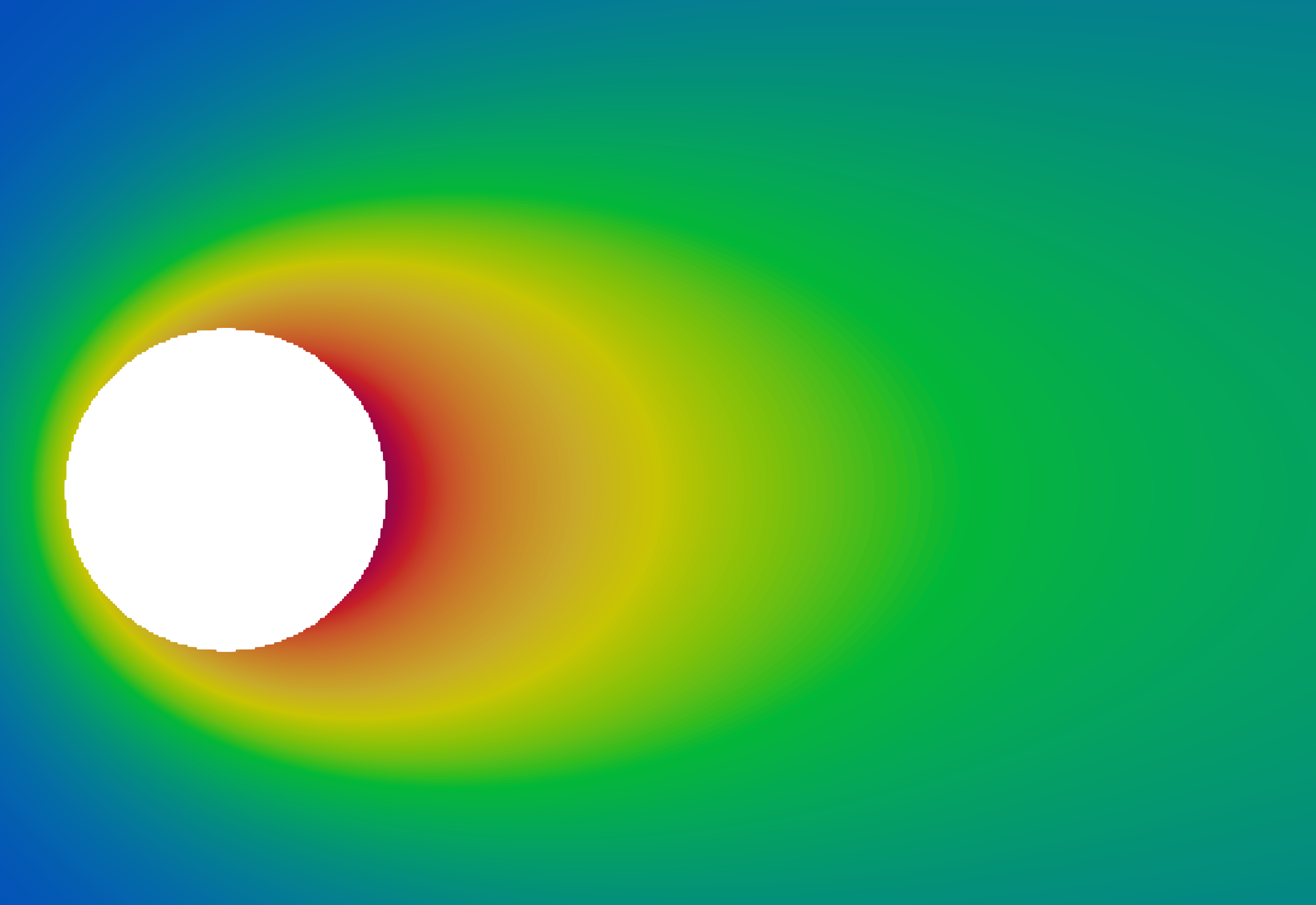}
\caption{Re = 10}
\label{fig:Re10_UHF_Contour}
\end{subfigure}
\qquad
\begin{subfigure}{0.33\textwidth}
\centering
\includegraphics[width=\linewidth]{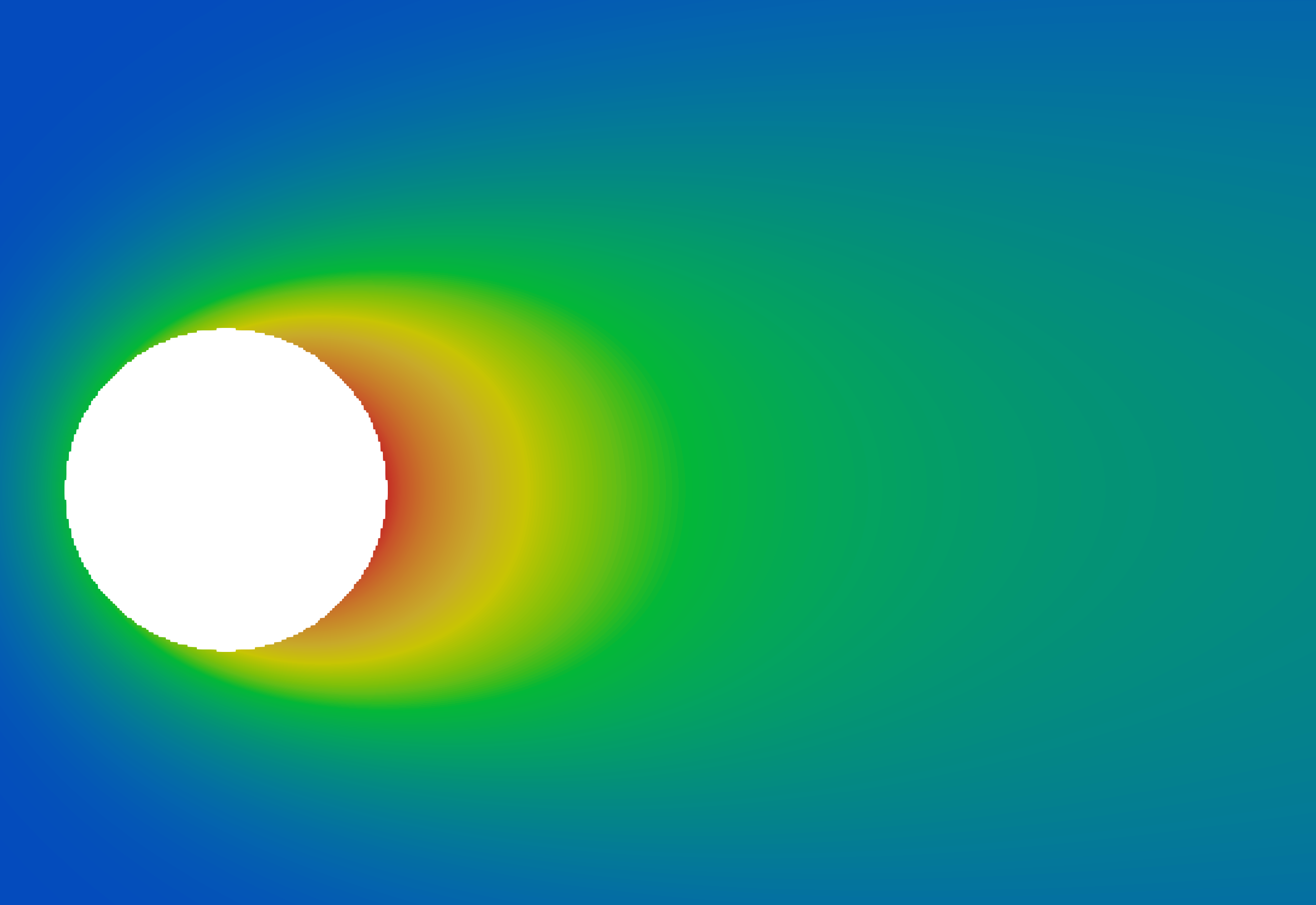}
\caption{Re = 20}
\label{fig:Re20_UHF_Contour}
\end{subfigure}
\\[.5cm]
\begin{subfigure}{0.33\textwidth}
\centering
\includegraphics[width=\linewidth]{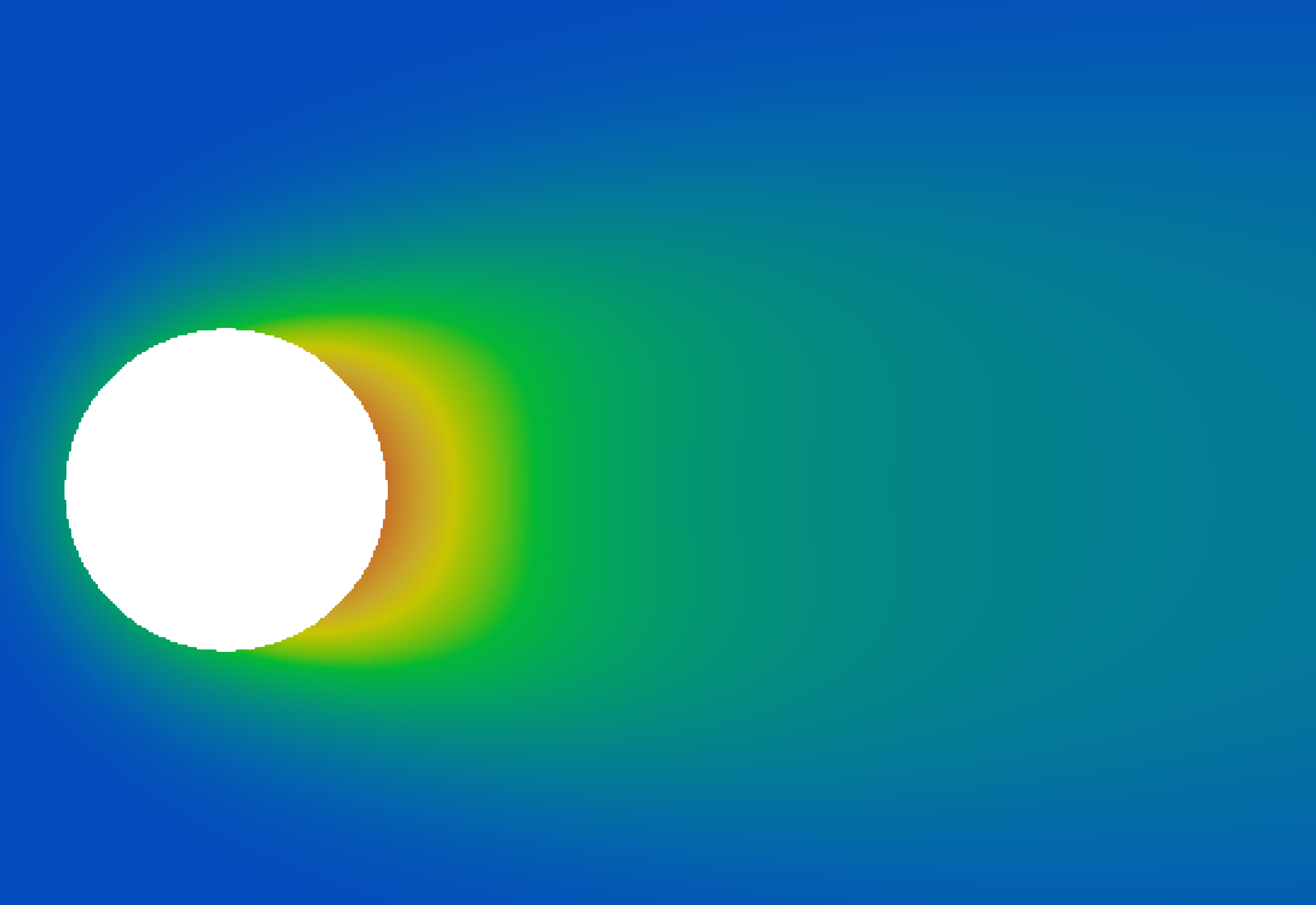}
\caption{Re = 40}
\label{fig:Re40_UHF_Contour}
\end{subfigure}
\qquad
\begin{subfigure}{0.33\textwidth}
\centering
\includegraphics[width=\linewidth]{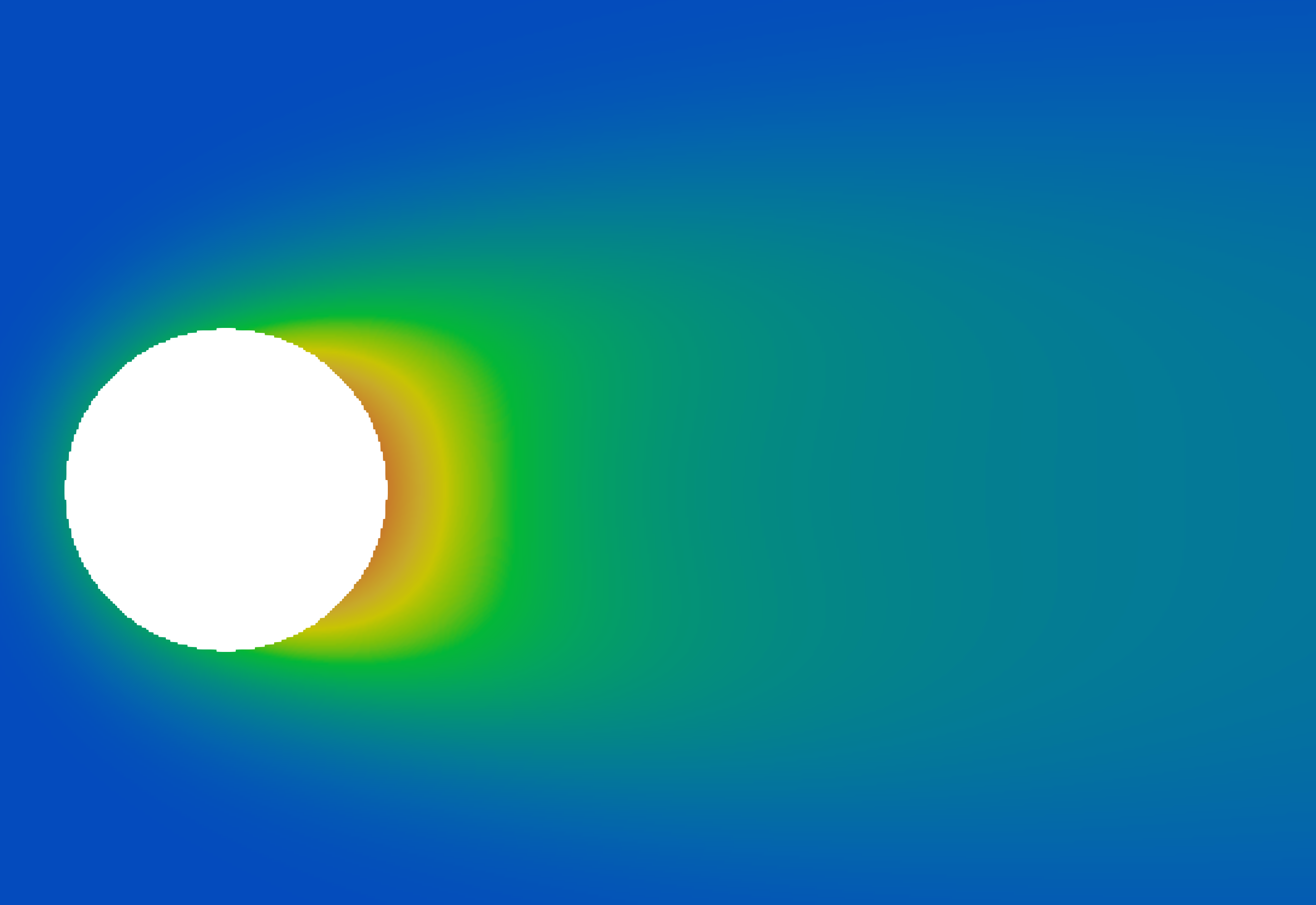}
\caption{Re = 45}
\label{fig:Re45_UHF_Contour}
\end{subfigure}
\\
\centering
\includegraphics[width=0.3\linewidth]{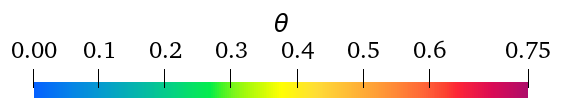}
\caption{Flow past a heated cylinder with constant wall heat (UHF, \secref{subsub:UHF}): temperature contours at steady-state for various Reynolds numbers.}
\label{fig:NeumannHT_TemperatureContour}
\end{figure}

{We conducted a mesh convergence study on the global Nusselt number ($Nu = P^{-1} \int_\Gamma \theta^{-1} d\Gamma$) and the global non-dimensional flux ($P^{-1} \int_\Gamma \nabla \theta \cdot \bs{n} d\Gamma$) after the flow reached a steady state, as shown in \figref{fig:UHF_Nu_Global} and \figref{fig:UHF_Flux_Global}, respectively.
In \figref{fig:UHF_Nu_Global}, the results indicate that the Nusselt number converges with an order of accuracy close to 1.16, slightly exceeding the expected theoretical value of 1. Similarly, the non-dimensional global flux converges with an order of accuracy close to 1, as shown in \figref{fig:UHF_Flux_Global}. This result was somewhat surprising, given that the Neumann boundary conditions are enforced in our framework using a simplified version of the SBM (see \eqnref{eq:HT-SBM}). Specifically, we do not incorporate the Hessian term required to shift the Neumann flux, as our implementation uses only linear basis functions.
Under these conditions, characterized by the use of simplified Neumann SBM boundary conditions, the expected order of accuracy for the solution $\theta$ is approximately 1. Since the non-dimensional flux is essentially the first derivative evaluated on the cut (or \Intercepted{}) element, the inclusion of the Hessian term in the Neumann boundary condition would still limit the highest attainable order of accuracy to around 1. Considering these two factors -- the simplified Neumann boundary condition and the evaluation of the first derivative -- the order of accuracy for the non-dimensional flux in the UHF case could potentially be less than 1. However, the observed order of accuracy (1.02) demonstrates that our simplified Neumann SBM boundary condition is remarkably robust, yielding reliable and accurate simulation results even without the inclusion of higher-order terms.}

The steady-state temperature contours are illustrated in \figref{fig:NeumannHT_TemperatureContour} for different Reynolds numbers.
The results in \tabref{tab:Neumann_results} show that the SBM compares well with the existing literature.
We summarize in~\tabref{tab:Neumann_Area_Correction} the results of experiments aimed at quantifying the impact of the area correction term ($\bs{n} \cdot \ti{\bs{n}}$). Specifically, we performed computations with and without this factor in the Neumann SBM condition described in~\eqnref{eq:Neumann_derivation_SBM}.

Without including the area correction term ($\bs{n} \cdot \ti{\bs{n}}$), the errors on the global Nusselt number $Nu$ are $O(1)$. 
Hence, the SBM plays a crucial role in ensuring that the pixelated Octree-based mesh satisfies the boundary conditions on the true boundary. 
In fact, the surrogate area grossly overestimates the area of the true boundary, and this is the reason why $O(1)$-errors are obtained in the global Nusselt number, as shown in \tabref{tab:Neumann_Area_Correction}. 
{\it Note that the area mismatch is not reduced by refining the grid}, and this is the reason why we characterize these errors as $O(1)$.
{In fact, the ratio of the global Nusselt numbers obtained with and without the SBM area correction term is approximately $\pi/4$, which is the ratio of the perimeters of a circle and a square, where the square has side equal to the diameter of the circle.}
\tabref{tab:Neumann_Area_Correction} also shows that the ratio between the results with and without area correction is approximately constant as the Reynolds number changes, indicating a geometric inconsistency when the area correction is not applied.
This illustrates that the SBM strategy is a viable strategy for imposing Neumann boundary conditions.  
\begin{table}[h]
\centering
\caption{Comparison of the steady-state global Nusselt number ($Nu = P^{-1} \, \int_\Gamma \theta^{-1} \, d\Gamma$, where $P$ represents the perimeter of the circular cylinder) obtained using the Octree-Shifted Boundary Method (Octree-SBM) with literature values across different Reynolds numbers for flow past a UHF circular obstacle (\secref{subsub:UHF}).}
\label{tab:Neumann_results}
\setlength{\extrarowheight}{3pt}
\begin{tabular}{L{5cm}R{2cm}R{2cm}R{2cm}R{2cm}}
\hline
\textbf{Study} & \textbf{Re = 10} & \textbf{Re = 20} & \textbf{Re = 40} & \textbf{Re = 45} \\ 
\hline
\citet{bharti2007numerical} & 2.0400 & 2.7788 & 3.7755 & 3.9727 \\ 
\citet{ahmad1992laminar} & 2.0410 & 2.6620 & 3.4720 & - \\ 
\citet{dennis1968steady} & 2.1463 & 2.8630 & 3.7930 & - \\ 
\hline
Octree-SBM & 2.0365 & 2.7534 & 3.7640 & 3.9630 \\ \hline
\end{tabular}
\end{table}

\begin{figure}[!t]
\centering
\begin{subfigure}{0.5\textwidth}
\centering
\includegraphics[width=\linewidth]{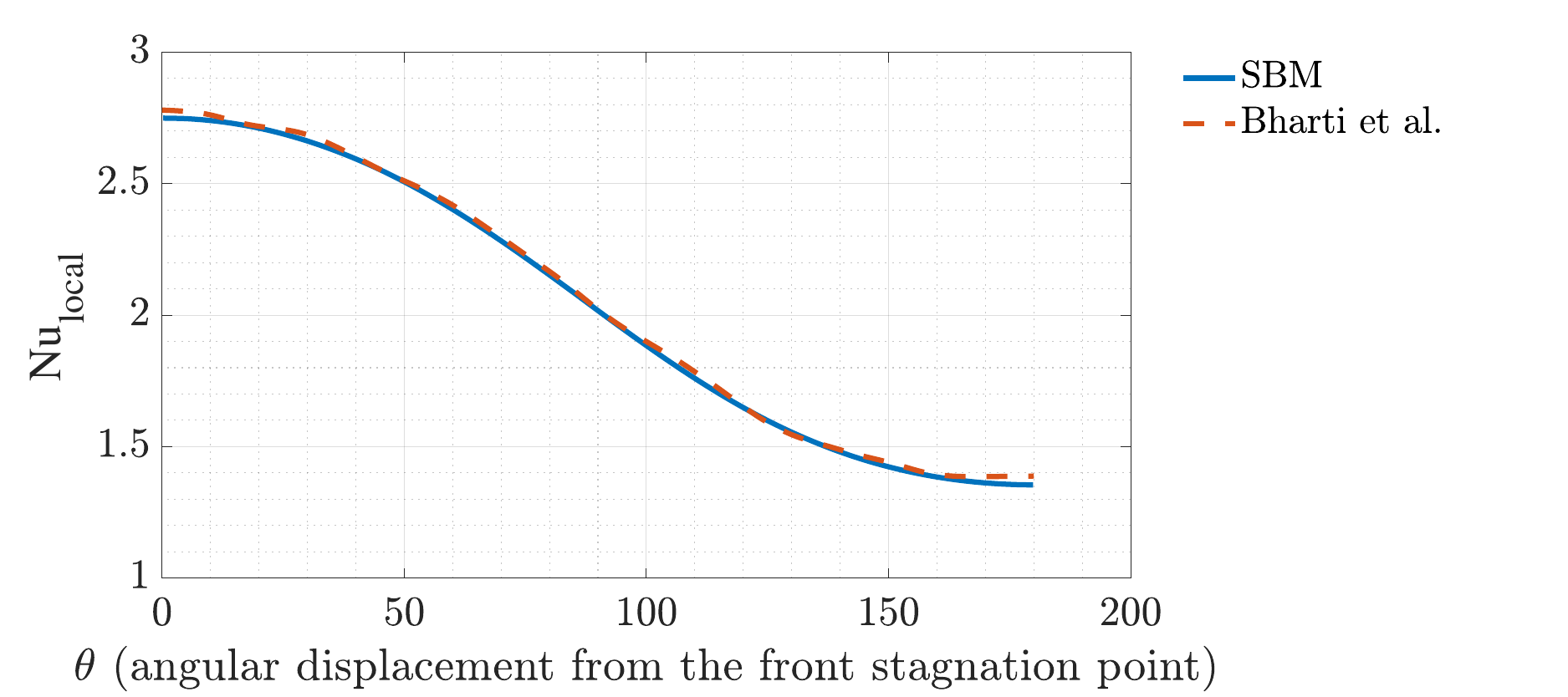}
\caption{$Re = 10$}
\label{fig:sub1Re}
\end{subfigure}%
\begin{subfigure}{0.5\textwidth}
\centering
\includegraphics[width=\linewidth]{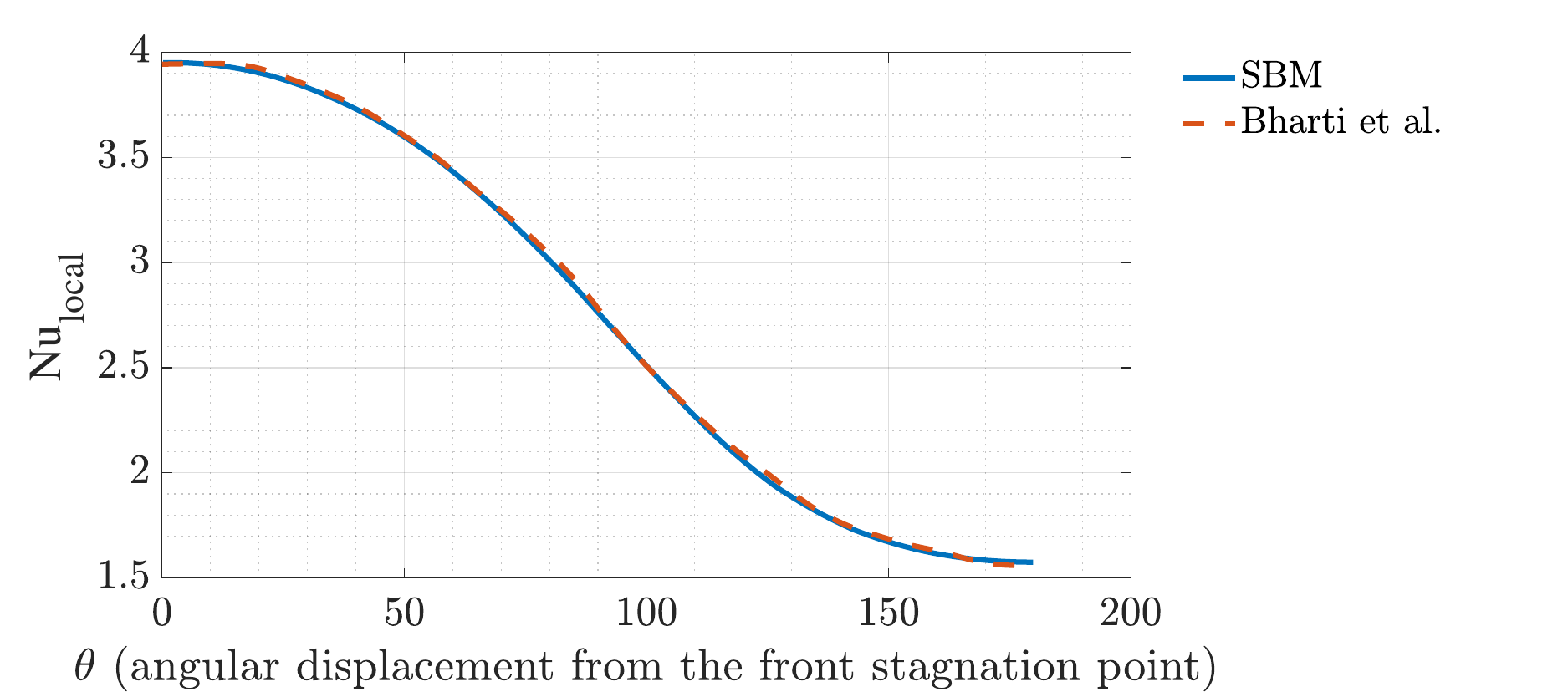}
\caption{$Re = 20$}
\label{fig:sub2}
\end{subfigure}
\caption{Flow past a heated cylinder with constant wall heat flux (UHF, \secref{subsub:UHF}): Comparison against~\citet{bharti2007numerical} of the local Nusselt number ($\theta^{-1}$) distribution at steady state, as a function of the angular position.}
\label{fig:NeumannHT_LocalNu}
\end{figure}

\begin{table}[!t]
    \centering
    \caption{Flow past a heated cylinder with constant wall heat (UHF, \secref{subsub:UHF}): comparison of the steady-state global Nusselt number ($Nu = P^{-1} \, \int_\Gamma \theta^{-1} \, d\Gamma$, where $P$ represents the perimeter of the circular cylinder) obtained with and without the area correction term ($\bs{n} \cdot \ti{\bs{n}}$). Observe the gross overestimation of the global Nusselt numbers when the correction term is not included. Note also that the SBM simulations have reached grid convergence.}
    \label{tab:Neumann_Area_Correction}
    \setlength{\extrarowheight}{3pt}
    \begin{tabular}{L{8cm}R{1.5cm}R{1.5cm}R{1.5cm}}
    \toprule
    \textbf{$Re$} & \textbf{10} & \textbf{20} & \textbf{40} \\
    \midrule
    \begin{filecontents*}{UHF.txt}
Re,Ten,Twenty,Fourty
Area Correction,2.0365,2.7534,3.7640
Without Area Correction (by setting $n \cdot \Tilde{n} = 1$) ,2.5794,3.5024,4.7362
\citet{bharti2007numerical},2.0400,2.7788,3.7755
\end{filecontents*}
\csvreader[
      separator=comma,
      head to column names,
      late after line=\\ 
    ]{UHF.txt}{}%
    {\csvcoli & \csvcolii & \csvcoliii & \csvcoliv} 
    \bottomrule
    \end{tabular}
\end{table}

\newpage

\subsection{Natural convection around a sphere in a cubic enclosure} \label{Natural-3D}

We conduct simulations similar to those presented in~\citep{yoon2010three,Chen2020}, where a heated sphere with a radius of 0.2 is placed inside a cold cubic domain of unit side.
This setup is used to study natural convection driven by temperature differences.
The boundary conditions are set as \(\theta = 1\) at the surface of the sphere and \(\theta = 0\) at the cube's walls. No-slip boundary conditions are applied to all walls and the sphere's boundary. Unlike the mixed convection problem discussed in \secref{subsub:Mixed_LDC} and the forced convection problems in \secref{subsub:CWT} and \secref{subsub:UHF}, this test focuses exclusively on natural convection. We simulate a wide range of Rayleigh numbers ranging from $10^3$ all the way to $10^8$. 

\begin{figure}[!h]
    \centering
    \begin{subfigure}{0.32\textwidth}
        \centering
        \includegraphics[width=\linewidth,trim=600 100 600 200,clip]{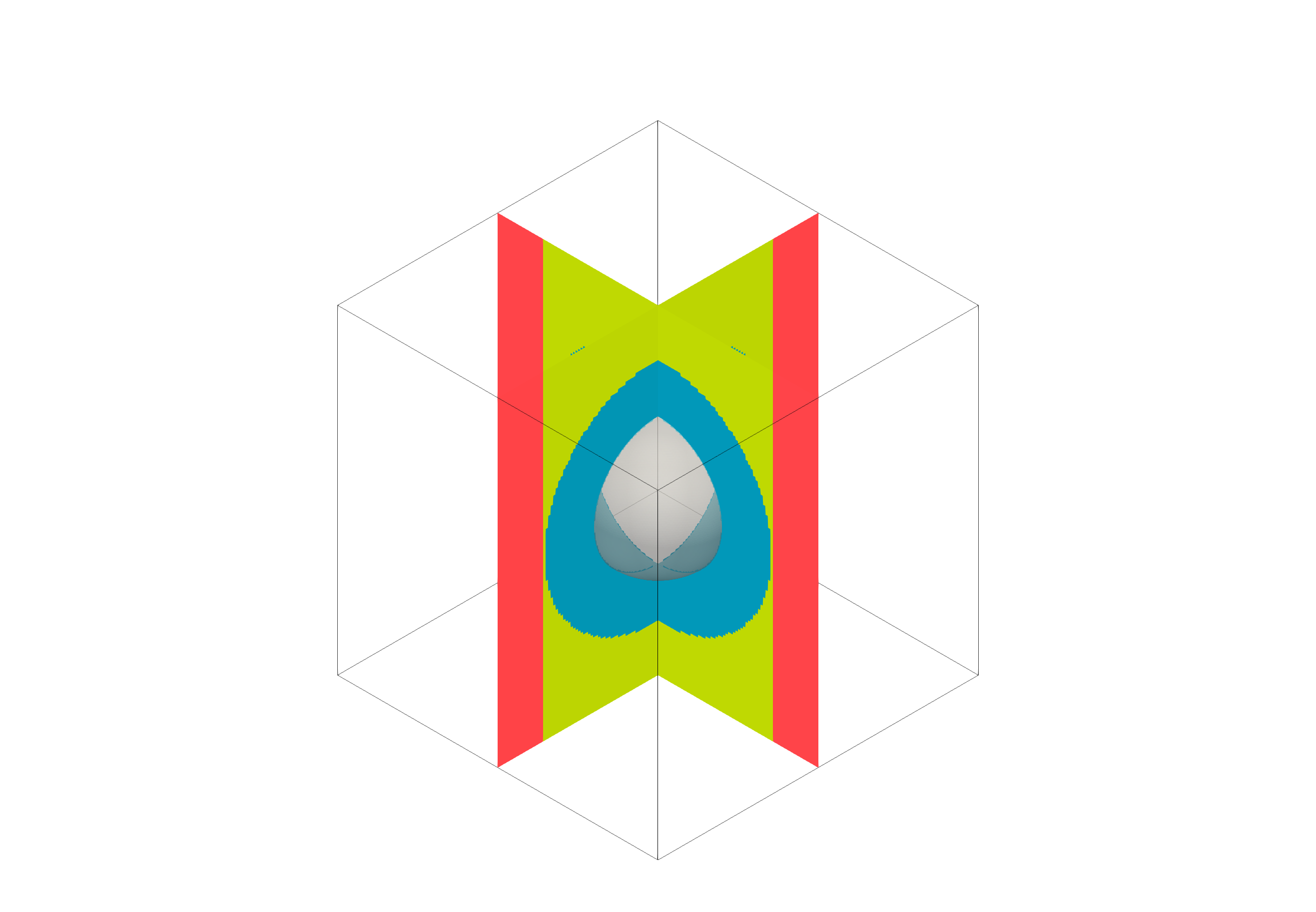}
        \caption{$Ra = 10^3, 10^4$}
        \label{fig:M1}
    \end{subfigure}
    \begin{subfigure}{0.32\textwidth}
        \centering
        \includegraphics[width=\linewidth,trim=600 100 600 200,clip]{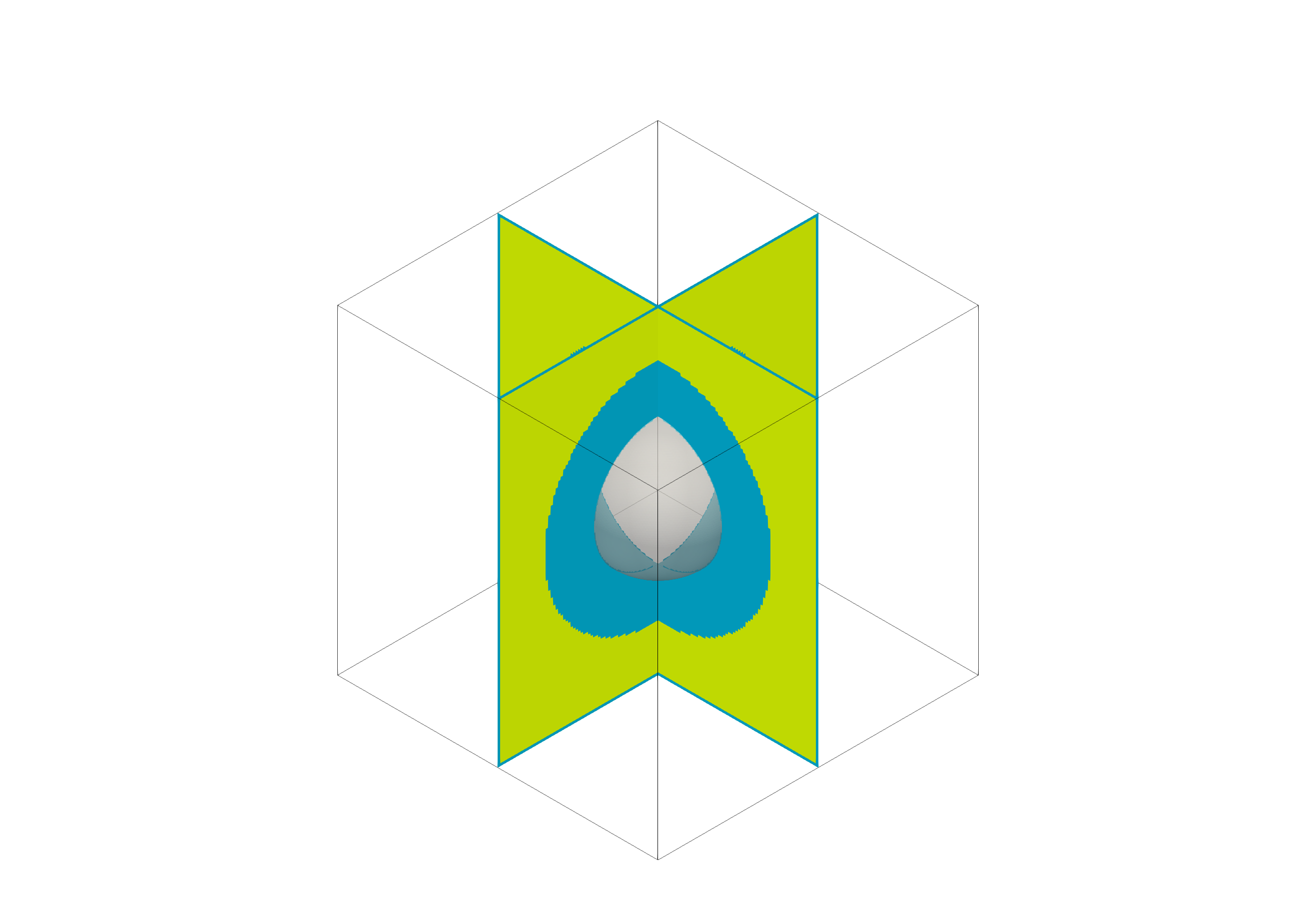}
        \caption{$Ra = 10^5, 10^6$}
        \label{fig:M2}
    \end{subfigure}
    \begin{subfigure}{0.32\textwidth}
        \centering
        \includegraphics[width=\linewidth,trim=600 100 600 200,clip]{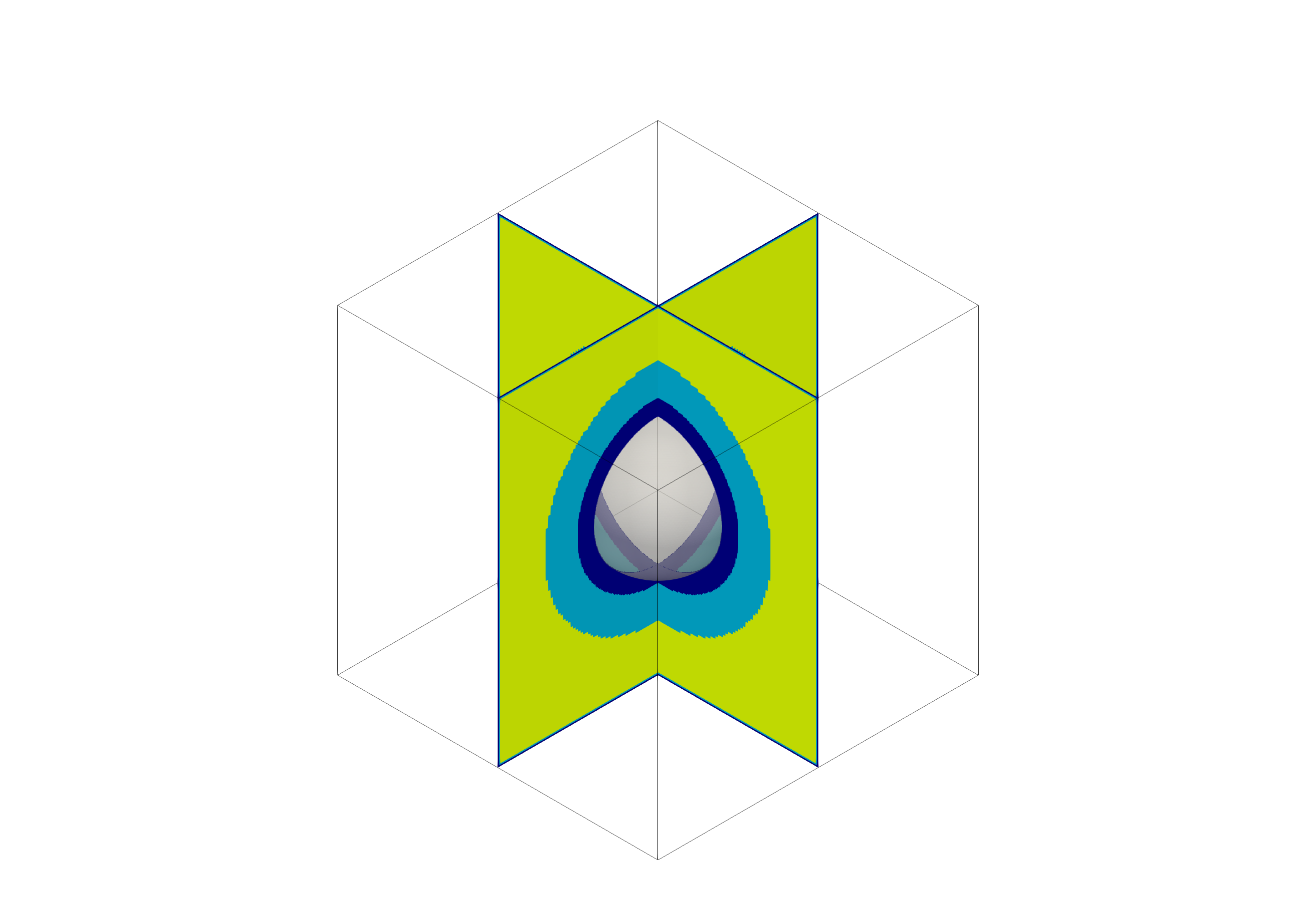}
        \caption{$Ra = 10^7, 10^8$}
        \label{fig:M3}
    \end{subfigure}
    \begin{subfigure}{0.7\textwidth}
        \centering
        \includegraphics[width=\linewidth,trim=0 0 0 0,clip]{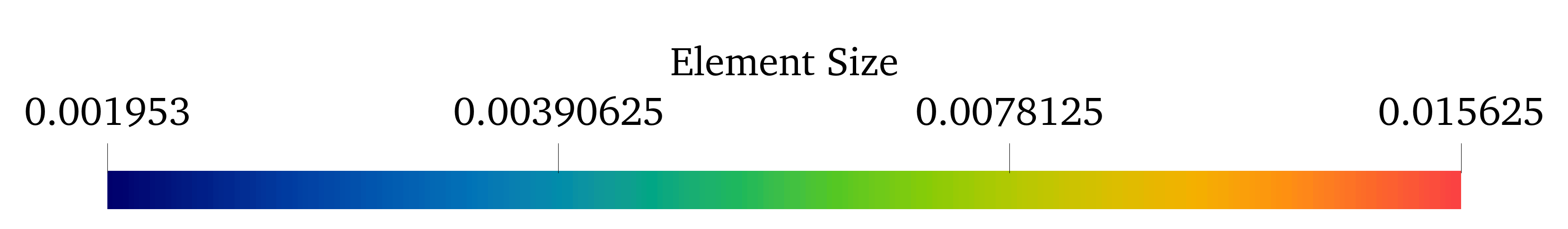}
    \end{subfigure}
    \caption{Natural convection around a sphere in a cubic enclosure (\secref{Natural-3D}): element sizes for various Rayleigh numbers.}
    \label{fig:Elementsize_3D_sphere}
\end{figure}

For $Ra = 10^3$ and $10^4$, a spherical refinement region with a radius of $0.35$, centered at the same location as the sphere, is refined to level $8$ (mesh size = $2^{-8}$), while the base refinement level is set to $6$ (mesh size = $2^{-6}$). Compared to the mesh size used in \citet{yoon2010three}, which is $201^{-1}$, our finest mesh is approximately $1.27$ times smaller, while our coarsest mesh is about $3.14$ times larger. Additionally, a cylindrical refinement region with a radius of 0.35 is applied, with its centerline aligned with that of the cube, extending from the bottom wall to the top wall. The mesh sizes for these two cases are shown in \figref{fig:M1}. For \(Ra = 10^5\) and \(10^6\), the spherical refinement region, with a radius of 0.35 and centered at the sphere, remains at level 8  (mesh size = $2^{-8}$), but the base refinement level is increased to 7 (mesh size = $2^{-7}$). The mesh sizes for these cases are shown in \figref{fig:M2}. For \(Ra = 10^7\) and \(10^8\), the refinement strategy is similar to that for \(Ra = 10^5\) and \(10^6\), with an additional level 9 refinement (mesh size = $2^{-9}$) applied at the boundary of the cube, and an additional spherical refinement region at level 9 with a radius of 0.25, centered at the sphere. The mesh sizes for these cases are shown in \figref{fig:M3}.

\begin{table}[h!]
    \centering
    \caption{Natural convection around a sphere in a cubic enclosure (\secref{Natural-3D}): results of a mesh convergence study, detailing the effect of different element sizes on the time-averaged Nusselt number values. The Nusselt numbers are defined as follows: $\overline{Nu}_{T}$ represents the average Nusselt number at the top boundary, $\overline{Nu}_{B}$ at the bottom boundary, $\overline{Nu}_{S}$ at the side boundaries, and $\overline{Nu}_{Sp}$ on the surface of the sphere.}
    \begin{tabular}{@{}P{1.5cm}P{4cm}P{1.5cm}P{1.5cm}P{1.5cm}P{1.5cm}@{}}
        \toprule
        \textit{Ra} & \text{Highest refinement level} & \textbf{$\overline{Nu}_{T}$} &\textbf{$\overline{Nu}_{B}$}&\textbf{$\overline{Nu}_{S}$} &\textbf{$\overline{Nu}_{Sp}$} \\ 
        \midrule
        \multirow{3}{*}{$10^{7}$}  
         & 7           &  10.51   &  0.02 & 1.27 & 29.05 \\      
         & 8           &  11.25   &  0.02 & 1.32 & 31.63 \\      
         & 9           &  11.57   &  0.02  & 1.35 & 33.05  \\      
        \bottomrule
    \end{tabular}
    \label{tab:SphereInCube_ConvergenceStudy}
\end{table}

\begin{figure}[!t]
    \centering
    \begin{subfigure}{0.22\textwidth}
    \centering
        \includegraphics[width=\linewidth,trim=600 100 600 200,clip]{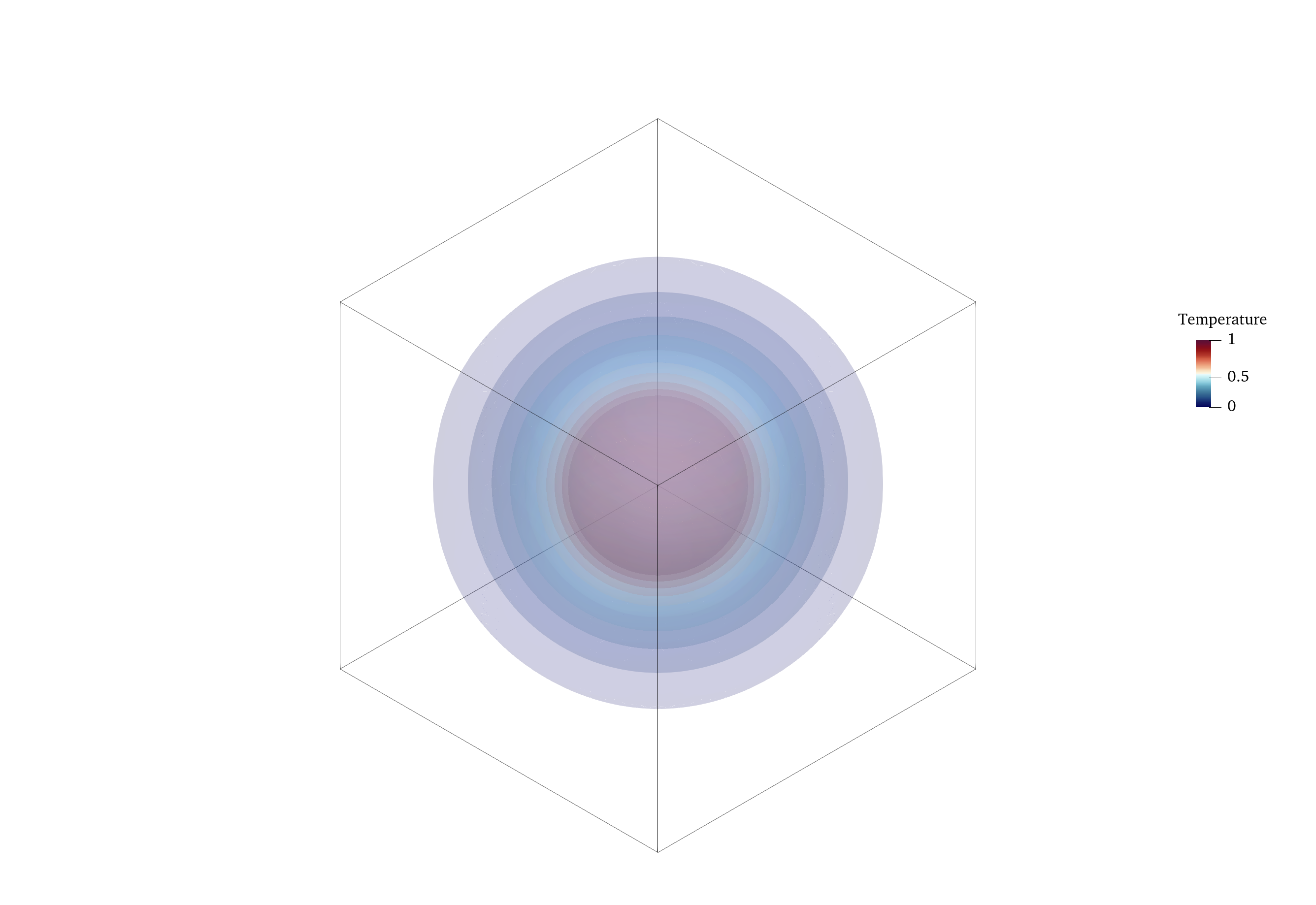}
        \caption{$Ra = 10^3$}
    \end{subfigure}
    \begin{subfigure}{0.22\textwidth}
    \centering
        \includegraphics[width=\linewidth,trim=600 100 600 200,clip]{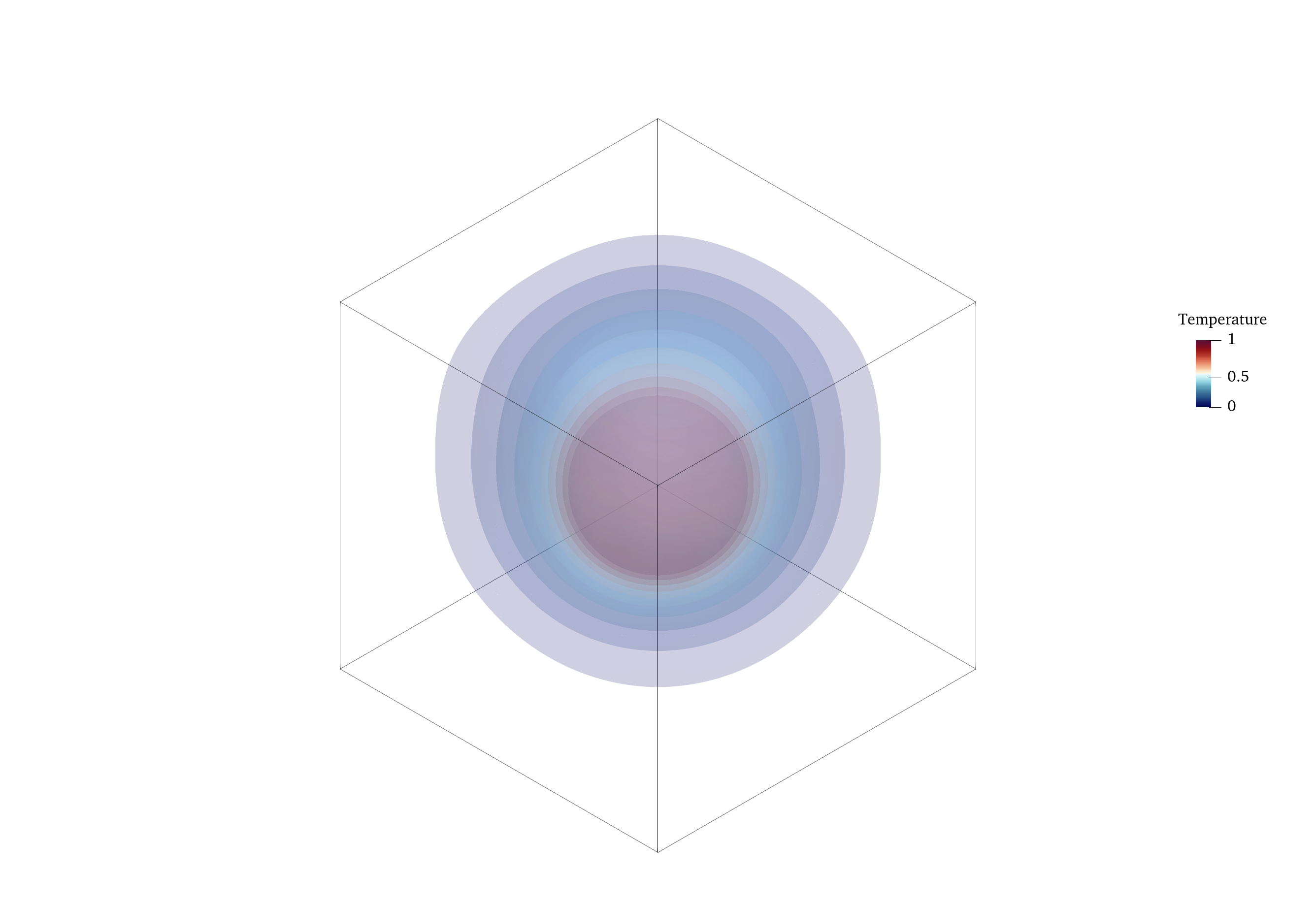}
        \caption{$Ra = 10^4$}
    \end{subfigure}
        \begin{subfigure}{0.22\textwidth}
        \centering
        \includegraphics[width=\linewidth,trim=600 100 600 200,clip]{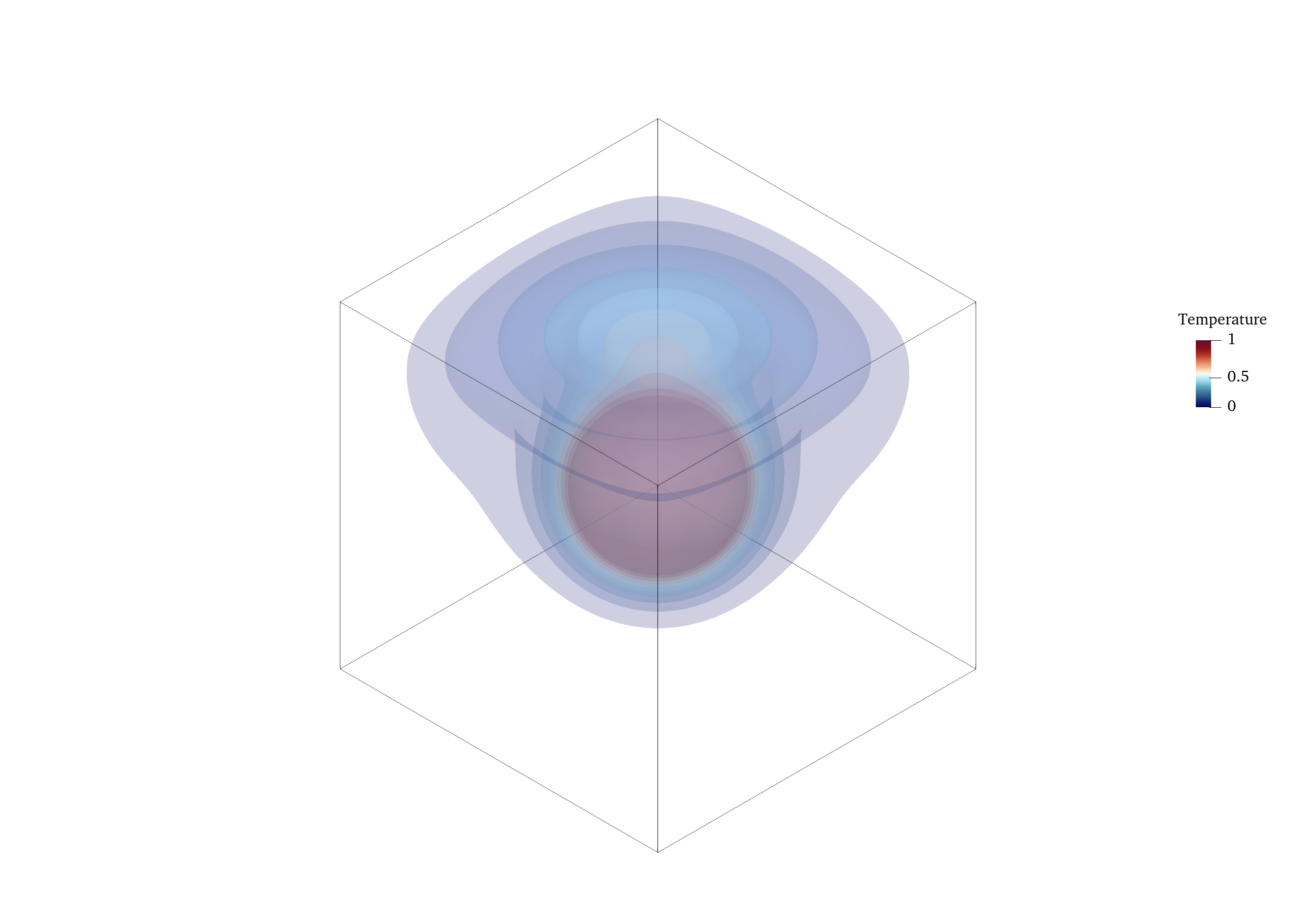}
        \caption{$Ra = 10^5$}
    \end{subfigure}
    \\
    \begin{subfigure}{0.22\textwidth}
    \centering
        \includegraphics[width=\linewidth,trim=600 100 600 200,clip]{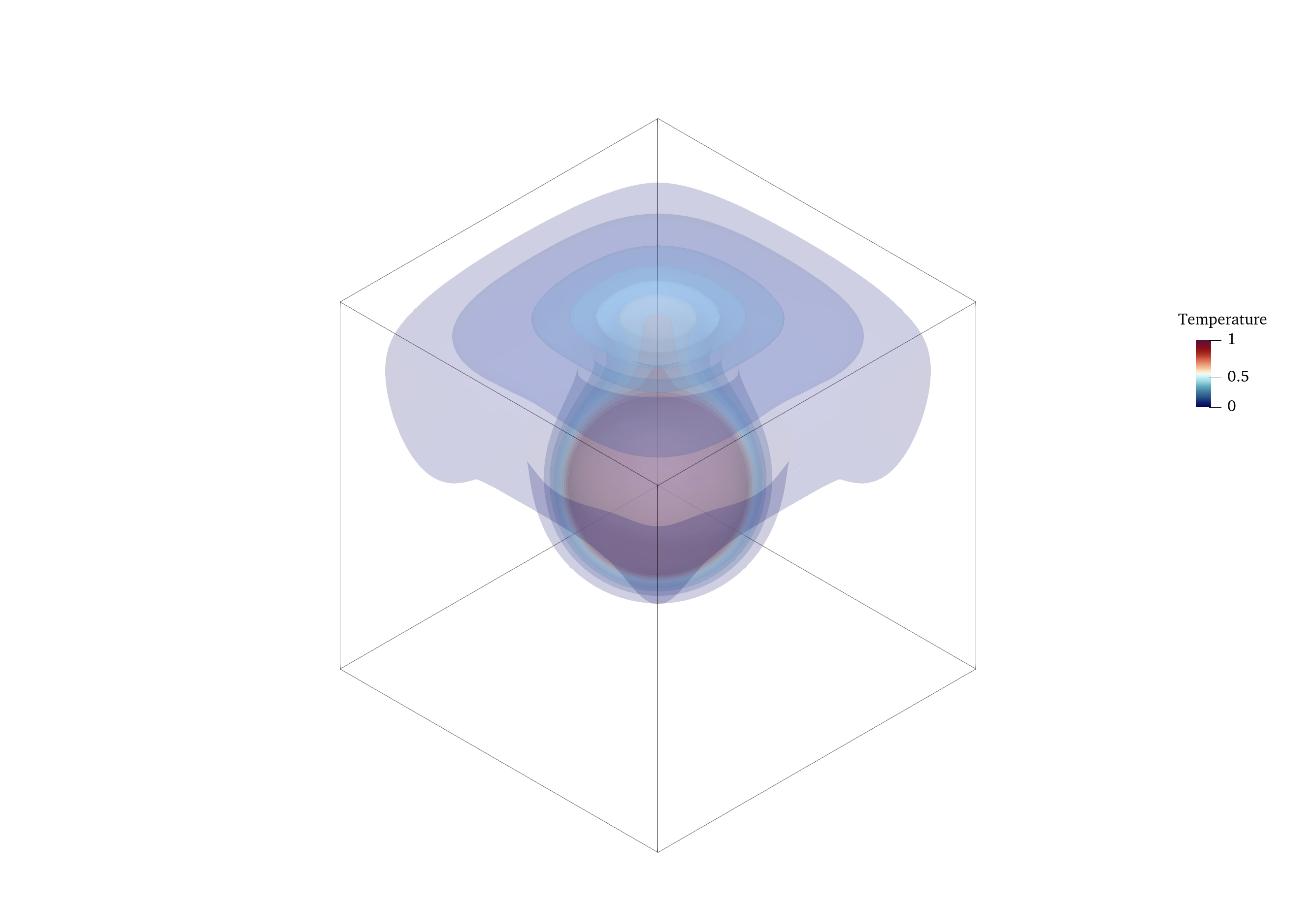}
        \caption{$Ra = 10^6$}
    \end{subfigure}
    \begin{subfigure}{0.22\textwidth}
    \centering
        \includegraphics[width=\linewidth,trim=600 100 600 200,clip]{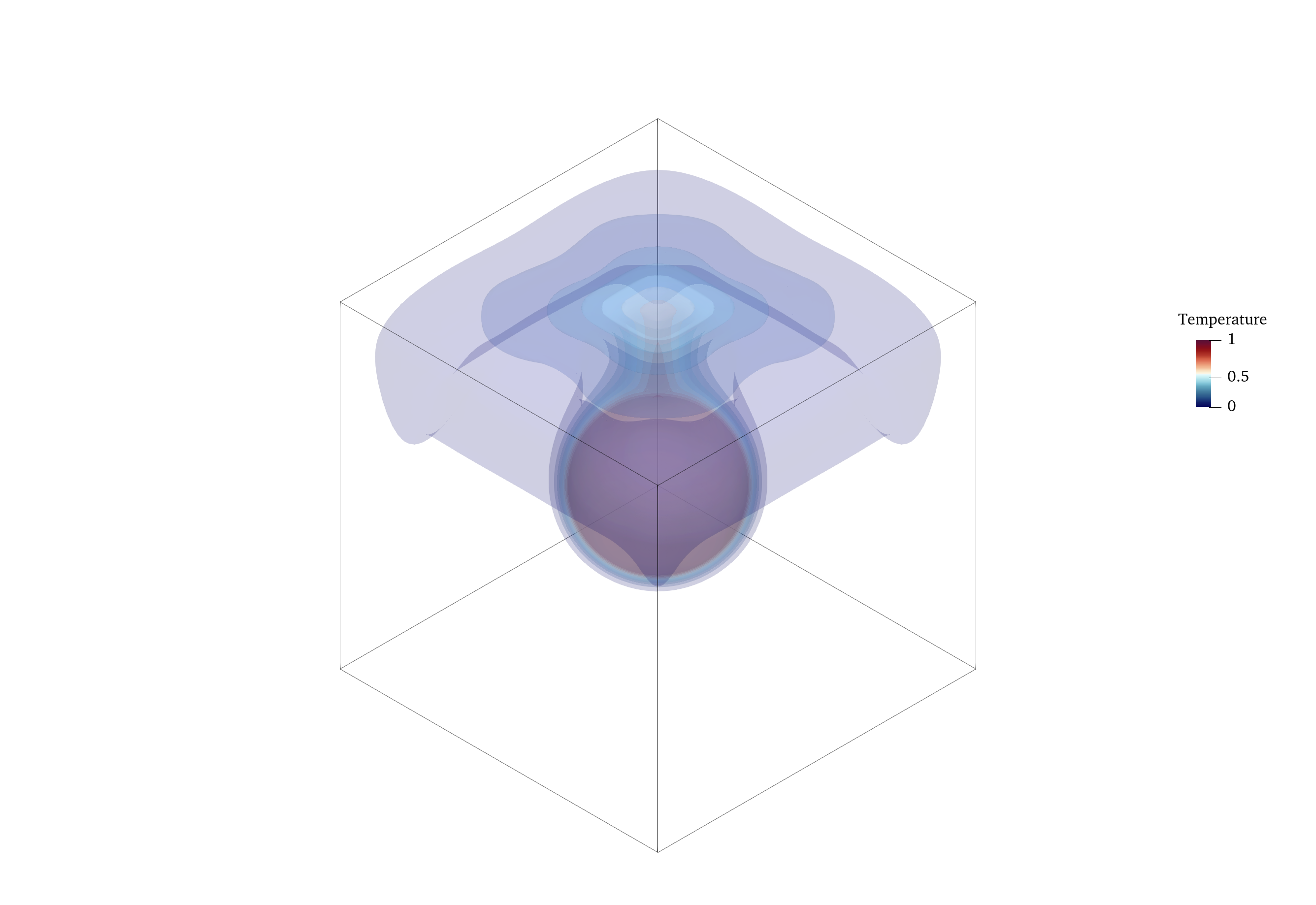}
        \caption{$Ra = 10^7$}
    \end{subfigure}
    \begin{subfigure}{0.22\textwidth}
    \centering
        \includegraphics[width=\linewidth,trim=600 100 600 200,clip]{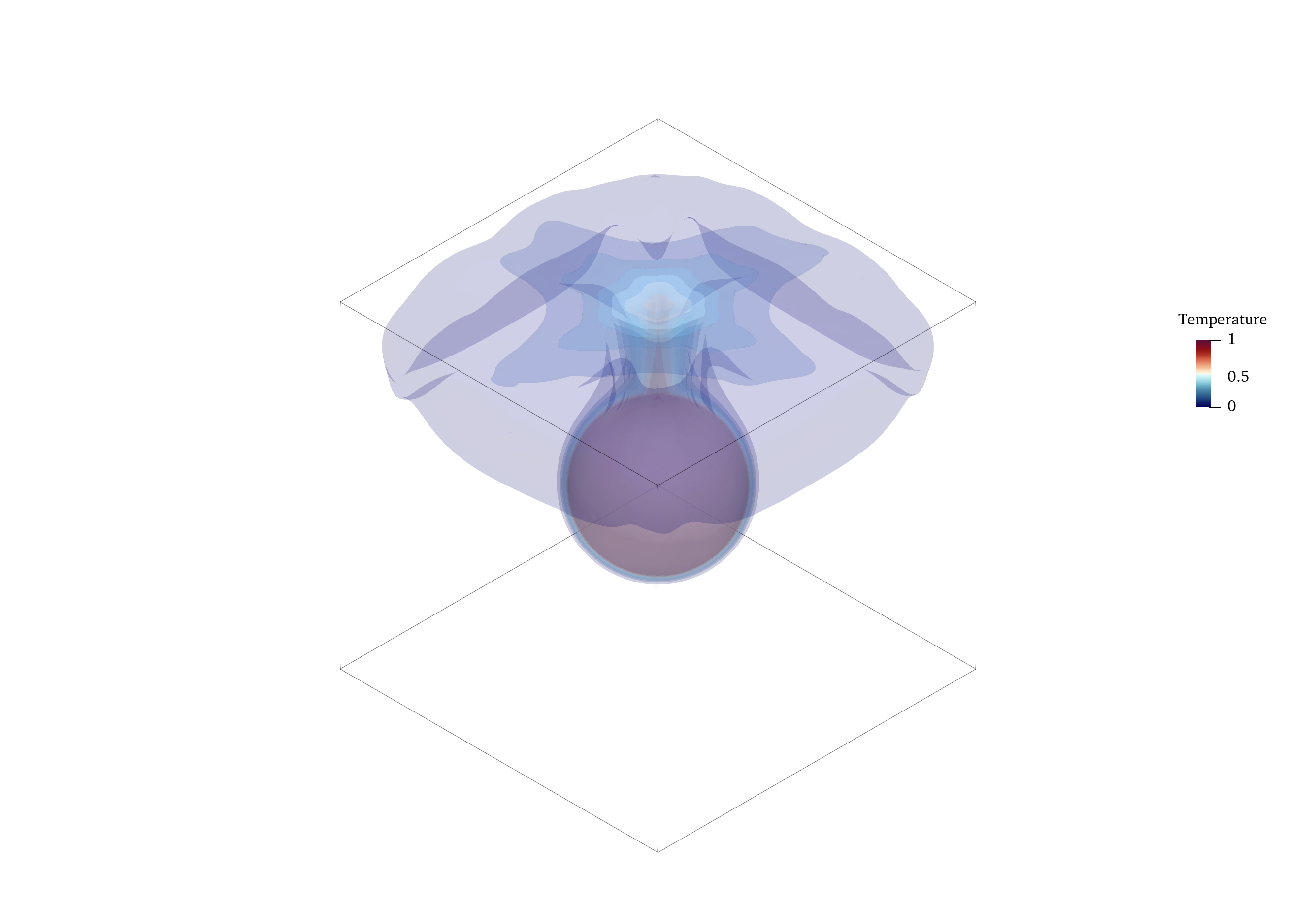}
        \caption{$Ra = 10^8$}
    \end{subfigure}%
    \caption{Natural convection around a sphere in a cubic enclosure (\secref{Natural-3D}): instantaneous (non-dimensional) temperature contours for various Rayleigh numbers. Contours are plotted from 0 to 1 at regular intervals of 0.1.}
    \label{fig:HT_SphereInCube}
\end{figure}

\begin{figure}[!t]
    \centering
    \begin{subfigure}{0.25\textwidth}
    \centering
        \includegraphics[width=\linewidth,trim=600 200 600 60,clip]{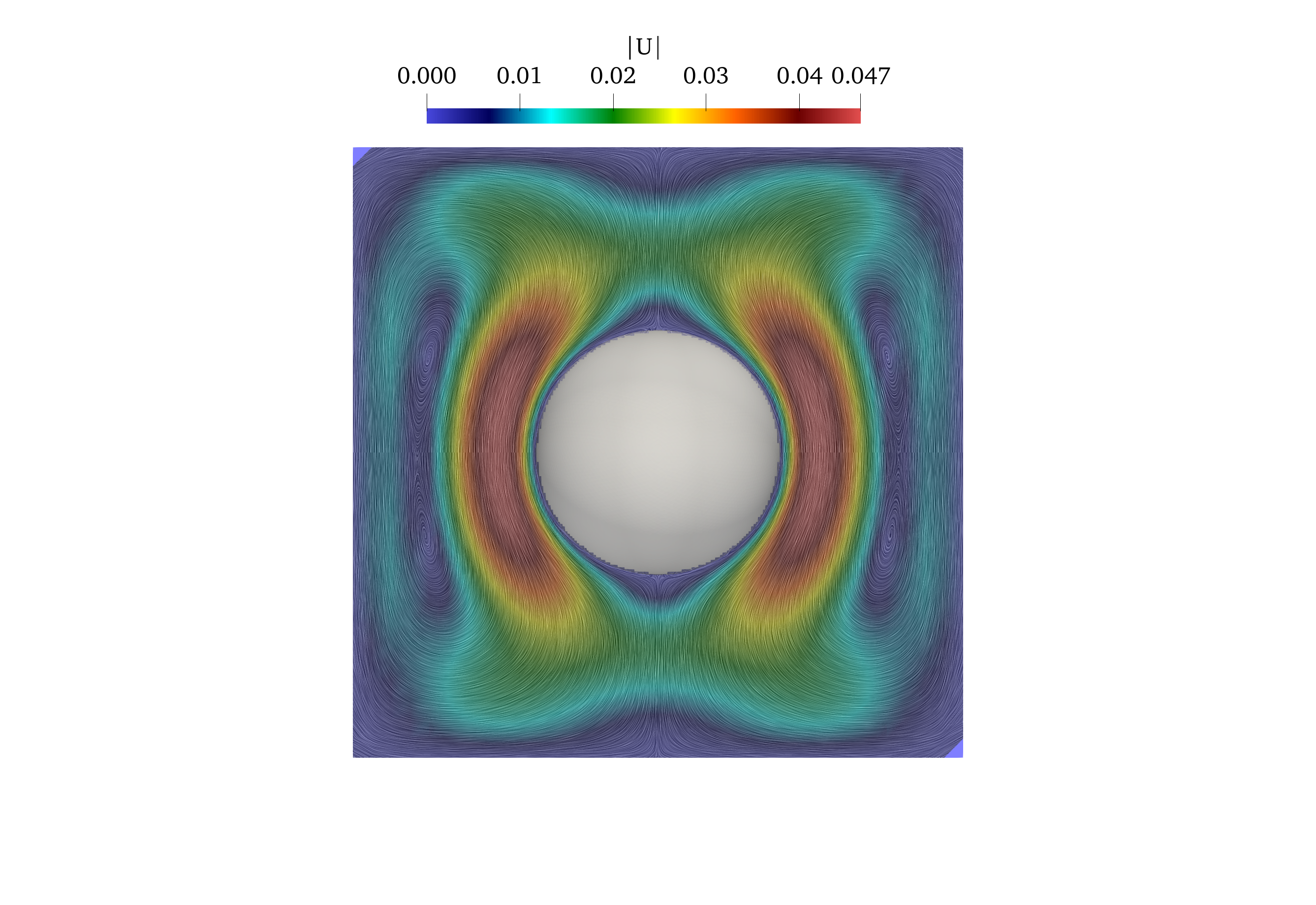}
        \caption{$Ra = 10^3$}
    \end{subfigure}%
    \begin{subfigure}{0.25\textwidth}
    \centering
        \includegraphics[width=\linewidth,trim=600 200 600 60,clip]{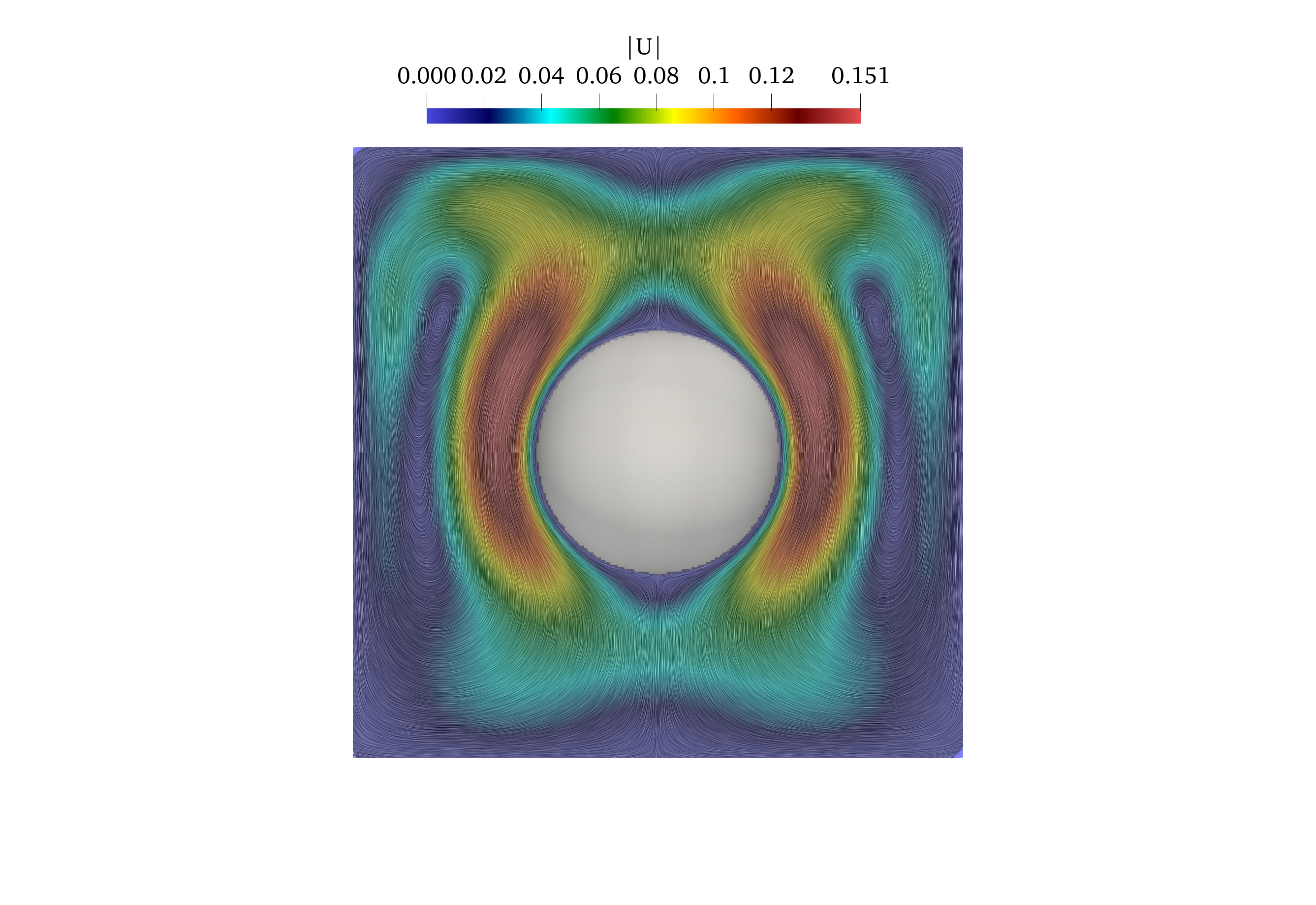}
        \caption{$Ra = 10^4$}
    \end{subfigure}%
        \begin{subfigure}{0.25\textwidth}
        \centering
        \includegraphics[width=\linewidth,trim=600 200 600 60,clip]{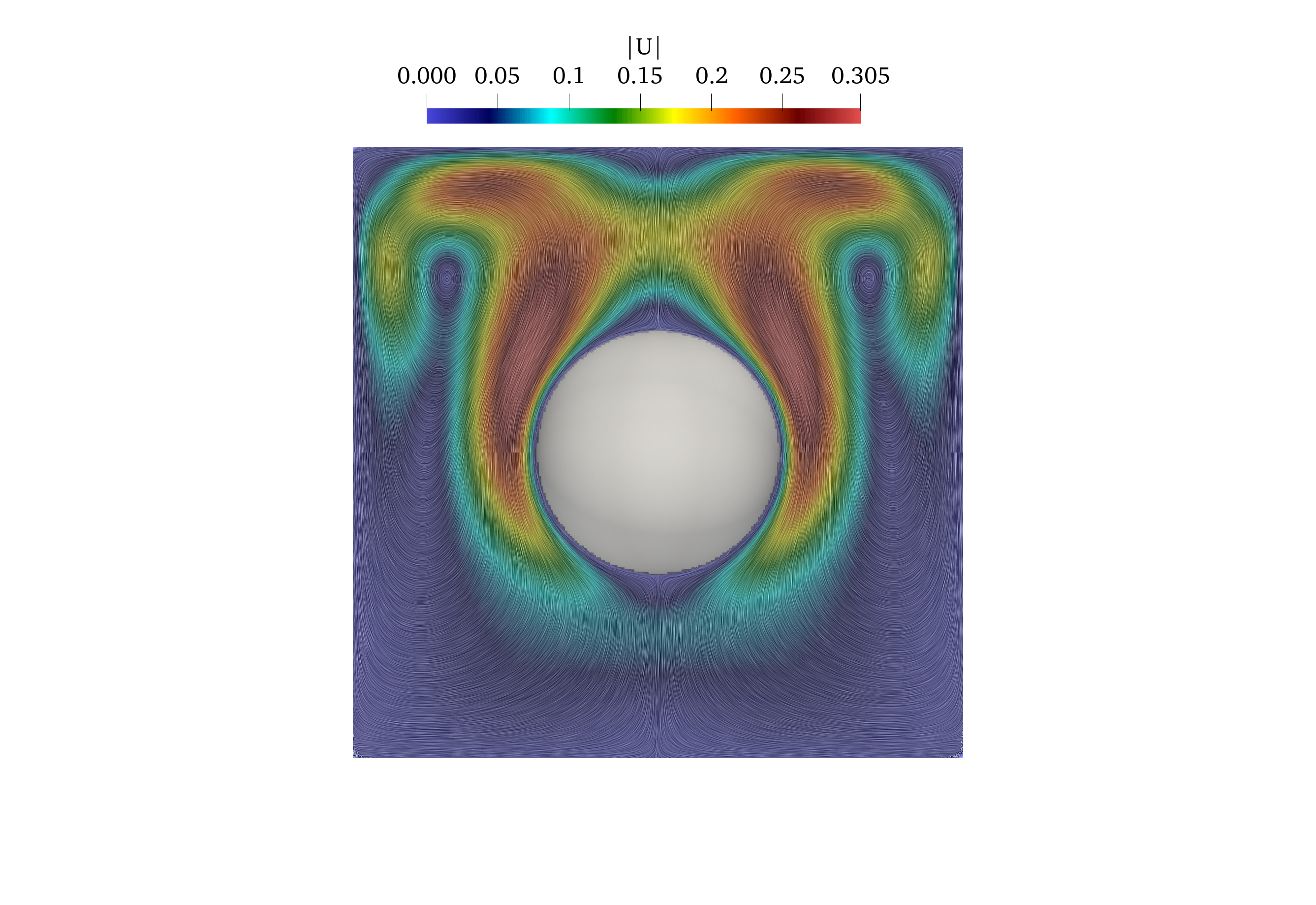}
        \caption{$Ra = 10^5$}
    \end{subfigure}%
    \\
    \vspace{0.2in}
    \begin{subfigure}{0.25\textwidth}
    \centering
        \includegraphics[width=\linewidth,trim=600 200 600 60,clip]{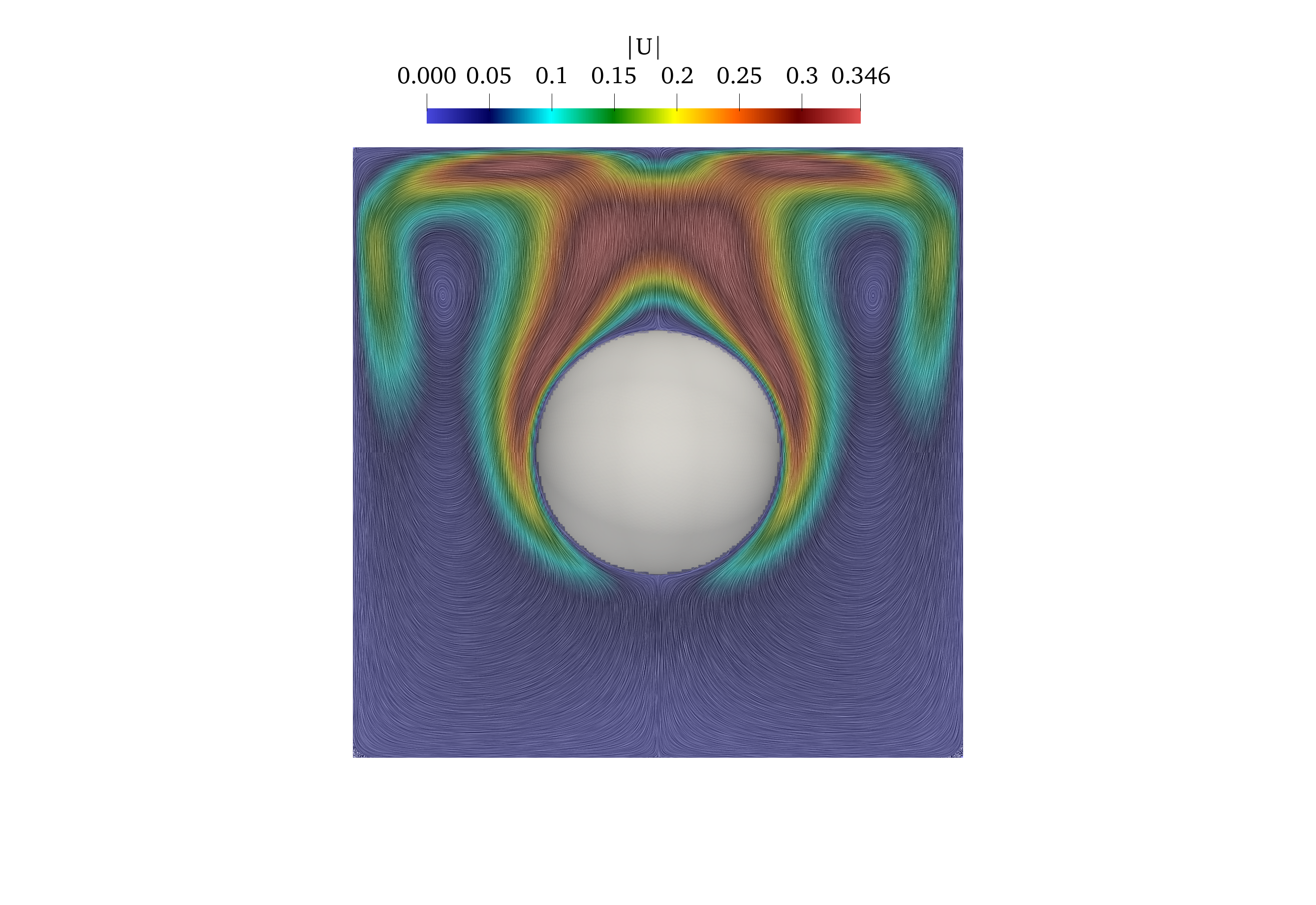}
        \caption{$Ra = 10^6$}
    \end{subfigure}%
    \begin{subfigure}{0.25\textwidth}
    \centering
        \includegraphics[width=\linewidth,trim=600 200 600 60,clip]{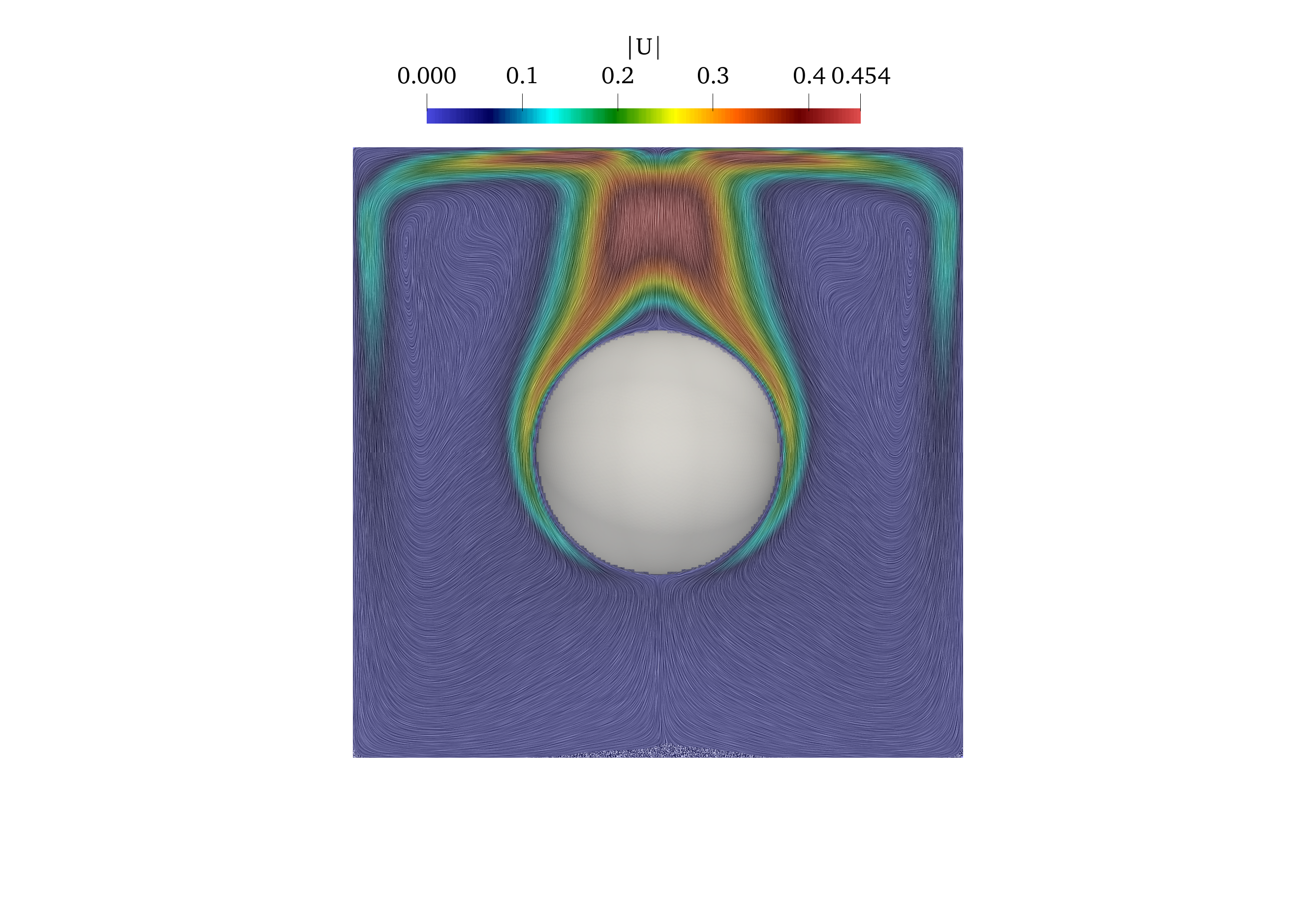}
        \caption{$Ra = 10^7$}
    \end{subfigure}%
    \begin{subfigure}{0.25\textwidth}
    \centering
        \includegraphics[width=\linewidth,trim=600 200 600 80,clip]{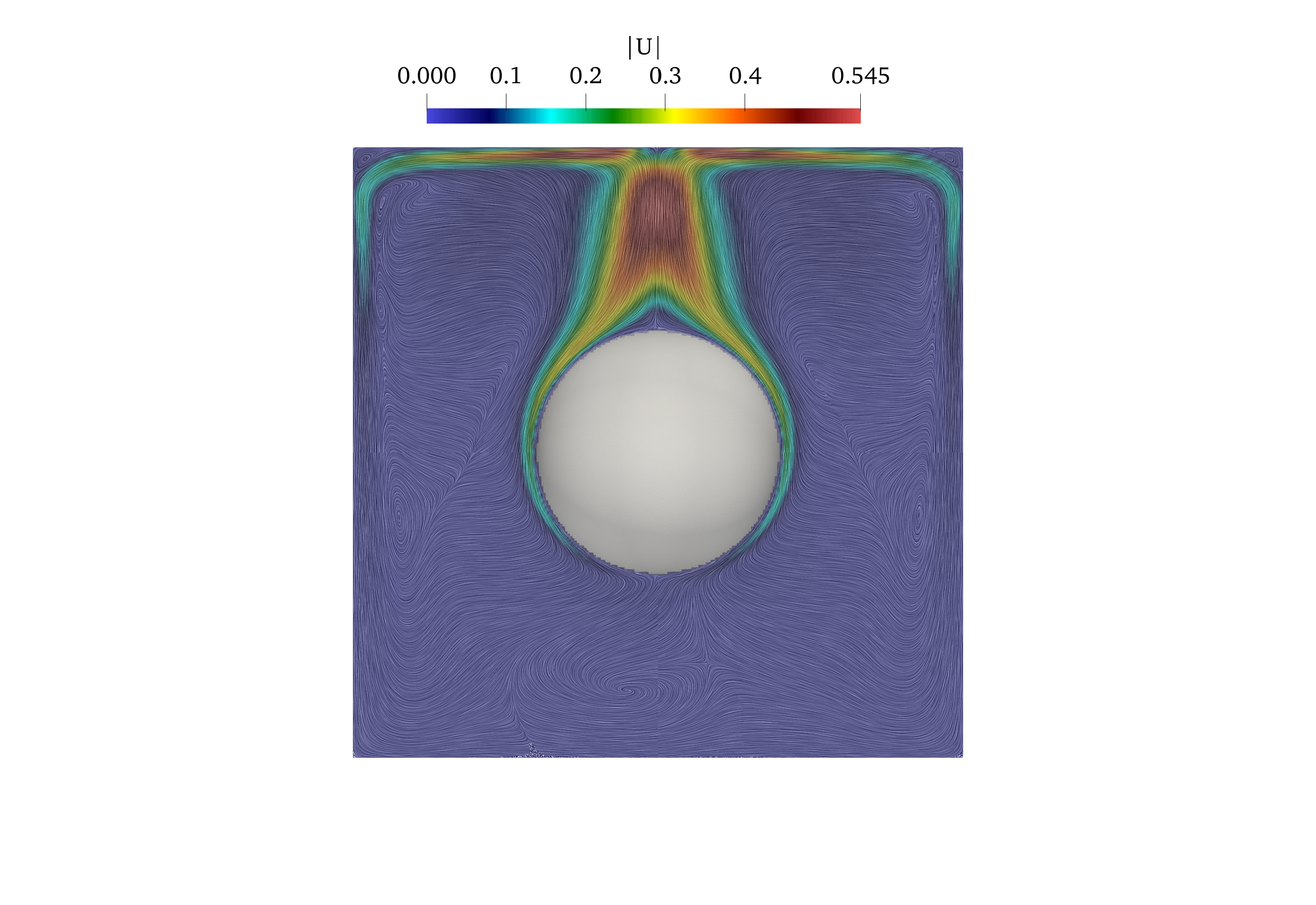}
        \caption{$Ra = 10^8$}
    \end{subfigure}%
    \caption{Natural convection around a sphere in a cubic enclosure (\secref{Natural-3D}): instantaneous velocity Line Integral Convolution (LIC) visualizations at various Rayleigh numbers.}
    \label{fig:LIC_SphereInCube}
\end{figure}

For various Reynolds numbers, instantaneous (non-dimensional) temperature contours are shown in~\figref{fig:HT_SphereInCube} and instantaneous velocity Line Integral Convolution (LIC) visualizations are shown in~\figref{fig:LIC_SphereInCube}.

For \(Ra = 10^7\), we conduct a mesh convergence study by testing three different refinements in the spherical refinement region near the sphere and along the cube's boundary, as shown in \tabref{tab:SphereInCube_ConvergenceStudy}. 
In \tabref{tab:SphereInCube}, we present the Nusselt numbers on the walls and the surface of the sphere for \(Ra = 10^3\) through \(Ra = 10^8\). For \(Ra = 10^3\) to \(Ra = 10^6\), we also compare with the existing literature, and we show that the proposed Octree-SBM framework is in good agreement with previous references. In \tabref{tab:SphereInCube_Cxyz}, we list the drag force coefficients acting on the sphere. Specifically, the drag force coefficients are calculated as the ratio of the force to the cross-section area of the sphere ($\pi \times 0.2^2$).

\pgfplotstableread{SphereInCube.txt}\SphereInCube

\begin{table}[t!]
    \centering
    \setlength{\extrarowheight}{2pt}
    \caption{Natural convection around a sphere in a cubic enclosure (\secref{Natural-3D}): time-averaged Nusselt number comparisons at various Rayleigh numbers.}
    \begin{tabular}{P{1.5cm}L{3.8cm}R{1.5cm}R{1.5cm}R{1.5cm}R{1.5cm}}
        \toprule
        \textit{Ra} & \textit{Study} & \textbf{$\overline{Nu}_{T}$} &\textbf{$\overline{Nu}_{B}$}&\textbf{$\overline{Nu}_{S}$} &\textbf{$\overline{Nu}_{Sp}$} \\ 
        \midrule
        \multirow{3}{*}{$10^{3}$}  & \citet{yoon2010three} (FVM) & 0.69 & 0.62 & 0.66 & 7.42 \\
        & \citet{Chen2020} (LBM) &  0.71   &  0.63 & 0.67 & 7.97 \\
        & Octree-SBM           &  \pgfplotstablegetelem{0}{Nu_T}\of\SphereInCube \pgfplotsretval
        &   \pgfplotstablegetelem{0}{Nu_B}\of\SphereInCube \pgfplotsretval &  \pgfplotstablegetelem{0}{Nu_s}\of\SphereInCube \pgfplotsretval & \pgfplotstablegetelem{0}{Nu_sp}\of\SphereInCube \pgfplotsretval \\ \hline
        \multirow{3}{*}{$10^{4}$}  & \citet{yoon2010three} (FVM)           & 1.23 & 0.38 & 0.64 & 7.80 \\
        & \citet{Chen2020} (LBM)  &  1.23   &  0.39 & 0.66 & 8.46 \\
        & Octree-SBM           &  \pgfplotstablegetelem{1}{Nu_T}\of\SphereInCube \pgfplotsretval
        &   \pgfplotstablegetelem{1}{Nu_B}\of\SphereInCube \pgfplotsretval &  \pgfplotstablegetelem{1}{Nu_s}\of\SphereInCube \pgfplotsretval & \pgfplotstablegetelem{1}{Nu_sp}\of\SphereInCube \pgfplotsretval \\ \hline
        \multirow{3}{*}{$10^{5}$}  & \citet{yoon2010three} (FVM)         & 3.87 & 0.08 & 0.67 & 12.61 \\
        & \citet{Chen2020} (LBM)  &  3.91   &  0.09 & 0.69 & 13.42 \\
        & Octree-SBM           &  \pgfplotstablegetelem{2}{Nu_T}\of\SphereInCube \pgfplotsretval
        &   \pgfplotstablegetelem{2}{Nu_B}\of\SphereInCube \pgfplotsretval &  \pgfplotstablegetelem{2}{Nu_s}\of\SphereInCube \pgfplotsretval & \pgfplotstablegetelem{2}{Nu_sp}\of\SphereInCube \pgfplotsretval \\ \hline
        \multirow{3}{*}{$10^{6}$}  & \citet{yoon2010three} (FVM)        & 6.97 & 0.04 & 0.97 & 20.64 \\
        & \citet{Chen2020} (LBM)  &  6.94   &  0.04 & 1.02 & 22.58 \\
        & Octree-SBM           &  \pgfplotstablegetelem{3}{Nu_T}\of\SphereInCube \pgfplotsretval
        &   \pgfplotstablegetelem{3}{Nu_B}\of\SphereInCube \pgfplotsretval &  \pgfplotstablegetelem{3}{Nu_s}\of\SphereInCube \pgfplotsretval & \pgfplotstablegetelem{3}{Nu_sp}\of\SphereInCube \pgfplotsretval \\ \hline
        \multirow{1}{*}{$10^{7}$}  
        & Octree-SBM           &  \pgfplotstablegetelem{4}{Nu_T}\of\SphereInCube \pgfplotsretval
        &   \pgfplotstablegetelem{4}{Nu_B}\of\SphereInCube \pgfplotsretval &  \pgfplotstablegetelem{4}{Nu_s}\of\SphereInCube \pgfplotsretval & \pgfplotstablegetelem{4}{Nu_sp}\of\SphereInCube \pgfplotsretval \\ \hline
            \multirow{1}{*}{$10^{8}$}  
        & Octree-SBM           &  \pgfplotstablegetelem{5}{Nu_T}\of\SphereInCube \pgfplotsretval
        &   \pgfplotstablegetelem{5}{Nu_B}\of\SphereInCube \pgfplotsretval &  \pgfplotstablegetelem{5}{Nu_s}\of\SphereInCube \pgfplotsretval & \pgfplotstablegetelem{5}{Nu_sp}\of\SphereInCube \pgfplotsretval \\
        \bottomrule
    \end{tabular}
    \label{tab:SphereInCube}
\end{table}

\begin{table}[t!]
    \centering
    \caption{Natural convection around a sphere in a cubic enclosure (\secref{Natural-3D}): comparison of time-averaged drag force coefficients at different Rayleigh numbers (the direction of gravity is $-y$).}
    \begin{tabular}{P{3cm}R{3cm}R{3cm}R{3cm}}
        \toprule
        \textit{Ra} & \textbf{$\overline{C_{x}}$} &\textbf{$\overline{C_{y}}$}&\textbf{$\overline{C_{z}}$}  \\ 
        \midrule
        $10^{3}$  &   \pgfplotstablegetelem{0}{Cx}\of\SphereInCube \pgfmathprintnumber[sci]{\pgfplotsretval}
   &   \pgfplotstablegetelem{0}{Cy}\of\SphereInCube \pgfplotsretval &  \pgfplotstablegetelem{0}{Cz}\of\SphereInCube \pgfmathprintnumber[sci]{\pgfplotsretval} \\
        $10^{4}$  &  \pgfplotstablegetelem{1}{Cx}\of\SphereInCube \pgfmathprintnumber[sci]{\pgfplotsretval}
   &   \pgfplotstablegetelem{1}{Cy}\of\SphereInCube \pgfplotsretval &  \pgfplotstablegetelem{1}{Cz}\of\SphereInCube \pgfmathprintnumber[sci]{\pgfplotsretval} \\
        $10^{5}$  & \pgfplotstablegetelem{2}{Cx}\of\SphereInCube \pgfmathprintnumber[sci]{\pgfplotsretval}
   &   \pgfplotstablegetelem{2}{Cy}\of\SphereInCube \pgfplotsretval &  \pgfplotstablegetelem{2}{Cz}\of\SphereInCube \pgfmathprintnumber[sci]{\pgfplotsretval} \\
        $10^{6}$  &  \pgfplotstablegetelem{3}{Cx}\of\SphereInCube \pgfmathprintnumber[sci]{\pgfplotsretval}
   &   \pgfplotstablegetelem{3}{Cy}\of\SphereInCube \pgfplotsretval &  \pgfplotstablegetelem{3}{Cz}\of\SphereInCube \pgfmathprintnumber[sci]{\pgfplotsretval} \\
        $10^{7}$  &  \pgfplotstablegetelem{4}{Cx}\of\SphereInCube \pgfmathprintnumber[sci]{\pgfplotsretval}
   &   \pgfplotstablegetelem{4}{Cy}\of\SphereInCube \pgfplotsretval &  \pgfplotstablegetelem{4}{Cz}\of\SphereInCube \pgfmathprintnumber[sci]{\pgfplotsretval} \\
           $10^{8}$  &  \pgfplotstablegetelem{5}{Cx}\of\SphereInCube \pgfmathprintnumber[sci]{\pgfplotsretval}
   &   \pgfplotstablegetelem{5}{Cy}\of\SphereInCube \pgfplotsretval &  \pgfplotstablegetelem{5}{Cz}\of\SphereInCube \pgfmathprintnumber[sci]{\pgfplotsretval} \\
        \bottomrule
    \end{tabular}
    \label{tab:SphereInCube_Cxyz}
\end{table}

\subsection{Natural convection around a sphere and gyroid in a cubic enclosure}\label{Natural-3D-Gyroid}
\begin{figure}[!t]
    \centering
    \begin{subfigure}{0.41\textwidth}
    \centering
    \includegraphics[width=1\linewidth,trim=0 0 0 0,clip]{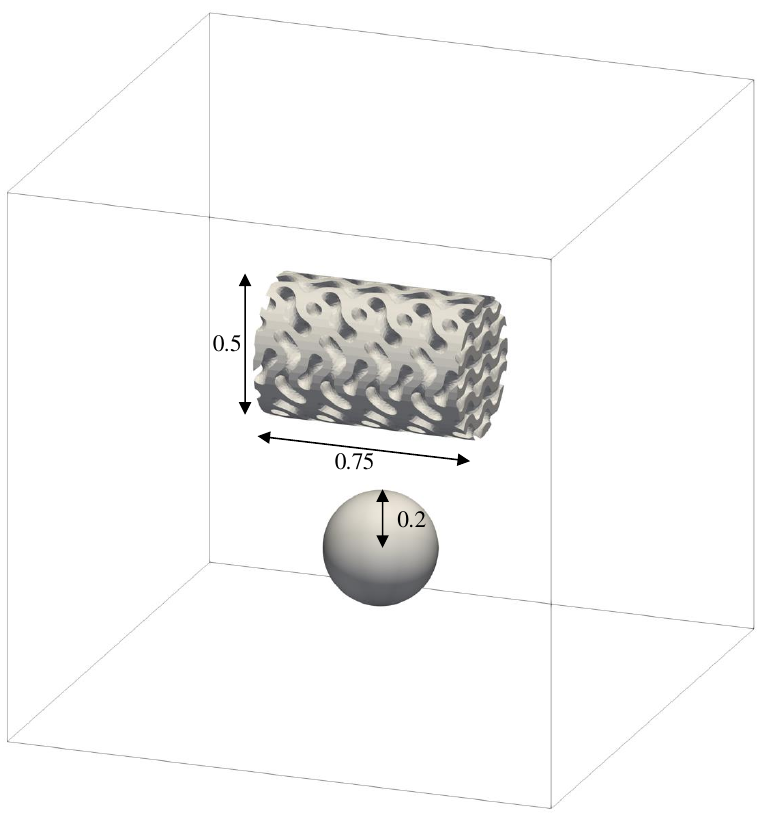}
    \caption{Schematic of the problem setup for natural convection around a three-dimensional sphere and a gyroid structure enclosed within a cubic domain.}
    \label{fig:ProblemSetup-Gyroid}
    \end{subfigure}%
    \hspace{0.05\textwidth} 
    \begin{subfigure}{0.49\textwidth}
    \centering
    \includegraphics[width=1\linewidth,trim=210 0 520 0,clip]{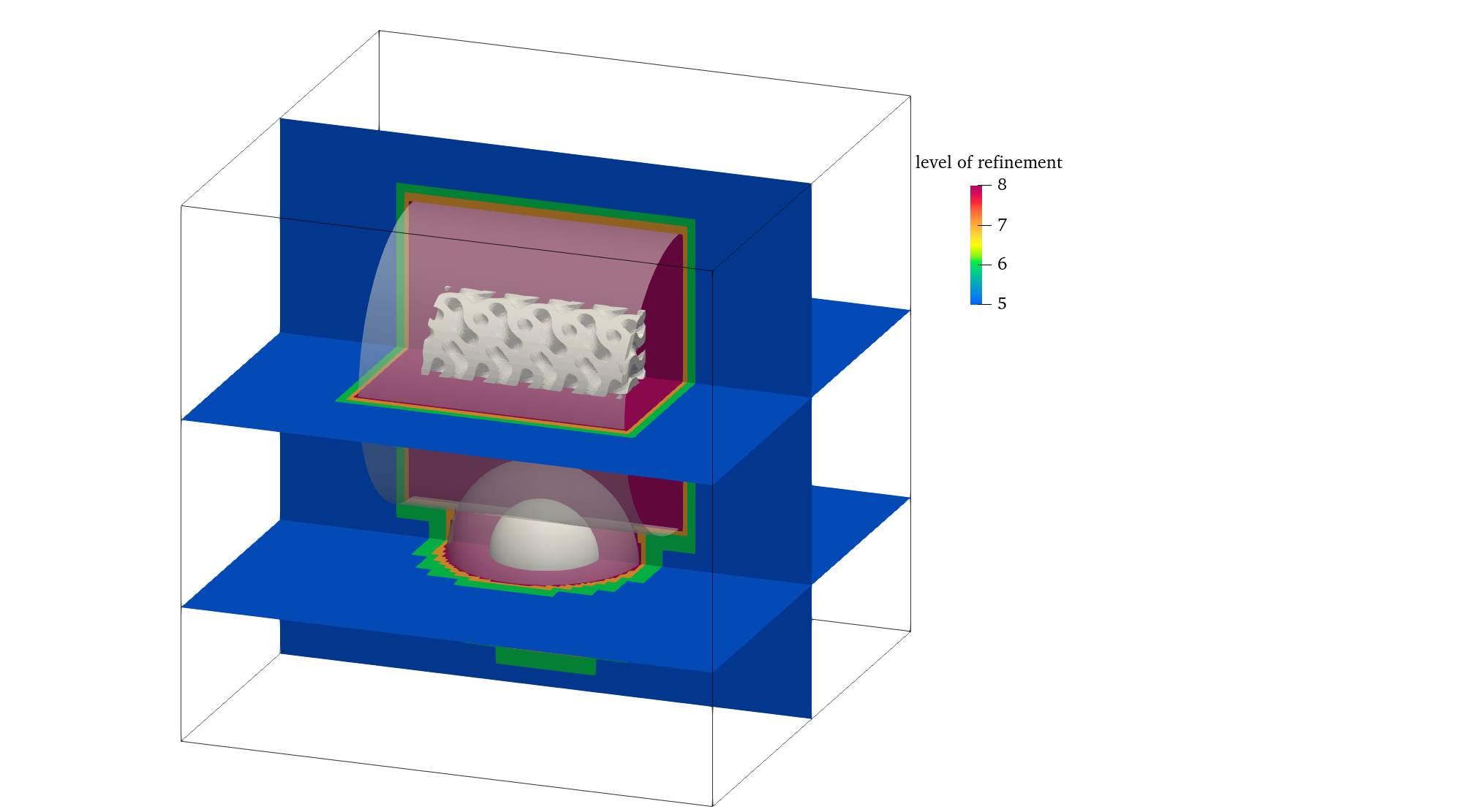}
    \caption{Refinement levels for the computational mesh, illustrating local mesh refinement in the vicinity of the larger cylindrical region aligned with the gyroid's centerline and the larger spherical region.}
    \label{fig:Lvl-Gyroid}
    \end{subfigure}%
    \caption{Natural convection around complex obstacles (\secref{Natural-3D-Gyroid}): illustration of the problem setup and mesh refinement strategy.}
    \label{fig:GyroidSphereInCube_ProblemSetup}
\end{figure}

We finally illustrate our framework on a case exhibiting a very complex geometry and coupled thermal-fluid phenomena. 
Specifically, we include a gyroid obstacle to the geometric domain of the previous tests, as illustrated in \figref{fig:GyroidSphereInCube_ProblemSetup}.
The gyroid has an intricate geometry, and body-fitted grids could be challenging to generate in this case.
Similarly, enforcing thermal and no-slip conditions in conventional immersed boundary approaches becomes non-trivial. 
These geometry can be easily described and handled by the proposed Octree-SBM approach. 

The fluid domain is a box with dimensions $[0, 2] \times [0, 2] \times [0, 2]$, containing a sphere with a radius of 0.2 and centered at $(1, 0.5, 1)$ and a gyroid structure.
The gyroid structure is bounded by a cylinder with a radius of 0.5 and a height of 0.75, centered at $(1, 1.2, 1)$. The mesh refinement strategy includes a base refinement level of 5 (mesh size = $2 \cdot 2^{-5}$). Additional local mesh refinements at a level of 8 (mesh size = $2 \cdot 2^{-8}$) are applied in two regions: (a) a larger cylinder with a radius of 0.55, aligned with the gyroid's centerline, and (b) a larger sphere with a radius of 0.55, centered at the same location as the sphere geometry. The mesh refinement levels around the geometries are illustrated in \figref{fig:Lvl-Gyroid}. 

The boundary conditions applied to the walls of the cube for both the Navier-Stokes and heat transfer equations in this problem are identical with those outlined in \secref{Natural-3D}. No-slip boundary conditions are applied to the flow velocity on the surface of the gyroid and the sphere.
The temperature on the surface of the gyroid is set as $\theta = -1$, while on the surface of the sphere, it is set as $\theta = 1$. $\theta = -1$  (non-dimensional temperature) represents a relatively lower temperature than other fluid regions within the thermal flow system being solved. Setting $\theta = -1$  generates a negative buoyancy force, causing the fluid to move downward relative to other parts of the fluid. Conversely, setting $\theta = 1$ generates a positive buoyancy force, causing the fluid to move upward relative to other parts of the fluid.

\begin{figure}[!t]
    \centering
    \begin{subfigure}{0.49\textwidth}
    \centering
    \includegraphics[width=\linewidth,trim=450 0 600 50,clip]{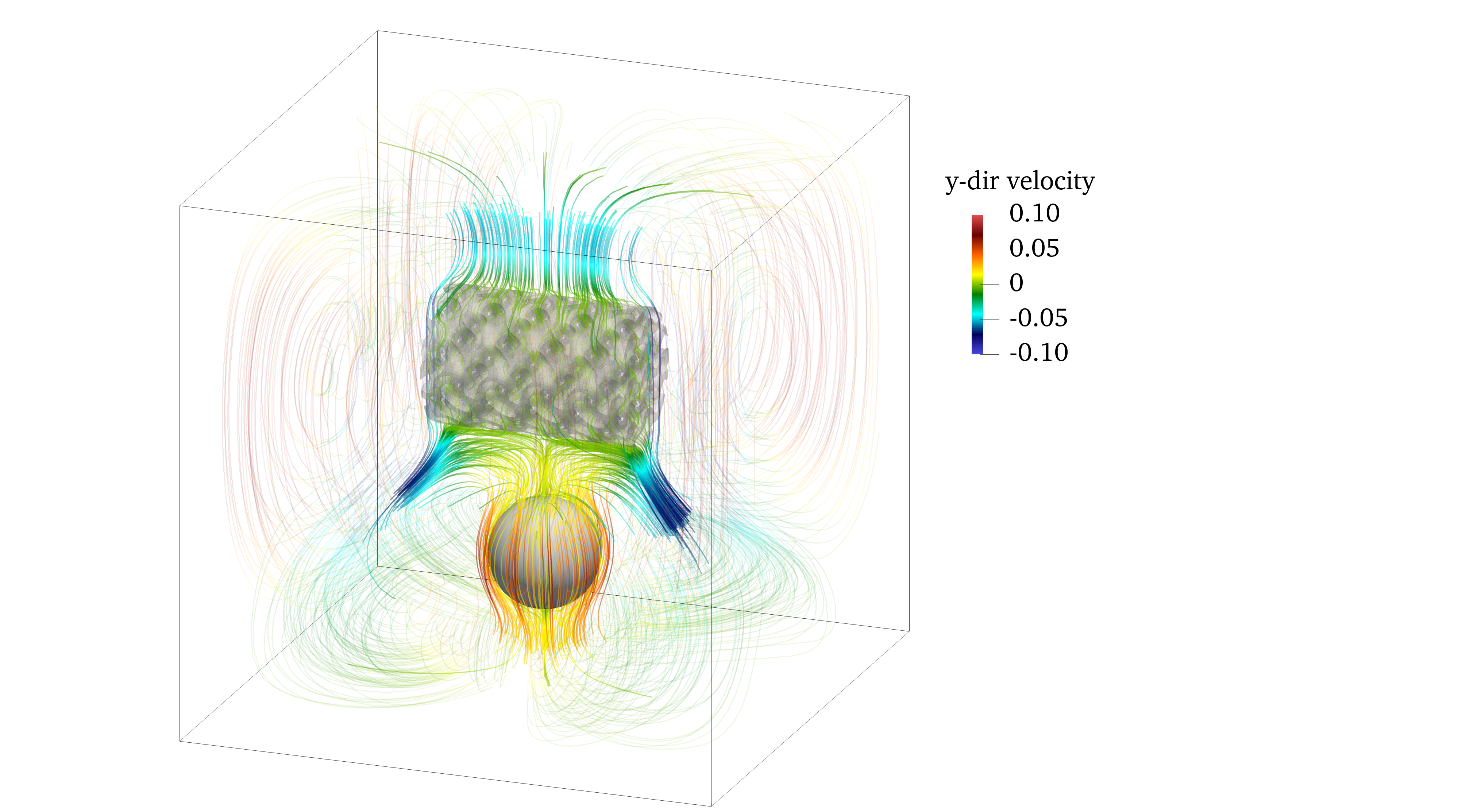}
    \caption{Stream lines}
    \label{fig:StreamLine-Gyroid}
    \end{subfigure}%
    \begin{subfigure}{0.49\textwidth}
    \centering
    \includegraphics[width=\linewidth,trim=450 0 600 50,clip]{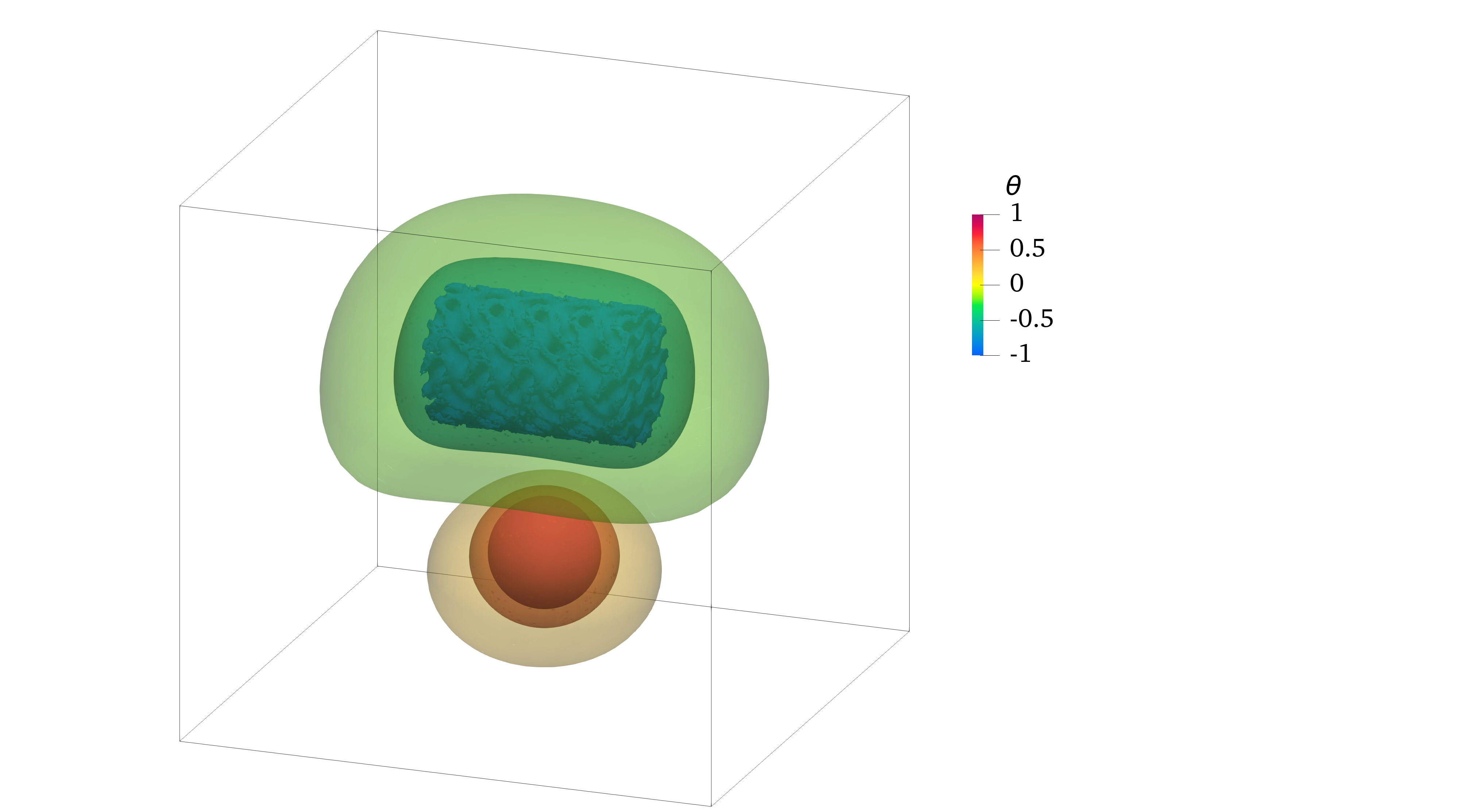}
    \caption{Temperature contours }
    \label{fig:TemperatureContour-Gyroid}
    \end{subfigure}%
    \caption{Natural convection around complex obstacles (\secref{Natural-3D-Gyroid}): streamlines and temperature contours after the flow reached steady-state. Temperature contour values are set to -1, -0.6, -0.2, 0.2, 0.6, and 1. Streamlines are colored by the y-direction velocity.}
    \label{fig:GyroidSphereInCube_Streamline_TemperatureContour}
\end{figure}

\begin{figure}[!t]
    \centering
    \includegraphics[width=0.95\textwidth,trim=0 0 0 0,clip]{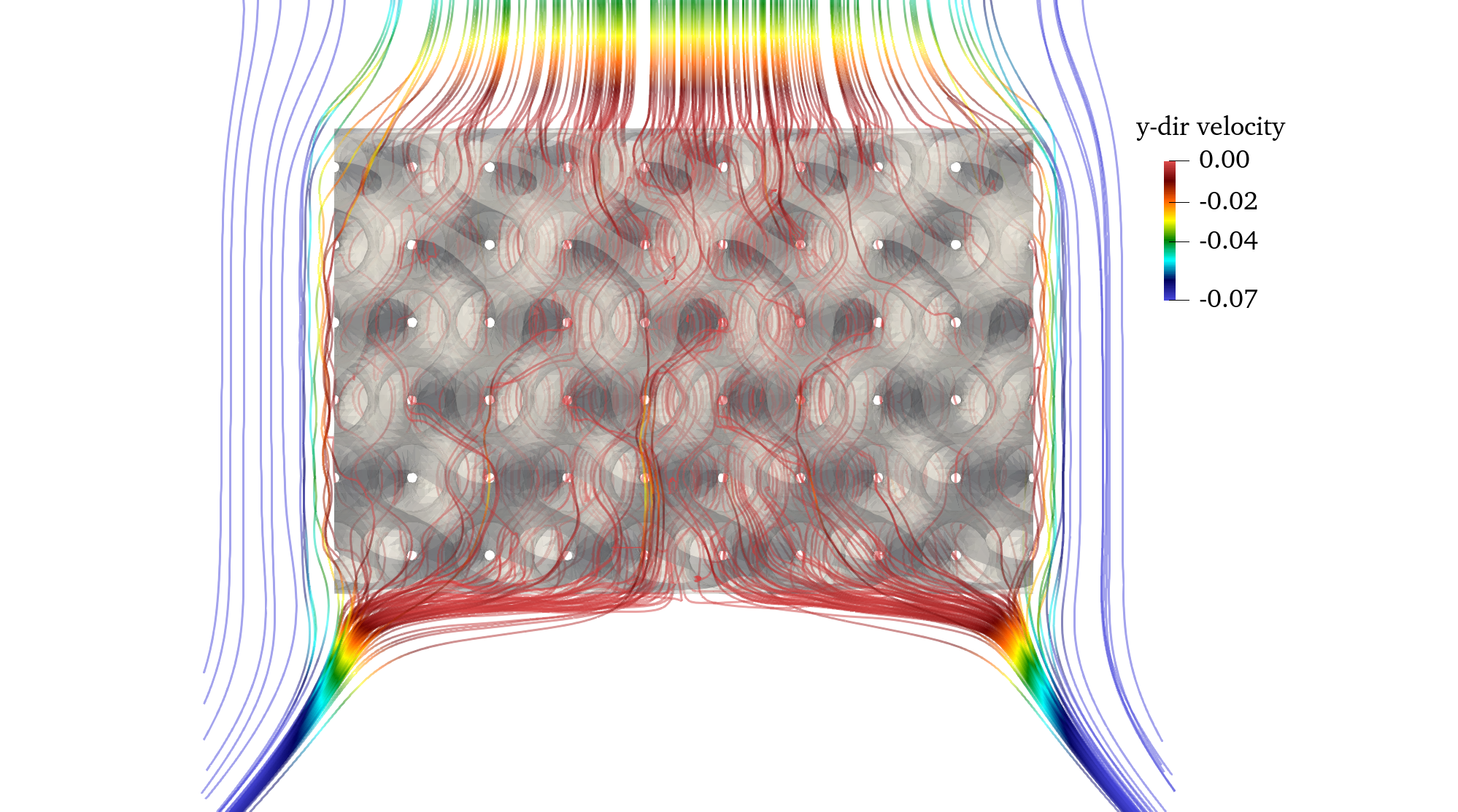}
    \caption{Natural convection around complex obstacles (\secref{Natural-3D-Gyroid}): streamlines of the flow at steady-state passing through the gyroid (front view). This image highlights the intricate streamline pattern as they navigate through the gyroid obstacle in a natural convection setting.}
    \label{fig:StreamLine-Gyroid-View-Front}
\end{figure}

We impose a Rayleigh number of $Ra = 10^3$ and a non-dimensional timestep equal to 1. Visualizations of the streamlines and temperature contours are shown in \figref{fig:StreamLine-Gyroid} and \figref{fig:TemperatureContour-Gyroid}.
The streamlines are colored by the y-direction of the flow velocity, indicating that the heated sphere generates an upward flow (positive y-velocity) in the surrounding region. Conversely, the y-direction velocity near the gyroid is negative, as the cooler temperature induces a downward flow. In \figref{fig:TemperatureContour-Gyroid}, the zero-temperature contour closely follows the shape of the Gyroid and demonstrates the proposed Octree-SBM's effectiveness in accurately enforcing boundary conditions. These visualizations provide a detailed depiction of the intricate flow behavior induced by the gyroid structure in a natural convection scenario. The front view (\figref{fig:StreamLine-Gyroid-View-Front}) captures the directional shifts of the streamlines as they interact with the gyroid, revealing the path of streamlines navigating through the gyroid's lattice-like geometry.

\subsection{Parallel performance of the Octree-SBM framework: strong scaling test}\label{subsec:scaling}
We present the scaling performance of our framework on the TACC~\Frontera~system. For this study, we analyze the problem outlined in \secref{subsub:UHF}. The simulations are conducted at a Reynolds number of 40 and a Peclet number of 28, using a computational mesh comprising 548,870 nodes. Starting from the initial condition, the simulations advance with a non-dimensional time step of 1 until reaching a total non-dimensional time of 5. A scaling analysis is performed, employing $56 \times n$ processors, where $n$ varies from 1 to 8.

\begin{figure}[!h]
\centering
\begin{subfigure}{.45\linewidth}
    \centering
    \begin{tikzpicture}
    \begin{loglogaxis}[ 
        width=0.99\linewidth,
        height=0.75\linewidth, 
        xtick={56,112,168,224,280,336, 392,448,504,560},  
        xticklabel style={rotate=45,font=\footnotesize},
        xticklabels={$56$,$112$, $168$, $224$,$280$, $336$, $392$,$448$,$504$,$560$}, 
        scaled y ticks=true,
        xlabel={\footnotesize Number of processors},
        ylabel={\footnotesize Time (s)},
        legend style={at={(0.37,0.2)}, anchor=north, nodes={scale=0.65, transform shape}}, 
        legend columns=3,
        xmin=50,
        xmax=500,
        ymin=10,
        ymax=1000,
        ]
        \addplot table [x={proc},y={T_1p0_total},col sep=comma] {Scaling_NSHT_UHF.txt};
        \addplot +[mark=none, black, dashed] [domain=50:700]{30000/x};
    \end{loglogaxis}
    \end{tikzpicture}
\end{subfigure}
\caption{Scaling performance of the Octree-SBM computation for two-dimensional flow past a cylinder with a uniform heat flux, as evaluated on TACC's~\Frontera~supercomputer (\secref{subsec:scaling}).} 
\label{fig:ScalingNSHTUHF}
\end{figure}
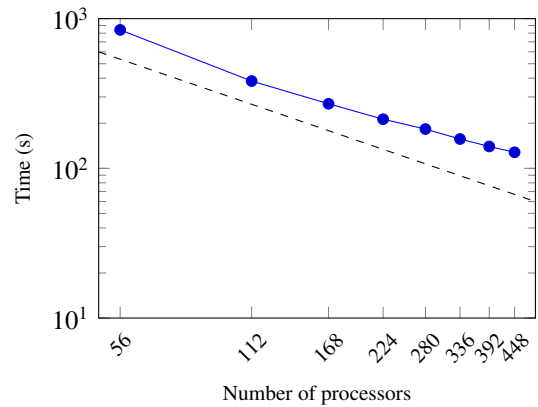

We utilize $\petsc$ solvers for both the Navier-Stokes (NS) and heat transfer (HT) equations in our simulations. The NS equations are solved using the GMRES (Generalized Minimal Residual) method with a restart value of 1000, coupled with the Additive Schwarz Method (ASM) preconditioner configured with an overlap of 10. Similarly, for the heat transfer (HT) equations, we use the GMRES solver with a restart value of 1000 and an ASM preconditioner configured with an overlap of 3. \figref{fig:ScalingNSHTUHF} illustrates the scalability of our approach, demonstrating the progression of total solution time as the processor count increases. The performance closely approximates an idealized scaling pattern, as represented by the reference dashed line in the plot.

\section{Conclusions and Future Work}
\label{sec:Conclusions}

The challenges posed by accurately simulating thermal incompressible flows in domains with complex geometries motivate the development of more versatile and efficient computational frameworks. Traditional boundary-fitted methods, though precise, struggle with scalability and preprocessing demands. Addressing these limitations, this study explored the potential of the Shifted Boundary Method (SBM) in combination with Octree-based grids. This approach aligns with the growing need for adaptable, high-performance tools capable of tackling multiphysics scenarios in diverse flow regimes and geometries.

Key contributions of this work include the application of the SBM for coupled thermal flow simulations using the linear {semi-implicit} Navier-Stokes and heat transfer equations, which significantly enhance computational efficiency. The use of Octree meshes facilitates accurate boundary condition enforcement in complex geometries, while comprehensive validation across two-dimensional and three-dimensional cases demonstrates robustness across laminar, transitional, and turbulent regimes. Furthermore, the framework's ability to handle Dirichlet and Neumann boundary conditions with high precision underscores its versatility and accuracy, paving the way for applications in a wide array of engineering and scientific domains.

Looking ahead, several exciting avenues for further exploration emerge. Extending the framework to encompass more complex multiphysics problems, such as fluid-structure interaction, could significantly broaden its applicability. Incorporating adaptive mesh refinement (AMR) techniques could enhance both accuracy and computational efficiency, while higher-order finite element basis functions may improve solution fidelity. 
Addressing these challenges will further solidify the Octree-SBM’s role as a powerful tool in the computational modeling of thermal flows.

\section*{Acknowledgements}
This work was partly supported by the National Science Foundation under the grants LEAP-HI 2053760 (BG, AK, CHY), DMREF 2323715/2323716 (BG, AK, CHY), and DMS 2207164 and DMS 2409919 (GS). BG, AK, and CHY are supported in part by AI Research Institutes program supported by NSF and USDA-NIFA under AI Institute for Resilient Agriculture, grant 2021-67021-35329. We gratefully acknowledge computing support through TACC via the NAIRR and ACCESS programs.

\bibliographystyle{elsarticle-num-names.bst}
\bibliography{Bibs} 

\appendix
\setcounter{figure}{0}
\setcounter{table}{0}

\section{Validation of simulation code}
\subsection{Two-dimensional manufactured solutions for {linear semi-implicit} Navier--Stokes}\label{sec:NS_MMS}

To assess the convergence behavior of the {linear semi-implicit} Navier-Stokes on Octree-SBM, we apply the technique of manufactured solutions. This approach involves selecting a solenoidal solution and substituting it into the Navier-Stokes equations. The resulting residual is then treated as a forcing term on the right-hand side. For our analysis, the manufactured solutions along with their corresponding forcing terms are defined as follows:

\begin{equation}
\begin{split}
\vec{v} &= \left( \pi \sin^2(\pi x)\sin(2 \pi y)\sin(t) , -\pi\sin(2\pi x)\sin^2(\pi y)\sin(t) \right), \\
p &= \cos(\pi x)\sin(\pi y)\sin(t).
\end{split}
\label{eq:manufac_exact}
\end{equation}

\begin{figure}[!h]
\centering
\begin{subfigure}{0.49\textwidth} 
        \centering
        \begin{tikzpicture}
        \begin{loglogaxis}[
            width=0.95\linewidth, 
            scaled y ticks=true,
            xlabel={Element size, h},
            ylabel={$||u - u_{\text{exact}}||_{L^2(\Omega)}$},
            legend entries={u, v, p, $slope = 2$},
            legend style={at={(0.5,-0.25)},anchor=north, nodes={scale=0.65, transform shape}}, 
            legend columns=2,
            xmin=3e-3,
            xmax=1e-1,
            grid=both, 
        ]
        \addplot table [x={h},y={u},col sep=comma] {L2errorMMS.txt};
        \addplot +[dashed] table [x={h},y={v},col sep=comma] {L2errorMMS.txt};
        \addplot table [x={h},y={p},col sep=comma] {L2errorMMS.txt};
        \addplot +[mark=none, red, dashed] [domain=0.001:0.1]{0.1*x^2};
        \end{loglogaxis}
        \end{tikzpicture}
    \caption{Mesh convergence study with constant timestep $k = 0.000628$.}
    \label{fig:const_k_spatial}
\end{subfigure}
\begin{subfigure}{0.49\textwidth} 
        \centering
        \begin{tikzpicture}
        \begin{loglogaxis}[
            width=0.95\linewidth, 
            scaled y ticks=true,
            xlabel={Element size, h},
            ylabel={$||u - u_{\text{exact}}||_{L^2(\Omega)}$},
            legend entries={u, v, p, $slope = 2$},
            legend style={at={(0.5,-0.25)},anchor=north, nodes={scale=0.65, transform shape}}, 
            legend columns=2,
            xmin=8e-4,
            xmax=5e-2,
            grid=both, 
        ]
        \addplot table [x={h},y={u},col sep=space] {MMS_error_temp_more_0p000628_withU_Star.txt};
        \addplot +[dashed] table [x={h},y={v},col sep=space] {MMS_error_temp_more_0p000628_withU_Star.txt};
        \addplot table [x={h},y={p},col sep=space] {MMS_error_temp_more_0p000628_withU_Star.txt};
        \addplot +[mark=none, red, dashed] [domain=0.0001:0.1]{0.1*x^2};
        \end{loglogaxis}
        \end{tikzpicture}
    \caption{Mesh convergence study with varying timestep ($k = 0.000628$).}
    \label{fig:variable_k_spatial}
\end{subfigure}
\caption{Mesh convergence results (spatial convergence).}
\label{fig:MMS}
\end{figure}
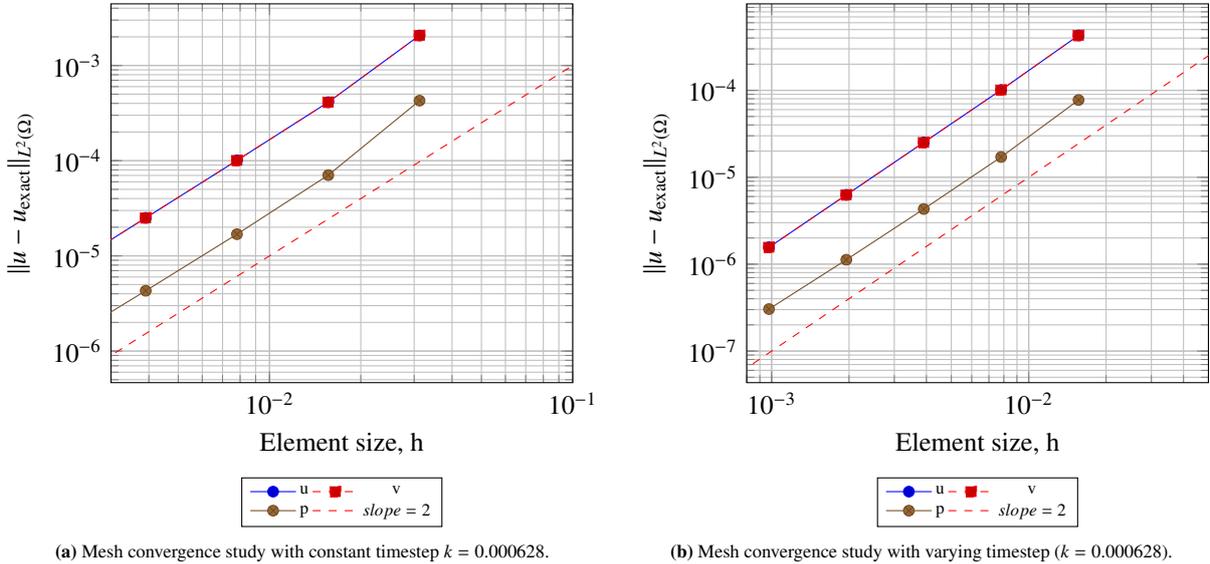

We solved the problem on the rectangle domain $[0, 1] \times [0, 1]$. The spatial convergence plot with constant timestep k are shown in \figref{fig:const_k_spatial}.

\begin{figure}[!h]
\centering
\includegraphics[width=0.6\linewidth]{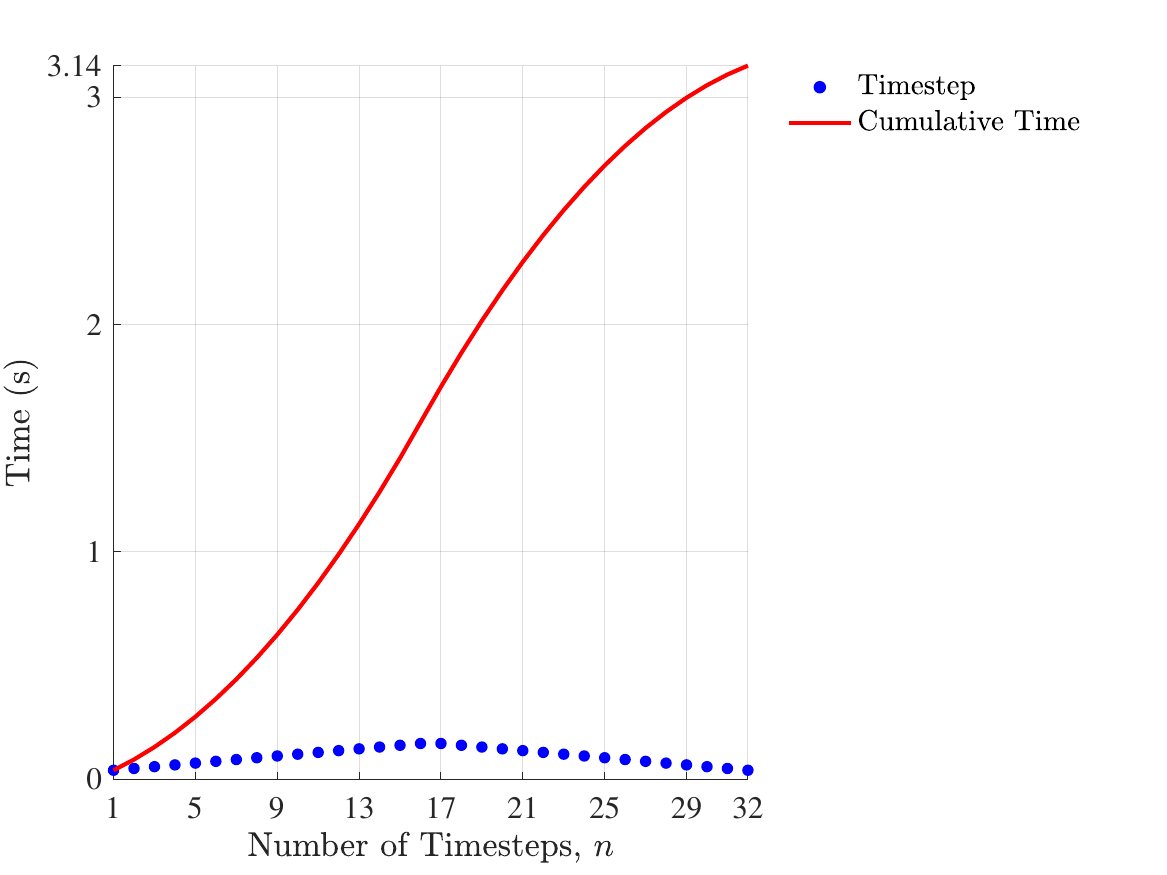}
\caption{Illustration of the variation in timestep during the MMS testing. The timestep increases linearly from $\frac{k}{4}$ to $k$, followed by a linear decrease from $k$ to $\frac{k}{4}$, with $k = 0.157$. This pattern is shown in the figure, providing a graphical representation of the time step variation throughout the simulation.}
\label{fig:VariableTime}
\end{figure}

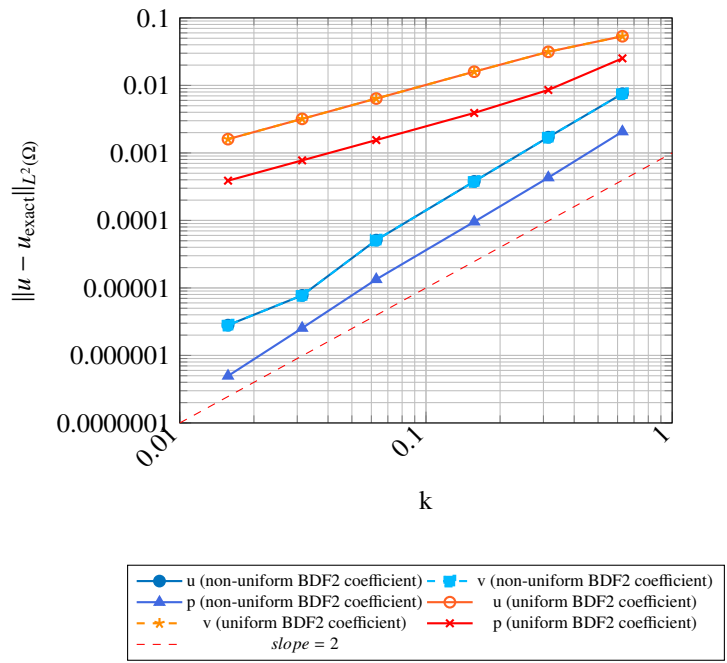
\begin{figure}[!t]
\centering
\begin{tikzpicture}
\begin{loglogaxis}[
    width=0.49\linewidth,
    scaled y ticks=true,
    xlabel={k},
    ylabel={$||u - u_{\text{exact}}||_{L^2(\Omega)}$},
    legend entries={u (non-uniform BDF2 coefficient), v (non-uniform BDF2 coefficient), p (non-uniform BDF2 coefficient), u (uniform BDF2 coefficient), v (uniform BDF2 coefficient), p (uniform BDF2 coefficient), $slope = 2$},
    legend style={at={(0.5,-0.35)}, anchor=north, nodes={scale=0.65, transform shape}},
    legend columns=2,
    xmin=0.01,
    xmax=1,
    ymin=1e-7,
    ymax=0.1,
    xtick={0.01,0.1,1},
    ytick={1e-7,1e-6,1e-5,1e-4,1e-3,1e-2,1e-1},
    x tick label style={rotate=45, anchor=east},
    log ticks with fixed point,
    grid=both,
]

\definecolor{myblue}{RGB}{0, 114, 189}
\definecolor{mycyan}{RGB}{0, 186, 255}
\definecolor{myroyalblue}{RGB}{65, 105, 225}
\definecolor{myorange}{RGB}{255, 97, 36}
\definecolor{mydarkorange}{RGB}{255, 140, 0}
\definecolor{myred}{RGB}{255, 0, 0}

\addplot[myblue, thick, mark=*] table [x={h},y={u},col sep=space] {MMS_error_temp_more_lvl9.txt};
\addplot[mycyan, dashed, thick, mark=square*] table [x={h},y={v},col sep=space] {MMS_error_temp_more_lvl9.txt};
\addplot[myroyalblue, thick, mark=triangle*] table [x={h},y={p},col sep=space] {MMS_error_temp_more_lvl9.txt};

\addplot[myorange, thick, mark=o] table [x={h},y={u},col sep=space] {MMS_error_temp_more_lvl9_WrongBDF2.txt};
\addplot[mydarkorange, dashed, thick, mark=star] table [x={h},y={v},col sep=space] {MMS_error_temp_more_lvl9_WrongBDF2.txt};
\addplot[myred, thick, mark=x] table [x={h},y={p},col sep=space] {MMS_error_temp_more_lvl9_WrongBDF2.txt};

\addplot[red, dashed, mark=none] [domain=0.01:1]{0.001*x^2};

\end{loglogaxis}
\end{tikzpicture}
\caption{Error norms for various BDF2 coefficients (see \tabref{tab:BDF_coefficients}) evaluated at different accuracy levels. The mesh size is uniformly set to $2^{-9}$.}
\label{fig:VariableTimeMMS}
\end{figure}

In addition to testing the case with a constant timestep \(k\), we have also evaluated the method of manufactured solutions (MMS) using a continuously varying timestep, based on the BDF2 coefficients shown in \tabref{tab:BDF_coefficients}. Specifically, we implemented a linear increase in the timestep from \(\frac{k}{4}\) to \(k\), followed by a linear decrease from \(k\) back to \(\frac{k}{4}\), as illustrated in \figref{fig:VariableTime}. The spatial convergence results for varying timestep are shown in \figref{fig:variable_k_spatial}. The results of the MMS temporal convergence are presented in \figref{fig:VariableTimeMMS}. As expected, the correct choice of coefficients produces the theoretical second-order convergence in time.

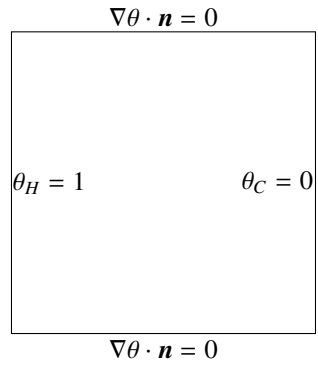
\begin{figure}[!t]
    \centering
\begin{tikzpicture}
\coordinate (A) at (0,0);
\coordinate (B) at (4,0);
\coordinate (C) at (4,4);
\coordinate (D) at (0,4);

\draw (A) -- (B) -- (C) -- (D) -- cycle;

\node at (0.5,2) {$\theta_H = 1$};

\node at (3.5,2) {$\theta_C = 0$};

\node at (2,4.2) {$\nabla \theta \cdot \bs{n} = 0$};
\node at (2,-0.2) {$\nabla \theta \cdot \bs{n} = 0$};
\end{tikzpicture}
    \caption{Boundary condition settings for Rayleigh-B\'{e}nard.}
    \label{fig:BC-RB}

\end{figure}

\begin{figure}[!t]
    \centering
        
    \begin{subfigure}[b]{0.36\textwidth}
        \centering
        \includegraphics[width=\linewidth,trim=700 290 450 290,clip]{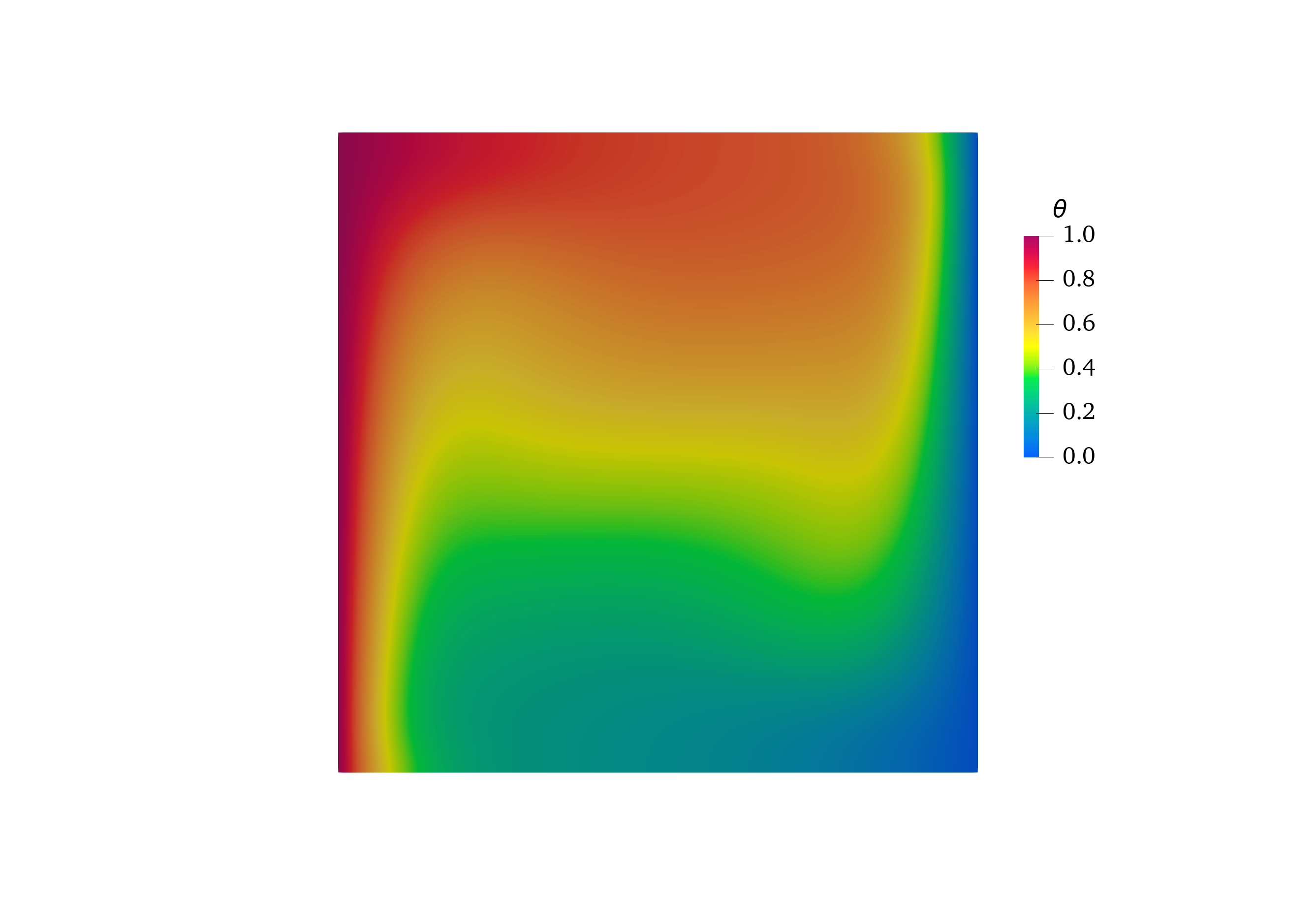}
        \caption{Temperature field ($\theta$) for Ra = $10^5$.}
    \end{subfigure}
    \hspace{0.05\textwidth} 
    \begin{subfigure}[b]{0.36\textwidth}
        \centering
        \includegraphics[width=\linewidth,trim=700 290 450 290,clip]{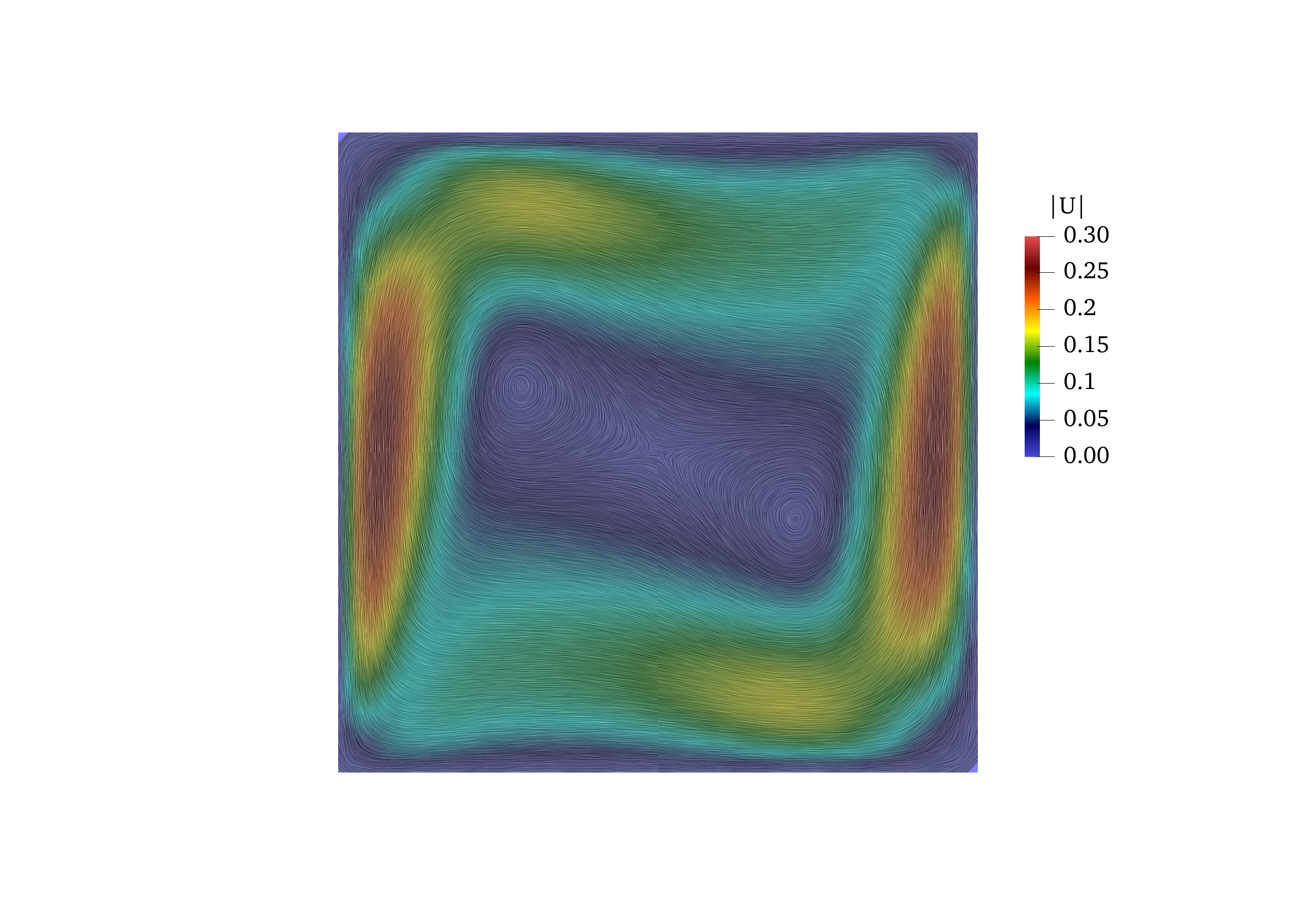}
        \caption{Velocity magnitude field ($|\mathbf{U}|$) for Ra = $10^5$.}
    \end{subfigure}
    
    \vspace{0.5cm} 
    \begin{subfigure}[b]{0.36\textwidth}
        \centering
        \includegraphics[width=\linewidth,trim=700 290 450 290,clip]{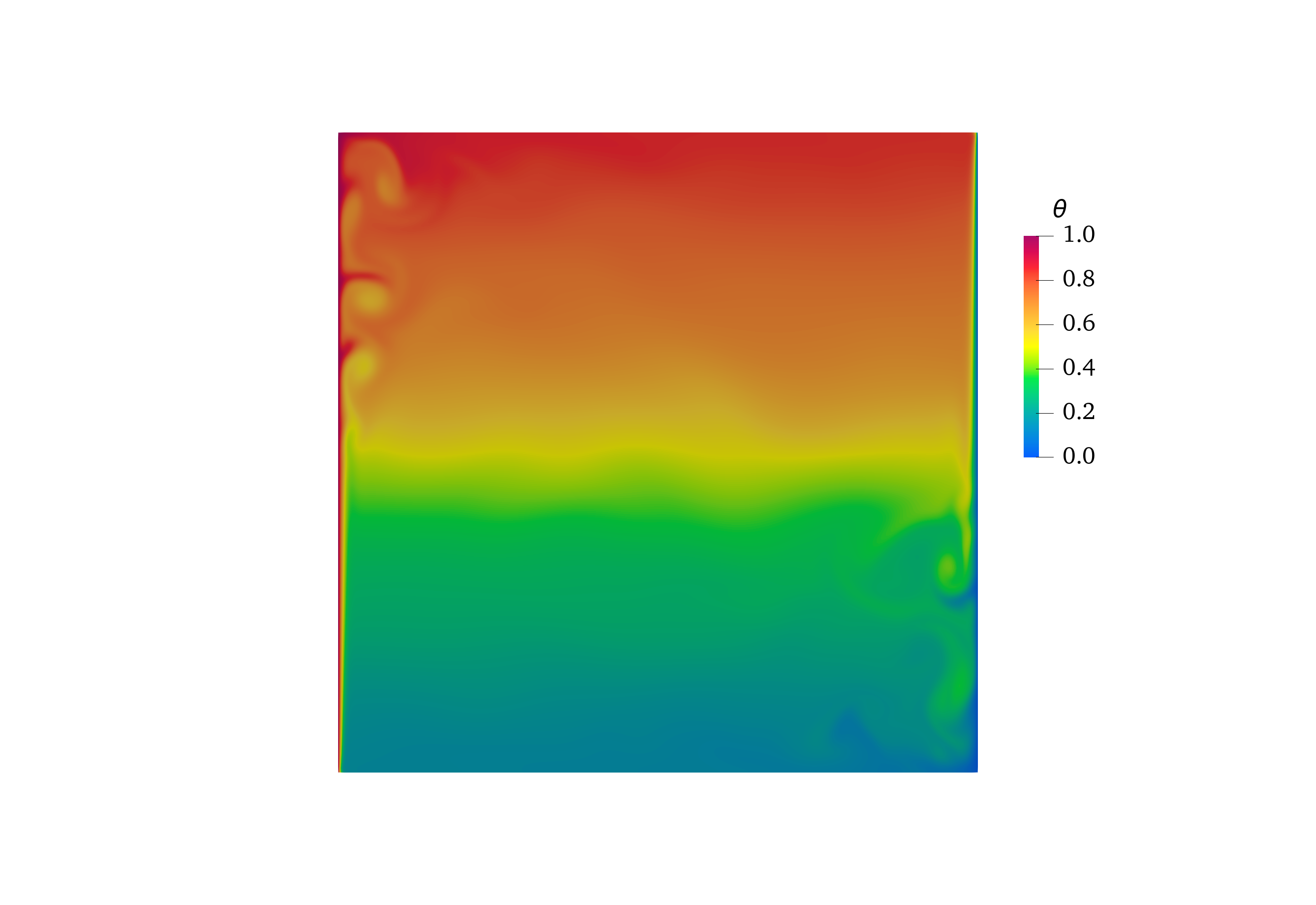}
        \caption{Temperature field ($\theta$) for Ra = $10^9$.}
    \end{subfigure}
    \hspace{0.05\textwidth} 
    \begin{subfigure}[b]{0.36\textwidth}
        \centering
        \includegraphics[width=\linewidth,trim=700 290 450 290,clip]{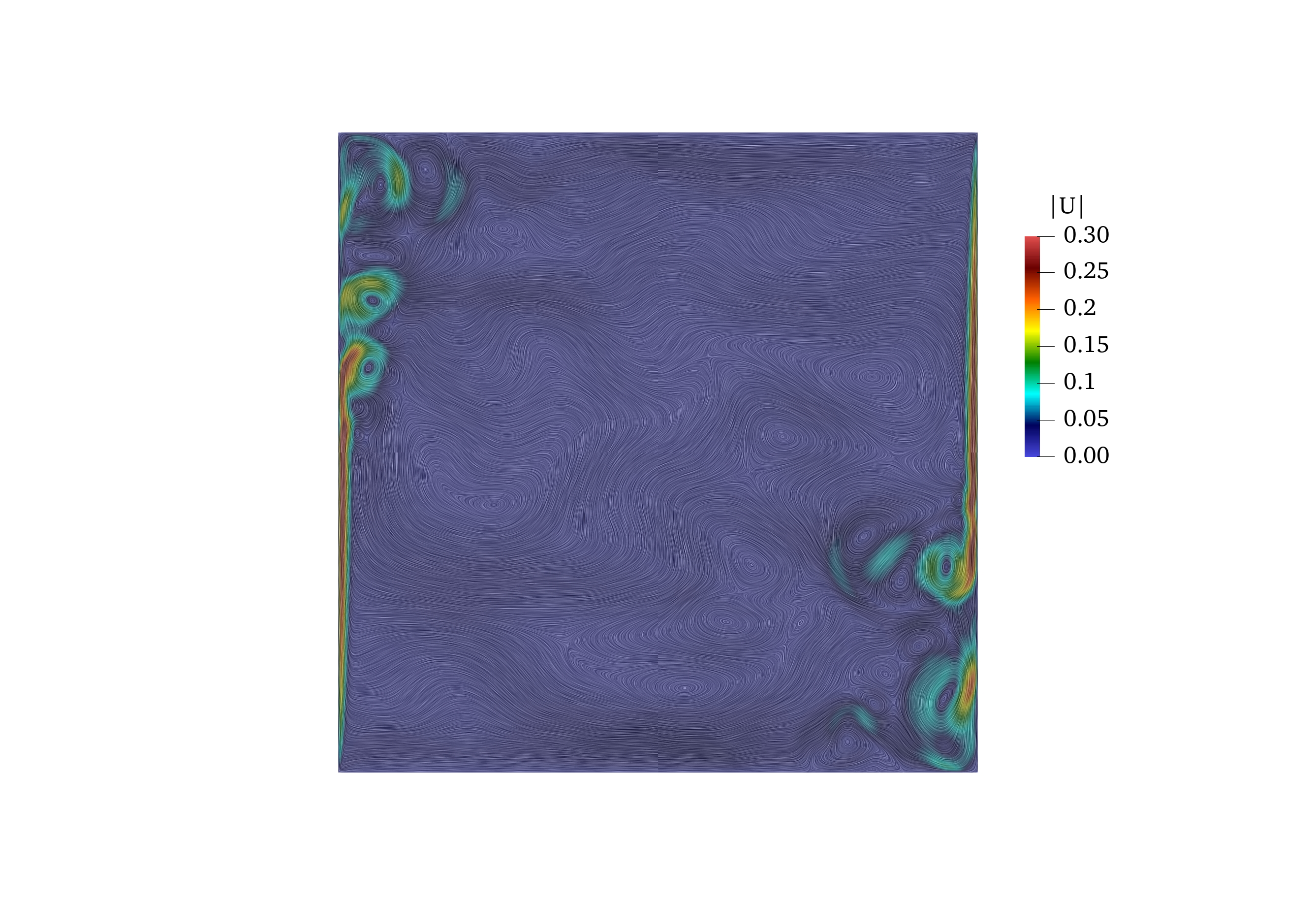}
        \caption{Velocity magnitude field ($|\mathbf{U}|$) for Ra = $10^9$.}
    \end{subfigure}

    \caption{Rayleigh-B\'{e}nard convection simulations for different Rayleigh numbers (Ra). The temperature field ($\theta$) is shown on the left and the velocity magnitude field ($|\mathbf{U}|$) is shown on the right. At lower Ra ($10^5$), the system exhibits steady convection, while at higher Ra ($10^9$), the system shows turbulent behavior with well-defined plumes and vortices.}
    \label{fig:RayleighBenard}
\end{figure}

 
\subsection{Rayleigh-B\'{e}nard convection problem}\label{sec:RB}

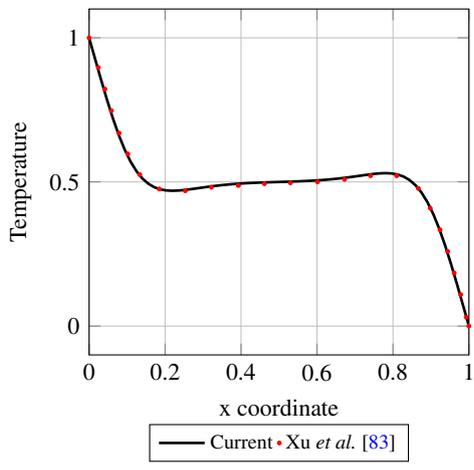
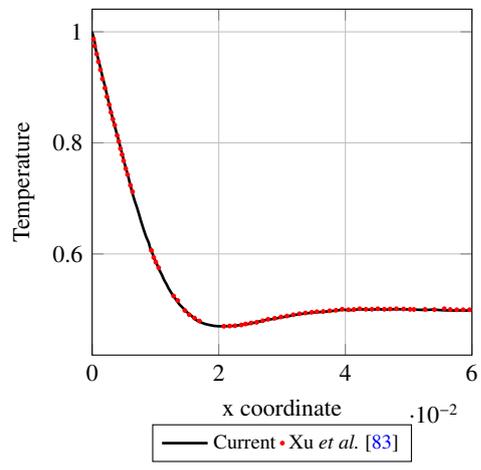
\begin{figure}[!t]
    \begin{subfigure}{0.5\textwidth}
        \begin{tikzpicture}
            \begin{axis}[
                width=0.8\linewidth, 
                height=0.75\linewidth,
                scaled y ticks=true,
                xlabel={x coordinate},
                ylabel={Temperature},
                legend entries={Current, Xu \textit{et al.}~\citep{xu2019residual}},
                legend style={at={(0.5,-0.2)},anchor=north, nodes={scale=0.85, transform shape}}, 
                legend columns=2,
                grid=both,
                grid style={line width=.1pt, draw=gray!10},
                major grid style={line width=.2pt,draw=gray!50},
                xmin=0,
                xmax=1,
                tick label style={font=\small},
                label style={font=\small},
                legend style={font=\small}
            ]
            \addplot[line width=1pt, mark=none] table [x={x},y={t},col sep=comma] {Ra1e5_Overleaf.csv};
            \addplot[red, only marks, mark size=0.7pt] table [x={x},y={t},col sep=comma] {SongZhe_Ra1e5_Overleaf.csv};
            \end{axis}
        \end{tikzpicture}
        \caption{$Ra = 10^5$}
        \label{fig:RB_Ra1e5}
    \end{subfigure}%
    \begin{subfigure}{0.5\textwidth}
        \begin{tikzpicture}
            \begin{axis}[
                width=0.8\linewidth, 
                height=0.75\linewidth,
                scaled y ticks=true,
                xlabel={x coordinate},
                ylabel={Temperature},
                legend entries={Current, Xu \textit{et al.}~\citep{xu2019residual}},
                legend style={at={(0.5,-0.2)},anchor=north, nodes={scale=0.85, transform shape}}, 
                legend columns=2,
                grid=both,
                grid style={line width=.1pt, draw=gray!10},
                major grid style={line width=.2pt,draw=gray!50},
                xmin=0,
                xmax=0.06,
                tick label style={font=\small},
                label style={font=\small},
                legend style={font=\small}
            ]
            \addplot[line width=1pt, mark=none] table [x={x},y={t},col sep=comma] {Ra1e9_Overleaf.csv};
            \addplot[red, only marks, mark size=0.7pt] table [x={x},y={t},col sep=comma] {SongZhe_Ra1e9_Overleaf.csv};
            \end{axis}
        \end{tikzpicture}
        \caption{$Ra = 10^9$}
        \label{fig:RB_Ra1e9}
    \end{subfigure}%
    \caption{Two-dimensional results for mean temperature profile.}
    \label{fig:RB_meanT}
\end{figure}

For the Rayleigh-B\'{e}nard convection problem (RB problem) with the boundary conditions showing in \figref{fig:BC-RB}, we run with two different Ra values: $10^5$ and $10^9$ with our {linear semi-implicit} Navier-Stokes framework. The temperature and velocity fields with LIC for $Ra = 10^5$ and $Ra = 10^9$ are shown in \figref{fig:RayleighBenard}. For $Ra = 10^5$, we keep our mesh size as $64 \times 64$; for $Ra = 10^9$, we keep our mesh size as $512 \times 512$, and we compare the mean temperature profile with literature~\citep{xu2019residual}. As seen in \figref{fig:RB_meanT}, we obtain excellent agreement with the literature.


\end{document}